# Discharge structure theory of highly electronegative plasma and its hierarchy and interdisciplinary meanings

Yuan-He Sun[1], Shu-Xia Zhao[1]*, and Yu Tian[2]

[1]School of Physics, Dalian University of Technology, Dalian, 116024, China

[2]Experimental training center, Dalian University of Science and Technology, Dalian, 116052, China

Correspondence: zhaonie@dlut.edu.cn

## Abstract

In this work the systematic theory of discharge structure is built for highly electronegative plasma, by means of self-consistent fluid model simulation that is based on the finite element method. The highly electronegative plasma is selected to be the inductively coupled Ar/$SF_6$ plasma source with 10% the reactive $SF_6$ concentration and in a pressure range of 10~90mTorr. The discharge structure is classified the transport dominated regime, transport and chemistry self-balanced regime and chemistry dominated regime. At low pressure of 10mTorr, the parabola feature of core plasma, stratification of whole discharge area, anion potential barrel, and the dipole and capacitor models of double layer characterize the discharge structure of transport dominated regime. At increasing the pressure, the recombination loss of ions becomes significant and the discharge structure is characterized by ellipse profile. Meanwhile, the regions of double layer and electropositive halo are strikingly shrunk, which means that the plasma of transport and chemistry self-balanced regime is a close system and probably do not need the shield of chamber anymore. The dimensional analysis shows the recombination can be transformed into drift flux, which balances the ambi-polar diffusion of plasma species. In the range of pressure considered, simulation shows astro-structures are inlayed in either the parabolic or elliptic profile. At observing the characteristics of the astro-structures, the self-coagulation theory and quasi-Helmholtz equation are built based on the free diffusion and negative chemical source. This chemistry dominated regime is defined as a tight type of self-balance since the inertia is lost automatically in the unsteady state continuity equations of anions after counteracting the diffusion and recombination. At high pressure of 90mTorr, the mixture of dispersed but closed ellipse background and astro-structure inlayed therein implies the wave-particle duality of quantum mechanics. Besides, the interdisciplinary meanings of discharge structure hierarchy of electronegative plasma with the astrophysics, geophysics, and nuclear and atomic physics are presented. The electrons that have negative chemical source in the periphery of astro-structure are self-coagulated at the collapse of ambi-polar diffusion potential that provides free diffusion component and the chemical potential is formed. The electron Boltzmann's balance does not exist anymore due to its weak self-coagulation behavior of periphery.

## I. Introduction

The low-pressure and radio frequency (RF) electronegative plasma sources are widely used in the Si-based material etching process and the functional thin film deposition [1,2]. In general, the capacitive [3] and inductive radio frequency sources are applied and the inductively coupled plasma [4,5] (abbreviated as the ICP) is characterized by its low-pressure discharge, high plasma density and simple device, etc. The discharge structure of plasma source at present characterizes the profiles of plasma species densities and the related theory illustrates the underlined physics and chemistry that give rise to the profiles. The physical process is about the ambi-polar diffusion and the chemistry is about the inelastic collisions of electron impact and the reactions of heavy species. Moreover, the collective interaction of plasma, the Boltzmann's balance and the Bohm's sheath criterion play important roles in forming the steady-state density profiles. Due to the attendance of anions in the discharge, the physical and chemical processes occurred in the electronegative plasma are complex and the discharge structure of it is hierarchical as compared to the electropositive plasma. The studies of discharge structure are important for they help people understand the mechanisms happened in the discharging plasmas and moreover for they open new frontiers for correlating certain disciplines that are presently isolated from each other, e.g., the low temperature plasma, astrophysics, nuclear and particle physics, quantum physics, and the geophysics.

The discharge structures of electronegative plasma have been extensively investigated from the last century [6-32, 35-44], among which many works are conducted through the theoretic analysis of simple fluid model. Based on the ambi-polar diffusion of highly electronegative plasma that consists of the electron, the cation, and the anion, the parabola discharge theory was established by Lieberman and Lichtenberg in Ref. [6]. Later in Ref. [7], they built the ellipse theory of electronegative plasma when the Boltzmann's balance is not reasonable for the anions and meanwhile constructed the flat-top structure of electronegative plasma when the electron density profile is not spatially uniform. In Ref. [8], they still found the internal sheaths existing in the electronegative discharges and the structures of these non-neutral regions varying significantly with both the ratio of anion and electron densities and the ratio between their temperatures. Besides, the stratification of electronegative plasma into the electronegative core and electropositive halo was recognized in Refs. [6-9], which is caused by the suppressing role of ambi-polar diffusion potential onto the anions due to the heavier mass of anions than the electrons. The terms of electronegative core and electropositive halo arise from the early studies of electronegative plasma in the positive column of normal glow discharge tube along the radial direction [10-13] and they are now used in the general electronegative plasma investigations that include the RF ones. At the approximation of electrical neutrality that is used in the fluid model, the transition region between the electronegative core and the electropositive halo is characterized by the swift anion density drop that is formed at the push of ambi-polar diffusion potential. When the Poisson's equation is included in the fluid model, the double layer structure that consists of two adjacent layers of charge with the different polarities was given by Sheridan at the interface of core and halo stratified and the condition of high electronegativity in Ref. [14]. The electronegative plasma that is stratified is hence called as the double-layer-stratified plasma. In addition, they found the stratification disappears when the electronegativity is higher due to the collapse of ambi-polar diffusion potential. Of more interest is that the double layer was accompanied by the oscillations of electropositive halo [8, 14] when the advection of ions is considered in the fluid model. This triggered the interests of many scholars, such as Chabert and Franklin in the Refs. [15,16]. Chabert *et al* utilized a kinetic model to investigate the double layer dynamics of electronegative plasma and generated the same discharge structures as Sheridan that had been presented by means of the fluid model, i.e., the stratified profile, the double-layer-stratified profile, and the uniform profile that is not stratified. Franklin *et al* concluded the oscillations that the double layer is accompanied with are artefacts since they disappear when the ion temperature effects are included in the fluid model. Besides for the numerical technique used to solve the simple fluid equations in Refs. [14, 16] and the kinetic model of Ref. [15], Kolobov and Economou [17, 18] utilized a semi-integer integral theory to solve the simple fluid equations and the structure of their calculated double layer is characterized by the swift drop of potential.

The above early studies of electronegative plasma structure are focused on the direct current glow discharges and meanwhile the plasmas are not self-consistently treated. For instance, in Refs. [6, 7, 11-13, 19, 20], the electropositive halo and electronegative core are first separately calculated and then matched via the transitional region. Moreover, in Refs. [9-13], the ion diffusion is neglected at the cold ion approximation and hence the flux of pure drift is assumed. As analyzed in Ref. [15, 16, 21], the inclusion of ion diffusion at finite ion temperatures smooths significantly the cliff border at the interface of core and halo stratified. Furthermore, in the above references, the self-consistent energy deposition process and the spatial variation of electron temperature are both omitted. It is shown later that the novel discharge structures of electronegative plasma other than the stratification are predicted by the completely self-consistent simulations that include both the particle and fluid models. In the RF electronegative plasma sources, such as the Ar/$CF_4$ capacitively coupled plasma (abbreviated as CCP), the famous drift-ambipolar (abbreviated as DA) discharge mode that is characterized by the strong bulk ionization and the striation discharge mode that the bulk plasma is striated were built By Schulze and Liu in Refs. [22, 23] through the particle simulation. It is noted that in the DA mode the electrons are coagulated and the density profile of them is a delta shape, which is different with the dispersed profiles of electrons in the $\alpha$ and $\gamma$ modes. Besides, our recent fluid simulations of Ar/$O_2$ ICP in Refs. [24, 25] revealed the delta profile as well, but for the anion not the electron. The delta structures of the Ar/$O_2$ ICP and the Ar/$CF_4$ CCP imply that the electronegative plasmas coagulate at the specific conditions and then the plasmas cannot be treated as the continuums anymore to some extent since the quasi- particles exist in the media. Another work of us characterizes the phenomena of the electronegative ICP coagulation in Ref. [26]. It is reported in this reference that a blue sheath surrounded the coagulated body and the coagulated structure can be dispersed at the expelling and Coulomb force. The blue sheath is formed by the advection of anions before they are chemically coagulated and it thereby consists of net charge with the electronegative polarity. This updates the contents of the Bohm's sheath theory. The scattering process between the separated coagulated bodies verifies the existence of particle model in the electronegative plasma. Moreover, the ambi-polar coagulation process of ions is constructed in Ref. [26], analogous to the ambi-polar diffusion process, and the electrons are hence decoupled from the ions and the ambi-polar diffusion potential collapses. It is shown in the Refs. [24-26] that the coagulated bodies are given by a self-coagulation theory that consists of free diffusion and purely negative chemical source term. The novel anion coagulated discharge structures of electronegative plasma are not achieved in the self-consistent simulations, such as in the Refs. [27-29], probably because either the anion kinetics of recombination and detachment [30-32] or the Poisson's equation that gives the ambi-polar diffusion potential [27, 33, 34] is not numerically addressed satisfyingly. On the contrast, although the bumped anion density structures are captured by the self-consistent simulations [35-37] they are more interpreted by the dynamics of physical coagulation that are given by the pressing role of ambi-polar potential, but are less interpreted by the spontaneous chemical coagulation type without the negative chemistry source and the self-generated chemical coagulation type with the negative chemistry source [26]. The coagulation of electrons in the DA mode of CCP is different to the anion coagulation of ICP due to the huge mass difference between the two species and the distinction between the two discharge reactors that respectively generate the capacitive and inductive RF plasmas. The investigation of electron coagulation [38-41] and its relation to the DA mode mechanism are out of the present article scope.

Besides for the above theoretical models and the self-consistent simulations, the experimental works are conducted to validate the classical discharge structures of electronegative plasma that are given in Refs. [6, 7]. In Ref. [42], the parabola profile in an asymmetric capacitively coupled oxygen plasma was measured by the photodetachment technique. In Ref. [43] both the parabola and flat-top structures of anion density in a symmetric oxygen discharge were obtained by means of the laser-induced photodetachment technique as well and in Ref. [44] the flat-top structure was obtained in the Ar/$CF_4$ mixture capacitively coupled plasma by means of the Langmuir probe based on the modified Druyvesteyn formula for the electronegative plasma. Both the structure and the concept of self-coagulation are newly introduced and the present experimental works are hence not conducted to validate them yet. In the analytical works of Ref. [6, 7], besides for the parabola structure given at the Boltzmann's balance of anions when neglecting the recombination loss, it is still mentioned when the spatial variation extent of electron density is larger than the cation, the profile of ions turns into the flat-top structure from the ellipse one. Nevertheless, our self-consistent fluid simulations in the $Ar/SF_6$ ICP revealed that the electron density profile is always more expansive than the ions due to the ambi-polar self-coagulation that decouples the electrons from the ions. Besides, the ellipse and flat-top profiles hold the same characteristics, i.e., the flat central part and the descending edge. Thus, we lump the flat-top structure into the ellipse structure and the above three classical discharge structures of electronegative plasma are reduced into the two formal types, parabola and ellipse, in this article.

In the present article, the systematic fluid simulations of $Ar/SF_6$ ICP reveal that the ratio of two times of the ionic recombination rate over the sum of generation rates for the ions that include the ionizations for cations and the attachments for anions somehow determines the discharge structures. Concretely, if (1) the recombination loss is negligible and the ratio tends to zero, the discharge profile is stratified as mentioned in Refs. [10-13]. The electronegative core is encompassed by the electropositive halo and the profile of core is parabolic as predicted by the theory in Refs. [6, 7]. At the interface of core and halo, the simulated double layer is modeled as the dipole microscopically and the capacitor macroscopically. The above discharging dynamics are occurred at the low pressure. As the ratio is increased by increasing the pressure the structure of electronegative core is evolved from the parabola to the ellipse and the trend is again predicted by the theory in Refs. [6, 7]. Meanwhile, the electropositive halo region is shrunk. If (2) the recombination loss is comparable to the generation of ions and the ratio tends to one, the stratification of profile disappears. The structure is solely determined by the ellipse profile and the electropositive halo is totally suppressed. After the profile transfers from the parabola to the ellipse, the ambi-polar diffusion potential collapses and the plasma is sustained by a relatively weak quasi-chemical potential, which is introduced due to the coexistence of the ambi-polar self-coagulation of ions in the RF-heating region near the coil and the monopolar self-coagulation of electrons away from the coil [26]. Since the weak chemical potential is unable to constrict the heavy anions, the stratification disappears, which is again in accord to the theoretic prediction of Sheridan in Ref. [14]. It is noted that the self-coagulation of peripherical electrons occurred in the fluid simulation of $Ar/SF_6$ ICP as reported here is different to the electron coagulation at the interface of sheath and bulk plasma in the DA mode of CCP of Ref. [22]. The self-coagulation of electrons in the ICP gives the quasi-chemical potential [26] and leads to the novel non-Boltzmann balance of electron statistics. Furthermore, if (3) the recombination loss rate is larger than the generation rate, i.e., the ratio is larger than one and the chemical source term is negative, the self-coagulated structure appears. The above three major discharging structures, i.e., the parabola profile with the stratification, the ellipse profile without the stratification, and the self-coagulation with the negative chemistry, correspond to the transport dominated regime of plasma, the transport and chemistry self-balanced regime of plasma, and the chemistry dominated regime of plasma, respectively. It is noted that in the third plasma regime the free diffusion and chemistry are self-balanced. Since the free diffusion is universal and not belonged to the plasma transport, this regime, even though accompanied by the free diffusion, is defined as the pure chemistry dominated regime. Besides, the self-coagulations cannot solely exist, but are imbedded in both the stratified parabolic profile of the low pressure and the unified ellipse profile of the high pressure. The coagulated body imbedded in the parabola at low pressure is analogous to the astro-structures, e.g., the white dwarf, the neutron star, and the fixed star that are formed from the self-balances of the universal gravitation with the electron degeneracy pressure [45], the neutron degeneracy pressure [46] and the radiation pressure [47, 48], respectively. Besides, the both coagulated and stratified discharge structure of low pressure is analogous to the earth structure that consists of the solid and liquid cores, the mantle, and the crust [49], which is helpful for people understand the origin of our earth. At high pressure, the blue sheath separates the self-balanced coagulated body and the self-balanced dispersed ellipse and the two self-balanced plasmas are treated as the particle and wave models of mass duality of quantum mechanics. Moreover, the coexistence of coagulated and dispersed plasmas is treated as the nuclear structure that consists of the neutron [50] and the proton [51]. The blue sheath is then treated as the mesotrons [52, 53] since the Debye's shielded potential of blue sheath that is generated from the collective interaction of plasma [54] corresponds well to the Yukawa's potential that is advised to introduce the short-range nuclear force [55, 56]. The shrunk double layer and electropositive halo are then treated as the extranuclear electrons of an atom model [57]. The structure of low-pressure electronegative plasma consists of the coagulated body, the parabola profile of electronegative core, the double layer, and the electropositive edge that carries the ambi-polar diffusion potential. It is an open system since there is mass exchange between the plasma and chamber wall, the surface loss of plasma species. The structure of high-pressure plasma consists of the both coagulated and dispersed plasmas and it is a close system since the two types of plasma are both self-balanced. The different levels of profile and model given by the fluid simulations at the low and high pressures compose the hierarchy of discharge structure of electronegative plasma and it is of interdisciplinary significance for the low temperature plasmas are the first time correlated to the celestial body model of astronomy, the wave-particle duality of quantum physics, the atomic and nuclear physics, and the geophysics. As analyzed, the correlations are natural since more than 99% universe masses are plasmas and they should include the celestial bodies and the micro-particles, and their existing states. Moreover, these new models given by the discharge structure hierarchy present the concrete characterizations for the wave-particle duality and the atomic structure and are hence helpful for solidifying the theoretic foundations for the modern physics disciplines that are more or less started from the experimental observations as well known, e.g., the diffraction of electrons of Davisson and Germer [58], the photoelectric effect of Hertz and Einstein [59, 60], the Compton's scattering effect [61, 62], and the measured black-body radiation spectrum from which the radiation formula of Planck is inferred [63-65].

This article is outlined as follows. The fluid model of ICP source, chemical reactions of highly electronegative Ar/$SF_6$ plasma, and the ICP reactor configuration are given in Sec. 2. The above three discharge regimes and the novel electron dynamics that deviate from the Boltzmann's balance are given in Sec. 3. The conclusion and further remarks are given in Sec. 4.

## II. Fluid model, plasma chemistry and reactor configuration

### (2.1) Fluid model formula

The formulae of fluid model used in the work are described in this section. They include the mass, momentum and energy equations of different plasma species, and the Poisson's and Maxwell's equations as well.

#### (2.1.1) Electrons equation

The equations of electron density and energy are given as follows.

$$\begin{aligned} &\frac{\partial n_e}{\partial t} + \nabla \cdot \mathbf{\Gamma}_e = R_e, \\ &\frac{\partial n_\varepsilon}{\partial t} + \nabla \cdot \mathbf{\Gamma}_\varepsilon + \mathbf{E} \cdot \mathbf{\Gamma}_e = R_\varepsilon + P_{ohm}. \end{aligned} \quad (1)$$

The fluxes of electron density and energy at the assumption of drift and diffusion are described in Eq. (2).

$$\begin{aligned} &\mathbf{\Gamma}_e = -(\boldsymbol{\mu}_e \cdot \mathbf{E}) n_e - \mathbf{D}_e \cdot \nabla n_e, \\ &\mathbf{\Gamma}_\varepsilon = -(\boldsymbol{\mu}_\varepsilon \cdot \mathbf{E}) n_\varepsilon - \mathbf{D}_\varepsilon \cdot \nabla n_\varepsilon. \end{aligned} \quad (2)$$

Herein, $n_e$ and $n_\varepsilon$ are the number density and energy density of electrons, respectively. $\boldsymbol{\mu}_e$ and $\boldsymbol{\mu}_\varepsilon$ are the electron mobility and electron energy mobility, respectively. $\mathbf{D}_e$ and $\mathbf{D}_\varepsilon$ are electron diffusivity and electron energy diffusivity, respectively. The relations among the above mass and energy transport coefficients are $\mathbf{D}_e = \boldsymbol{\mu}_e T_e$, $\mathbf{D}_\varepsilon = \boldsymbol{\mu}_\varepsilon T_e$, and $\boldsymbol{\mu}_\varepsilon = \frac{5}{3}\boldsymbol{\mu}_e$.

$R_e$ and $R_\varepsilon$ in Eq. (1) are the respective source terms of number density and energy density of electrons. Their expressions are stated in Eq. (3).

$$R_e = \sum_{j=1}^{M} l_j k_j \prod_{m=1}^{P} n_m^{\nu_{jm}},$$
$$R_\varepsilon = \sum_{j=1}^{M} \varepsilon_j l_j k_j \prod_{m=1}^{P} n_m^{\nu_{jm}}. \quad (3)$$

Herein, $l_j$ is the number of electrons created (lost) per electron-impact reaction that generates (depletes) electrons. $M$ is the number of these reactions. $k_j$ is the rate coefficient of reaction $j$, which are expressed in the Table 1. $n_m$ is the number density of reactant m of reaction $j$. $\nu$ is the stoichiometric coefficient of the reaction and $P$ is the number of reactants. $\varepsilon_j$ is the electron energy loss per electron-impact reaction.

$P_{ohm}$ is the deposited power density via the Ohm's heating scheme, illustrated in Eq. (4).

$$P_{ohm} = \frac{1}{2}\mathrm{Re}(\sigma |E_\theta|^2). (4)$$

Herein, $\sigma$ is the electron conductivity. $E_\theta$ is the radio frequency azimuthal field, which is calculated from the Maxwell's equation; see next the Sec. 2.1.3. Without the secondary electron emissions, the boundary conditions for the above equation are set in Eq. (5).

$$\mathbf{n}\cdot\Gamma_e = \frac{1-r_e}{1+r}\left(\frac{1}{2}v_{e,th}n_e\right),$$
$$\mathbf{n}\cdot\Gamma_\varepsilon = \frac{1-r_e}{1+r_e}\left(\frac{5}{6}v_{e,th}n_e\right). \quad (5)$$

Herein, $v_{e,th}$ is the thermal velocity of electrons. $r_e$ is the reflection coefficient of electrons from the reactor wall, which is set to 0.2 in this model.

### (2.1.2) Heavy species equation

Heavy species are supposed to comprise a reacting flow that consists of k= 1,2,...,Q species, except for the electron. The mass transport equations of these heavy species are summarized in Eq. (6).

$$\rho\frac{\partial w_k}{\partial t} = \nabla\cdot\mathbf{j}_k + R_k. (6)$$

Herein, $\rho$ is the total mass density of heavy species and $w_k$ is the mass fraction of species $k$. $\mathbf{j}_k$

is the diffusive and drift flux of species $k$ and expressed in Eq. (7).

$$\mathbf{j}_k = \rho w_k \mathbf{V}_k, \quad \mathbf{V}_k = D_{k,m} \nabla \ln(w_k) - z_k \mu_m \mathbf{E}. \quad (7)$$

Herein, $\mathbf{V}_k$ is the velocity of species $k$, $z_k$ is elementary charge number that species $k$ carries, $\mu_m$ is the mobility, and $\mathbf{E}$ is the electrostatic field vector, calculated from the Poisson's equation (see next). It is noted that the multi-component mass diffusion is considered and $D_{k,m}$ is the averaged diffusion coefficient of mixture, expressed in Eq. (8).

$$D_{k,m} = \frac{1-w_k}{\sum_{j \neq k}^{Q} x_j / D_{kj}}. (8)$$

Herein, $x_j$ is the number fraction of species $j$ and $D_{k,j}$ is just the Chapman-Enskog *binary diffusion* coefficient [66, 67].

The source term of Eq. (6), $R_k$, is expressed in Eq. (9).

$$R_k = M_k \sum_{j=1}^{N} l_{k,j} r_j. (9)$$

Herein, $M_k$ is the molecular weight, $r_j$ is the rate of reaction j that creates or consumes species $k$, $N$ is the reaction number, and $l_{k,j}$ is the particle number of species $k$ created or consumed for each reaction $j$. The reaction rate, $r_j$, is expressed in Eq. (10).

$$r_j = k_j \prod_{m=1}^{S} c_m^{\nu_{jm}}. (10)$$

Herein, $k_j$ is the rate coefficient, also given in Tab. 1, $S$ is the number of reactants, $\nu_{j,m}$ is the stoichiometric coefficient of reaction $j$ with the reactant $m$ and the reactant species $k$, and $c_m$ is the molar concentration of reactant $m$.

In sum, the heavy species equation, described in Eq. (6), sequentially describes the inertia term, diffusion, drift and chemical kinetic of heavy species. It is noted the inertia term considered herein is the inertia of mass, like the electron equation of Eq. (1) (not the flux inertia) which we believe is one possible origin of self-coagulation happened in radio frequency plasma sources, i.e., the inertial effect of density (see next Sec. 4.3). Besides, only Q-2 equations are used, since the mass fractions of feedstock gases of mixture, i.e., Ar and $SF_6$, are governed by the mass constraint condition, $\omega_{\mathrm{Ar,SF_6}} = 1 - \sum_{k}^{Q-2} \omega_k$, and meanwhile the assigned gas ratio between them.

The total mass density of heavy species, $\rho$, is obtained from the ideal gas law in Eq. (11).

$$\rho = \frac{P}{kT} \cdot \frac{M}{N_A}. (11)$$

Herein, $k$ is the Boltzmann's constant, $T$ is gas temperature, equal to 300K, $P$ is the fixed gas pressure, set as 10mTorr, and $N_A$ is the Avogadro's constant. $M$ is the mole averaged molecular weight, expressed in Eq. (12).

$$\frac{1}{M} = \sum_{k=1}^{Q} \frac{w_k}{M_k}. (12)$$

The mean molecular weight $M$ is generally not a constant, since it is a function of mass fractions and molecular weight of various species and the mass fractions, calculated from the Eq. (6), are both spatially and temporally varied.

The total flux boundary condition that includes both the diffusion and drift components is used onto the chamber wall, i.e., $\Gamma_k = -\mathbf{n} \cdot \rho \omega_k \mathbf{V}_k$, where the surface reaction kinetics of all species listed in Tab. 2 are considered.

### (2.1.3) Electromagnetic equation

To describe the electromagnetic field in the reactor, the Maxwell's equations are combined to express the Ampere's law in Eq. (13).

$$(j\omega\sigma - \omega^2 \varepsilon_0 \varepsilon_r)\mathbf{A} + \nabla \times (\mu_0^{-1} \mu_r^{-1} \nabla \times \mathbf{A}) = \mathbf{J}_a. \quad (13)$$

Herein, $j$ is the imaginary unit and $\omega$ is the angular frequency of power source, expressed as $2\pi f$ at $f = 13.56\mathrm{MHz}$. $\varepsilon_0$ and $\varepsilon_r$ are the vacuum permittivity and the relative permittivity of dielectric window material (quartz), respectively. $\mu_0$ and $\mu_r$ are the vacuum permeability and the relative permeability of coil that is made of copper, respectively. $\mathbf{A}$ is the magnetic vector potential, from which both the radio frequency (RF) magnetic and electric fields are calculated via the Coulomb's gauge, i.e., $\mathbf{B} = \nabla \times \mathbf{A}, \mathbf{E} = -\frac{\partial \mathbf{A}}{\partial t}$. $\mathbf{J}_a$ is the applied external coil current density and will be persistently varied in the simulation until the deposited total power approaches to the assigned value, i.e., 300W. When considering the azimuthal symmetry, only the azimuthal component of RF electric field, $E_\theta$, and the axial and radial components of RF magnetic field, $B_r, B_z$, need to be addressed. $\sigma$ is the electron conductivity, expressed in Eq. (14). It is given by analytically solving the Langevin's equation at the assumption of zero electron temperature (~ cold plasma), i.e., neglecting the diffusion [68].

$$\sigma = \frac{n_e q^2}{m_e(\nu_e + j\omega)}. \quad (14)$$

Herein, $n_e$, $m_e$ and $q$ are the number density, mass, and charge of electron, separately. $\nu_e$ is the elastic collision frequency of electrons with the neutral species. The magnetic insulation, i.e., $\mathbf{n} \times \mathbf{A} = 0$, is taken as the boundary condition for solving the Maxwell's equation.

### (2.1.4) Electrostatic equation

The Poisson's equation is used to calculate electrostatic field in Eq. (15).

$$\begin{aligned} \mathbf{E} &= -\nabla V, \\ \nabla \cdot \mathbf{D} &= \rho_V. \end{aligned} \quad (15)$$

Herein, $\mathbf{E}$ is the electrostatic field, $V$ is the potential, and $\rho_V$ is the charge density of space, separately. Zero potential boundary condition is used at both the chamber wall and the dielectric window underneath surface, since the Poisson's equation only needs to be solved in the discharge chamber. Our previous simulation of pure argon inductive discharge revealed the approximation of grounded dielectric window potential does not significantly change the simulation results, and the reasonability of this approximation is extended to the present article. As mentioned before, only the axial and radial electrostatic field components are considered at the assumption of azimuthal symmetry.

### (2.2) Chemistry of Ar/SF$_6$ plasma

The Ar/SF$_6$ gas-phase chemistry and surface kinetics are given in Tabs. 1 and 2. The electron-impact elastic collision, excitation and deexcitation, ionization, direct attachment and dissociative attachment, and dissociation are included. The heavy species reaction types considered are neutral and ionic recombination, detachment, Penning ionization and charge exchange. The rate coefficients and cross sections can be found in Refs. [69-72]. The surface kinetics of species considered in Tab. 2 include recombination and de-excitation.

**Table 1.** Ar/SF$_6$ chemical reaction set considered in the model

| **No.** | **Reaction** | **Rate coefficient[a]** | **Threshold (eV)** | **Ref.** |
|---|---|---|---|---|
| | Elastic collisions | | | |
| 1 | $e + Ar \rightarrow e + Ar$ | Cross Section | 0 | [69] |
| 2 | $e + SF_6 \rightarrow e + SF_6$ | Cross Section | 0 | [69] |
| 3 | $e + F_2 \rightarrow e + F_2$ | Cross Section | 0 | [69] |
| 4 | $e + F \rightarrow e + F$ | Cross Section | 0 | [69] |
| | Excitation and deexcitation reactions | | | |
| 5 | $e + Ar \rightarrow e + Ar^*$ | Cross Section | 11.6 | [69] |
| 6 | $e + Ar^* \rightarrow e + Ar$ | Cross Section | -11.6 | [69] |
| | Ionization reactions | | | |
| 7 | $e + Ar \rightarrow 2e + Ar^+$ | Cross Section | 15.76 | [69] |
| 8 | $e + Ars \rightarrow 2e + Ar^+$ | Cross Section | 4.43 | [69] |
| 9 | $e + SF_6 \rightarrow SF_5^+ + F + 2e$ | $1.2\times10^{-7}\exp(-18.1/T_e)$ | 16 | [70, 71] |
| 10 | $e + SF_6 \rightarrow SF_4^+ + 2F + 2e$ | $8.4\times10^{-9}\exp(-19.9/T_e)$ | 20 | [70, 71] |
| 11 | $e + SF_6 \rightarrow SF_3^+ + 3F + 2e$ | $3.2\times10^{-8}\exp(-20.7/T_e)$ | 20.5 | [70, 71] |
| 12 | $e + SF_6 \rightarrow SF_2^+ + F_2 + 2F + 2e$ | $7.6\times10^{-9}\exp(-24.4/T_e)$ | 28 | [70, 71] |
| 13 | $e + SF_6 \rightarrow SF^+ + F_2 + 3F + 2e$ | $1.2\times10^{-8}\exp(-26.0/T_e)$ | 37.5 | [70, 71] |
| 14 | $e + SF_6 \rightarrow F^+ + SF_4 + F + 2e$ | $1.2\times10^{-8}\exp(-31.7/T_e)$ | 29 | [70, 71] |
| 15 | $e + SF_6 \rightarrow S^+ + 4F + F_2 + 2e$ | $1.4\times10^{-8}\exp(-39.9/T_e)$ | 18 | [70, 71] |
| 16 | $e + SF_5 \rightarrow SF_5^+ + 2e$ | $1.0\times10^{-7}\exp(-17.8/T_e)$ | 11 | [70, 71] |
| 17 | $e + SF_5 \rightarrow SF_4^+ + F + 2e$ | $9.4\times10^{-8}\exp(-22.8/T_e)$ | 15 | [70, 71] |
| 18 | $e + SF_4 \rightarrow SF_4^+ + 2e$ | $4.77\times10^{-8}\exp(-16.35/T_e)$ | 13 | [70, 71] |
| 19 | $e + SF_4 \rightarrow SF_3^+ + F + 2e$ | $5.31\times10^{-8}\exp(-17.67/T_e)$ | 14.5 | [70, 71] |
| 20 | $e + SF_3 \rightarrow SF_3^+ + 2e$ | $1.0\times10^{-7}\exp(-18.9/T_e)$ | 11 | [70, 71] |
| 21 | $e + F \rightarrow F^+ + 2e$ | $1.3\times10^{-8}\exp(-16.5/T_e)$ | 15 | [70, 71] |
| 22 | $e + S \rightarrow S^+ + 2e$ | $1.6\times10^{-7}\exp(-13.3/T_e)$ | 10 | [70, 71] |
| 23 | $e + F_2 \rightarrow F_2^+ + 2e$ | $1.37\times10^{-8}\exp(-20.7/T_e)$ | 15.69 | [70, 71] |
| | Attachment and dissociative attachment reactions | | | |
| 24 | $e + SF_6 \rightarrow SF_6^-$ | Cross Section | 0 | [69] |
| 25 | $e + SF_6 \rightarrow SF_5^- + F$ | Cross Section | 0.1 | [69] |
| 26 | $e + SF_6 \rightarrow SF_4^- + 2F$ | Cross Section | 5.4 | [69] |
| 27 | $e + SF_6 \rightarrow SF_3^- + 3F$ | Cross Section | 11.2 | [69] |
| 28 | $e + SF_6 \rightarrow SF_2^- + 4F$ | Cross Section | 12 | [69] |
| 29 | $e + SF_6 \rightarrow F^- + SF_5$ | Cross Section | 2.9 | [69] |
| 30 | $e + SF_6 \rightarrow F_2^- + SF_4$ | Cross Section | 5.4 | [69] |
| 31 | $e + F_2 \rightarrow F^- + F$ | Cross Section | 0 | [69] |
| | <u>Dissociation reactions</u> | | | |
| 32 | $e + SF_6 \rightarrow SF_5 + F + e$ | $1.5\times10^{-7}\exp\left(-8.1/T_e\right)$ | 9.6 | [70, 71] |
| 33 | $e + SF_6 \rightarrow SF_4 + 2F + e$ | $9.0\times10^{-9}\exp\left(-13.4/T_e\right)$ | 12.4 | [70, 71] |
| 34 | $e + SF_6 \rightarrow SF_3 + 3F + e$ | $2.5\times10^{-8}\exp\left(-33.5/T_e\right)$ | 16 | [70, 71] |
| 35 | $e + SF_6 \rightarrow SF_2 + F_2 + 2F + e$ | $2.3\times10^{-8}\exp\left(-33.9/T_e\right)$ | 18.6 | [70, 71] |
| 36 | $e + SF_6 \rightarrow SF + F_2 + 3F + e$ | $1.5\times10^{-9}\exp\left(-26.0/T_e\right)$ | 22.7 | [70, 71] |
| 37 | $e + SF_5 \rightarrow SF_4 + F + e$ | $1.5\times10^{-7}\exp\left(-9.0/T_e\right)$ | 5 | [70, 71] |
| 38 | $e + SF_4 \rightarrow SF_3 + F + e$ | $6.2\times10^{-8}\exp\left(-9.0/T_e\right)$ | 8.5 | [70, 71] |
| 39 | $e + SF_3 \rightarrow SF_2 + F + e$ | $8.6\times10^{-8}\exp\left(-9.0/T_e\right)$ | 5 | [70, 71] |
| 40 | $e + SF_2 \rightarrow SF + F + e$ | $4.5\times10^{-8}\exp\left(-9.0/T_e\right)$ | 8 | [70, 71] |
| 41 | $e + SF \rightarrow S + F + e$ | $6.2\times10^{-8}\exp\left(-9.0/T_e\right)$ | 7.9 | [70, 71] |
| 42 | $e + F_2 \rightarrow 2F + e$ | $1.2\times10^{-8}\exp\left(-5.8/T_e\right)$ | 1.6 | [70, 71] |
| | <u>Neutral / neutral recombination reactions</u> | | | |
| 43 | $S + F \rightarrow SF$ | $2\times10^{-16}$ | 0 | [70, 71] |
| 44 | $SF + F \rightarrow SF_2$ | $2.9\times10^{-14}$ | 0 | [70, 71] |
| 45 | $SF_2 + F \rightarrow SF_3$ | $2.6\times10^{-12}$ | 0 | [70, 71] |
| 46 | $SF_3 + F \rightarrow SF_4$ | $1.6\times10^{-11}$ | 0 | [70, 71] |
| 47 | $SF_4 + F \rightarrow SF_5$ | $1.7\times10^{-11}$ | 0 | [70, 71] |
| 48 | $SF_5 + F \rightarrow SF_6$ | $1.0\times10^{-11}$ | 0 | [70, 71] |
| 49 | $SF_3 + SF_3 \rightarrow SF_2 + SF_4$ | $2.5\times10^{-11}$ | 0 | [70, 71] |
| 50 | $SF_5 + SF_5 \rightarrow SF_4 + SF_6$ | $2.5\times10^{-11}$ | 0 | [70, 71] |
| 51 | $SF + SF \rightarrow S + SF_2$ | $2.5\times10^{-11}$ | 0 | [70, 71] |
| 52 | $SF_x + F_2 \rightarrow SF_{x+1} + F$ [b] | $7.0\times10^{-15}$ | 0 | [70, 71] |
| | Ion / ion recombination reactions | | | |
| 53 | $X^+ + Y^- \rightarrow X + Y$ [c] | $5.0\times10^{-9}$ | 0 | [70, 71] |
| | Detachment reactions | | | |
| 54 | $Z + Y^- \rightarrow Z + Y + e$ [d] | $5.27\times10^{-14}$ | 0 | [70, 71] |
| | Other reactions | | | |
| 55 | $Ars + Ars \rightarrow e + Ar + Ar^+$ | $6.2\times10^{-10}$ | 0 | [70, 71] |
| 56 | $Ars + Ar \rightarrow Ar + Ar$ | $3.0\times10^{-15}$ | 0 | [70, 71] |
| 57 | $Ar^+ + SF_6 \rightarrow SF_5^+ + F + Ar$ | $9.0\times10^{-10}$ | 0 | [70, 71] |
| 58 | $SF_5^+ + SF_6 \rightarrow SF_3^+ + SF_6 + F_2$ | $6.0\times10^{-12}$ | 0 | [70, 71] |

[a] The unit of the rate coefficient is $cm^3s^{-1}$.

[b] $x$ stands for the number 1-5.

[c] X = $SF_5$, $SF_4$, $SF_3$, $SF_2$, SF, F, S or $F_2$ and Y = $SF_6$, $SF_5$, $SF_4$, $SF_3$, $SF_2$, F or $F_2$.

[d] Z = $SF_6$, $SF_5$, $SF_4$, $SF_3$, $SF_2$, SF, F, S or $F_2$ and Y = $SF_6$, $SF_5$, $SF_4$, $SF_3$, $SF_2$, F or $F_2$.

**Table 2.** Ar/$SF_6$ Surface reaction set considered in the model

| **No.** | **Surface reaction** | **Sticking coefficient** | **Ref.** |
|---|---|---|---|
| 1 | $SF_x^+ + wall \rightarrow SF_x$ ; $x = 1-5$ | 1 | [70, 71] |
| 2 | $F^+ + wall \rightarrow F$ | 1 | [70, 71] |
| 3 | $F_2^+ + wall \rightarrow F_2$ | 1 | [70, 71] |
| 4 | $S^+ + wall \rightarrow S$ | 1 | [70, 71] |
| 5 | $F + wall \rightarrow 1/2\,F_2$ | 0.02 | [72] |
| 6 | $Ar^+ + wall \rightarrow Ar$ | 1 | [70, 71] |
| 7 | $Ars + wall \rightarrow Ar$ | 1 | [70, 71] |

### (2.3) ICP reactor configuration

The reactor used consists of the vacuum chamber (also called as matching box) and the discharge chamber, which are separated by the dielectric window. The discharge chamber is 15 cm in radius and 13 cm in height. The dielectric window and vacuum chamber hold the same radius as the discharge chamber, and the heights of them are 1 cm and 3 cm, respectively. A substrate with the radius of 12 cm and the height of 4 cm is seated at the bottom center of discharge chamber. A two-turn coil is installed above the dielectric window, with the radial locations of 8 cm and 10 cm, respectively. The coil is square in cross section, with the side-length of 0.6 cm. More detail about the reactor can be found in Ref. [73].

## III. Results and analysis

### (3.1) Transport dominated regime

### (3.1.1) Parabola theory

In the parabola theory, the ambi-polar diffusion process is adopted. It is arisen from the electropositive plasma that consists of electron and ion. The electronegative plasma is made of cation, anion and electron and it is a triple-species system. Here, the ambi-polar diffusion process of cation, anion and electron is constructed for the transport of electronegative plasma.

Whatever the transport scheme is, in the bulk plasma, the electrical neutrality is required. So, the flux equilibrium and charge density equilibrium exist in the electronegative plasma. In the Eqs. (16, 17), the flux and density balances between the cation, anion and electron are expressed. Herein, $\Gamma_+$, $\Gamma_-$ and $\Gamma_e$ are the fluxes of cation, anion and electron, respectively. $n_+$, $n_-$ and $n_e$ are the densities of cation, anion and electron, respectively. The electronegativity $\alpha$ is given in Eq. (18), defined as the ratio of anion density over the electron density. Next, the fluxes of cation, anion and electron calculated at the drift and diffusion approximation of momentum balance equation are given in Eqs. (19-21). Herein, $D_+$, $D_-$ and $D_e$ are the diffusion coefficients of cation, anion and electron, respectively. $\mu_+$, $\mu_-$ and $\mu_e$ are the mobilities of cation, anion and electron, respectively. The electric field $E$ in the drift term is electrostatic and determined by the ambi-polar diffusion potential gradient of electronegative plasma. These previous six equations, i.e., Eqs. (16-21), are correlated and the flux of cation is accordingly rewritten as a function of the quantities, $\mu_+$, $\mu_-$, $\mu_e$, $D_+, D_-, D_e$, $n_+, n_-, n_e$ and $\alpha$, illustrated in Eq. (22). Like the electropositive plasma, the ambi-polar diffusion coefficient of electronegative plasma, $D_{a+}$, is introduced in Eq. (23). The cation flux is therefore re-expressed as a product of this newly introduced coefficient and the cation density gradient, illustrated in Eq. (24).

$$\Gamma_+ = \Gamma_- + \Gamma_e, \quad (16)$$

$$n_+ = n_- + n_e, \quad (17)$$

$$\alpha = n_- / n_e, \quad (18)$$

$$\Gamma_+ = -D_+ \nabla n_+ + n_+ \mu_+ E, \quad (19)$$

$$\Gamma_- = -D_- \nabla n_- - n_- \mu_- E, \quad (20)$$

$$\Gamma_e = -D_e \nabla n_e - n_e \mu_e E, \quad (21)$$

$$\Gamma_+ = -\frac{(\mu_e + \mu_- \alpha) D_+ + \mu_+ (1+\alpha) D_e (\nabla n_e / \nabla n_+) + \mu_+ (1+\alpha) D_- (\nabla n_- / \nabla n_+)}{\mu_e + \mu_- \alpha + \mu_+ (1+\alpha)} \nabla n_+, \quad (22)$$

$$D_{a+} = \frac{(\mu_e + \mu_- \alpha) D_+ + \mu_+ (1+\alpha) D_e (\nabla n_e / \nabla n_+) + \mu_+ (1+\alpha) D_- (\nabla n_- / \nabla n_+)}{\mu_e + \mu_- \alpha + \mu_+ (1+\alpha)}, \quad (23)$$

$$\Gamma_+ = -D_{a+} \nabla n_+ . \quad (24)$$

As seen above, the deduced ambi-polar diffusion coefficient of electronegative plasma in Eq. (23) is quite complicated. So, reasonable approximations need to be introduced to reduce its complexity and more clear physics can thereby emerge. To arrive at this goal, the significant difference between the electron and anion temperatures is first noticed, which characterizes the non-thermal equilibrium plasma that is generated through the radio frequency glow discharge. In Eq. (25), the parameter, $\gamma$, is introduced, which is defined as the ratio of electron temperature over the anion temperature. Herein, $T_e$ and $T_i$ are the electron and anion temperatures, respectively. It is noticed that the anion and cation are in the thermal equilibrium and hence their temperatures are the same. Then, the Boltzmann's balance is adopted for both the electron and anion and a relation between the electron and anion densities is therefore found, as shown in Eq. (26), which is expressed by means of the relative changes of anion and electron densities and the $\gamma$ parameter. The gradient operator is added onto the electrical neutrality condition, in Eq. (27). Utilizing the Eqs. (26, 27), the ratios between the gradients of electron density and cation density and between the gradients of anion density and cation density are obtained in Eq. (28), sequentially. As seen, the ratios of these gradients of densities are all expressed as a function of $\gamma$ and $\alpha$. Furthermore, the correlations between the diffusion coefficients and mobilities of different species in Eq. (29), based on the Einstein's formula, are noticed. Lastly, at the helps of Eqs. (28, 29), the ambi-polar diffusion coefficient of electronegative plasma of Eq. (23) is significantly simplified in Eq. (30).

$$\gamma = T_e / T_i, \quad (25)$$

$$\frac{\nabla n_-}{n_-} = \gamma \frac{\nabla n_e}{n_e}, \quad (26)$$

$$\nabla n_+ = \nabla n_- + \nabla n_e, \quad (27)$$

$$\frac{\nabla n_e}{\nabla n_+} = \frac{1}{1+\gamma\alpha}, \frac{\nabla n_-}{\nabla n_+} = \frac{\gamma\alpha}{1+\gamma\alpha}, \quad (28)$$

$$\frac{D_-}{D_+} = \frac{\mu_-}{\mu_+}, \frac{D_e}{D_+} = \gamma \frac{\mu_e}{\mu_+}, \quad (29)$$

$$D_{a+} = D_+ \frac{(1+\gamma+2\gamma\alpha)\left(1+\alpha\frac{\mu_-}{\mu_e}\right)}{(1+\gamma\alpha)\left(1+\frac{\mu_+}{\mu_e}(1+\alpha)+\frac{\mu_-}{\mu_e}\right)}. \quad (30)$$

After the present simplification, the ambi-polar diffusion coefficient is now a function of quantities, $D_+$, $\alpha$, $\gamma$, and the mobilities, $\mu_+, \mu_-, \mu_e$. At noticing the significant difference between the mobilities of the heavy cation and anion and the electron's mobility that is illustrated in Eq. (31), the coefficient can be further simplified in Eq. (32). Moreover, if the electronegativity $\alpha$ is rather high, which is just the assumed condition of this article, it is again simplified to be a constant, $2D_+$, as illustrated in Eq. (33). As seen next, the constant ambi-polar diffusion coefficient value, which is obtained after a set of simplifications, is very helpful for us to analyze the transport of electronegative plasma and the structure that it then forms.

$$\mu_- / \mu_e, \mu_+ / \mu_e \ll 1, \quad (31)$$

$$D_{a+} \cong D_+ \frac{1+\gamma+2\gamma\alpha}{1+\gamma\alpha}, \quad (32)$$

$$\alpha \gg 1, D_{a+} \cong 2D_+. \quad (33)$$

Still, at the Boltzmann's balances of both the electron and anion, a relation between the anion and electron densities (not their gradients anymore) is obtained, illustrated in Eq. (34). Herein, $n_{e0}$ and $n_{-0}$ are the fixed densities of electron and anion at the center, respectively. In the non-thermal equilibrium plasma, the electron temperature $T_e$ is high, with an amplitude of several electron Volts; but the anion temperature $T_i$ is low, which is about hundreds of Kelvins. So, the value of $\gamma$ is in an order of 100. At such a high value of $\gamma$, the term at the right side of Eq. (34), $\left(\frac{n_-}{n_{-0}}\right)^{1/\gamma}$, tends to one and then the electron density at the left side becomes spatially unchangeable. This process is expressed in Eq. (35).

$$\frac{n_e}{n_{e0}} = \left(\frac{n_-}{n_{-0}}\right)^{1/\gamma}, (34)$$

$$\gamma \cong 100.0, \left(\frac{n_-}{n_{-0}}\right)^{1/\gamma} \cong 1.0, n_e \cong n_{e0}. (35)$$

In Eq. (36), the continuity equation (i.e., the mass balance equation) of cation is given. At the left side of equation, the flux is expressed as a product of the ambi-polar diffusion coefficient and the density gradient of cation. At the right side of equation, the chemical sources include the ionization reactions for the creation of cation and the recombination reactions for the depletion of cation. Herein, $n_0$ is the density of target neutral atom for the electron collisions. $K_{iz}$ and $K_{rec}$ are the rate coefficients of the ionization and recombination reactions, respectively. In Eq. (37), the continuity equation is significantly simplified as follows. Firstly, at the left side of continuity equation, the ambi-polar diffusion coefficient that is originally a function of electronegativity is replaced by the constant value given above in Eq. (33), $2D_+$. Then, at the right side of continuity equation and in the ionization reaction term, the electron density that is originally spatially varied is replaced by the constant value given above in Eq. (35), $n_{e0}$. Furthermore, the recombination loss term for the cation still at the right side of equation is neglected, assuming that the influence of recombination is insignificant. The simplified continuity equation of cation in Eq. (37) after the set of above operations can be analytically solved, which is a parabola function as shown in Eq. (38). Herein, $\alpha_0$ defined in Eq. (39) is the electronegativity of center and $l$ is the nominal position where the electronegativity is zero and the anions disappear. Besides, by utilizing the Eq. (17), we can reform the cation density function in Eq. (38) into an electronegativity function in Eq. (40). As seen, the $\alpha(x)$ function that is deduced from the parabola theory can express the densities of both the cation and anion, in an assistance of the constant electron density (see further the Eqs. (68)), and it will be compared to the same function of ellipse theory in Fig. 9.

$$-\frac{d}{dx}\left(D_{a+}(\alpha)\frac{dn_+}{dx}\right) = K_{iz}n_0n_e - K_{rec}n_+n_-, (36)$$

$$-2D_+ \frac{d^2 n_+}{dx^2} = K_{iz} n_0 n_{e0}, \ (37)$$

$$\frac{n_+}{n_{e0}} = \alpha_0 \left(1 - \frac{x^2}{l^2}\right) + 1, \ (38)$$

$$\alpha_0 = n_{-0} / n_{e0}, \ (39)$$

$$\alpha = \alpha_0 \left(1 - \frac{x^2}{l^2}\right). \ (40)$$

### (3.1.2) Simulated parabola profiles by fluid model

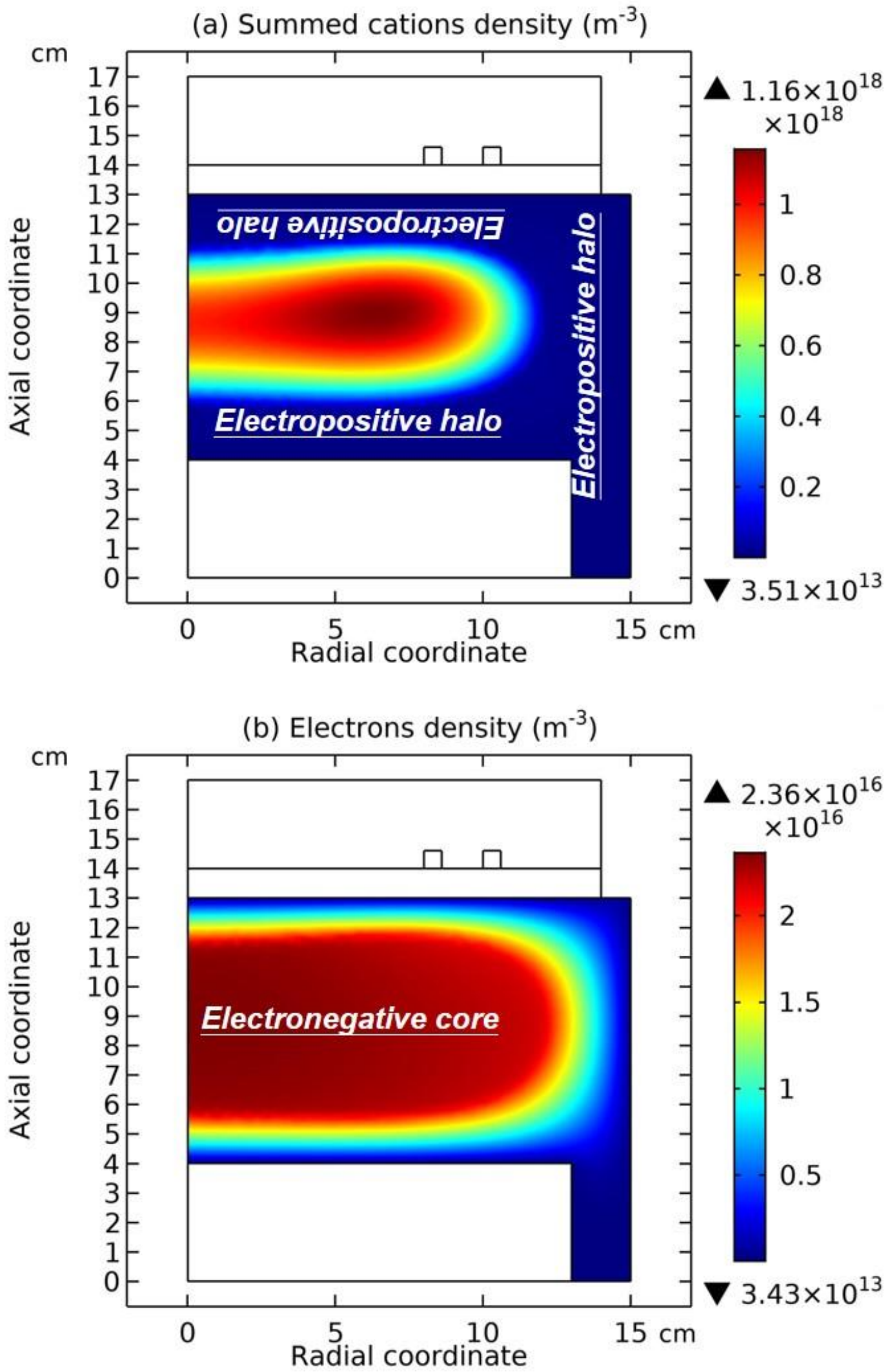

Figure 1. Simulated summed cations density (a) and electrons density (b) two-dimensional profiles by fluid model. The discharge conditions are 300W, 10mTorr and 10% $SF_6$ content.

In Fig. 1, the simulated summed cations density and electrons density two dimensional profiles given by the fluid model are presented. The discharge conditions are 300W, 10mTorr and 10% $SF_6$ content.

In Fig. 2, the corresponding axial and radial density profiles of the cations, anions (also summed) and electrons are presented. It is seen from the Figs. 1(a) and 2(a) that the whole density profile is stratified into the electronegative core and electropositive halo. In the electronegative core, the axial profiles of summed cations and anions are both parabolic due to the requirement of plasma neutrality. They are compared to the Fig. 3(b) where a parabolic function is constructed based on the two data points that are sampled from the simulated density curve of cations. In Fig. 1(b), the electrons density of electronegative core is flattened and it decreases in the electropositive halo. The constant electron density simulated in the core agrees with the prediction of Eq. (35) of parabola theory that is arisen from the assumptions that the electron and anion are in the Boltzmann's balance. Since the electron density in the core is constant, the electric potential of core plasma is flattened as well in an order of Volt, as determined by the Boltzmann's balance of electron. However, in the Fig. 4 of sub-Sec. (3.1.3), it will be shown when reducing the potential order from the Volt to the Kelvin, the weight ambi-polar diffusion potential with the certain distribution emerges. In addition, it is seen from the Figs. 1(a) and 2(b), along the radial profile the cations and anions densities are not monotonic and the bump appears in the core. As seen next, the bump is caused by the self-coagulated behavior that will be reported in Sec. (3.3). Besides, since the simulated bump appears transversely at the edge of electronegative core, as seen from Fig. 1(a), it does not affect the parabolic profiles of the cation and anion in Fig. 2(a) that are longitudinally plotted along the discharge axis.

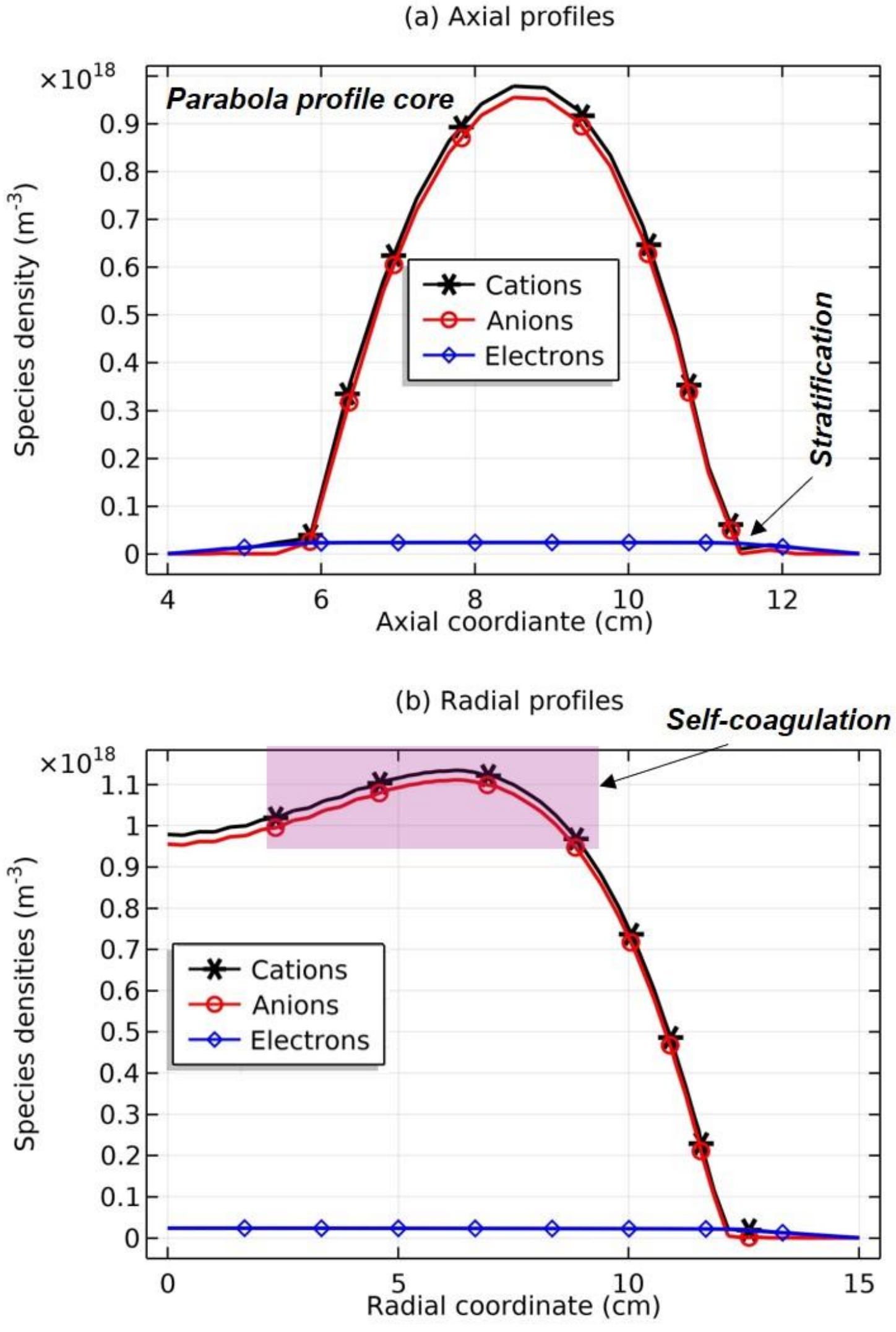

Figure 2. Simulated axial (a) and radial (b) profiles of the cations, anions, and electrons densities by the fluid model. The discharge conditions are 300W, 10mTorr and 10% $SF_6$ content.

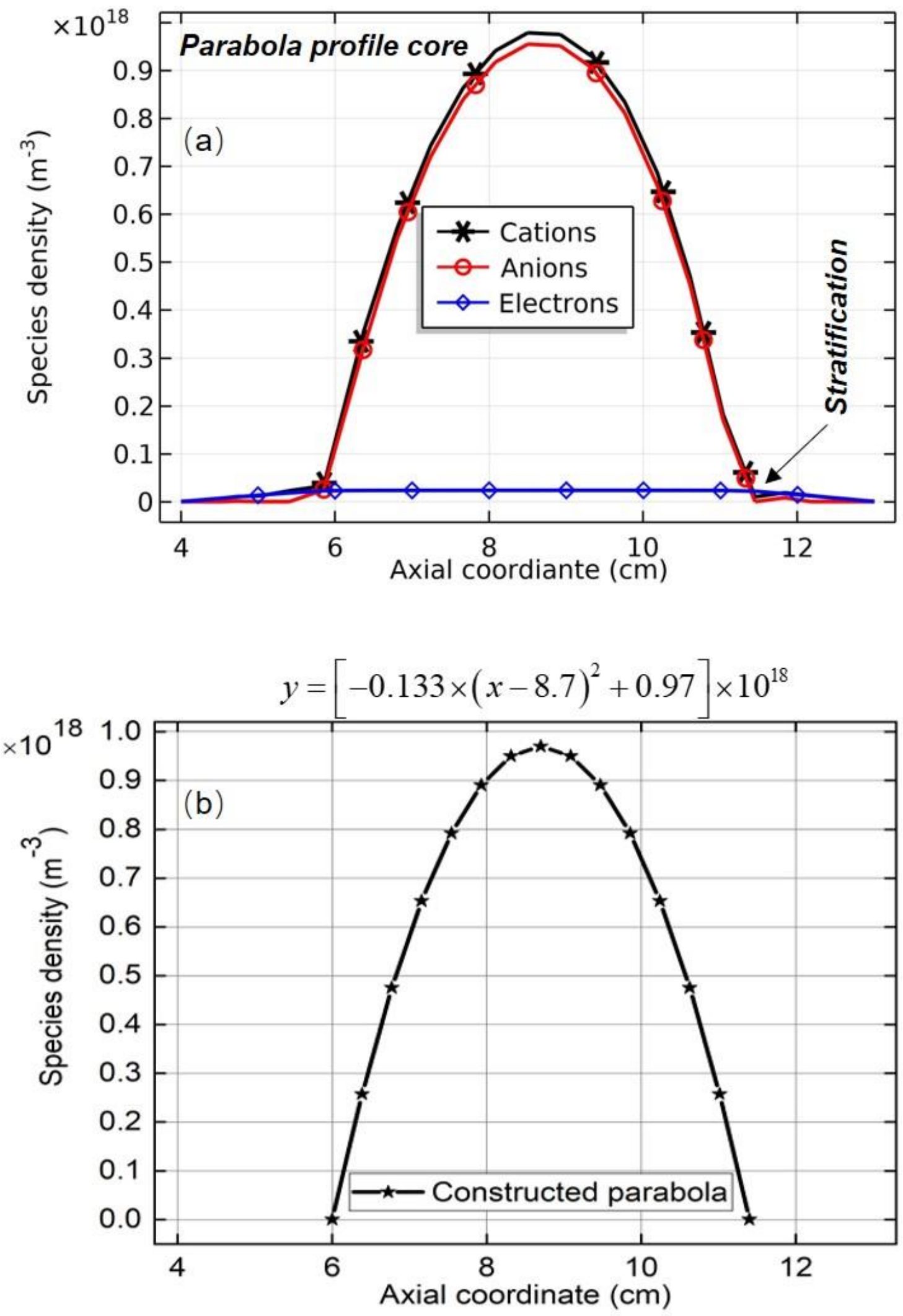

Figure 3. (a) Simulated axial profiles of species density and (b) constructed parabola function based on two critical points from the simulated cations density curve, i.e., the peaked point with its coordinates, $(8.7, 0.97\times10^{18})$, and the truncated close-zero point with its coordinates, $(11.4, 0)$.

### (3.1.3) Theory of weighted ambi-polar diffusion potential

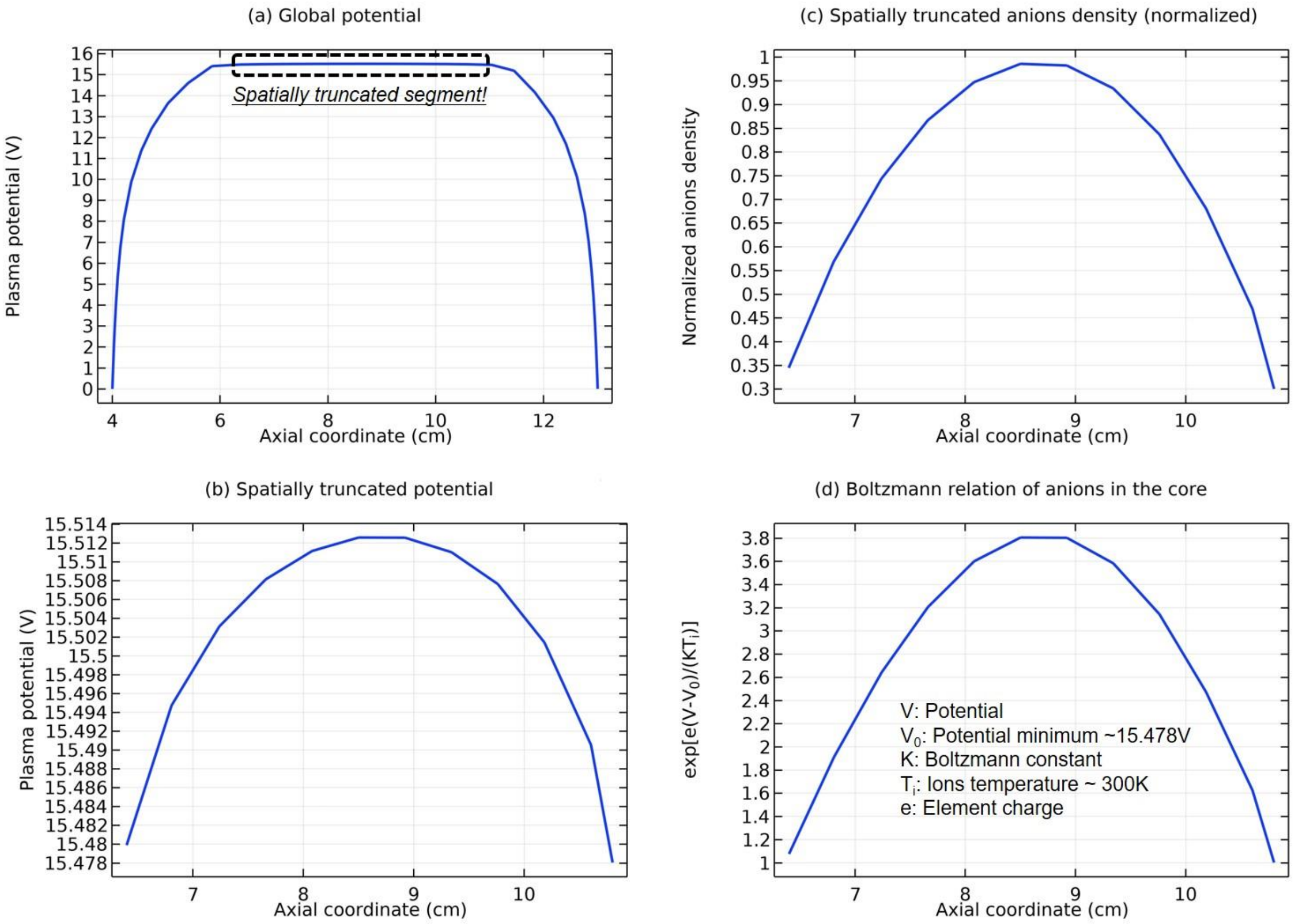

Figure 4. Simulated global and axial electrical potential of plasma (a), zoom in exhibition of spatially truncated plasma potential that is originally flattened (b), spatially truncated and normalized anions density of the same path (c) and the Boltzmann's relation of anions in this path (d).

As mentioned before, since the electron density is in the Boltzmann's balance, the electrical potential of plasma that is plotted globally and axially in Fig. 4(a) has the similar distribution as the electron density in Fig. 1(b), i.e., spatially unchangeable in the core and continually decreasing in the halo, but in a large potential unit, Volt, as noticed from the vertical axis. As zoomed in the potential and seen from the Figs. 4(a,b), along a spatially truncated path the originally constant electrical potential in the core now has definite profile, in a small range of potential from $15.478\,\mathrm{V}$ to $15.514\mathrm{V}$. In Fig. 4(c), the normalized anions density in the path is plotted, i.e., $\frac{n_-}{n_{-0}}$. In Fig. 4(d), a certain function, $\exp\left[e\left(V-V_0\right)/KT_i\right]$, is constructed, based on the tiny variation of potential and the anion's room temperature assumed, i.e., $T_i = 300\,\mathrm{K}$. Herein, $V_0$ is the potential minimum, $K$ is the Boltzmann's constant, and $e$ is the unit electron charge, which are also illustrated in Fig. 4(d). Upon comparing the Figs. 4(c,d), the similarity between the two profiles indicates that the constructed formula imitates the Boltzmann's relation of anions, i.e., $\frac{n_-}{n_{-0}} \sim \exp\left[e\left(V-V_0\right)/KT_i\right]$.

The Boltzmann's balance of anions herein given by the fluid simulation validates the base of parabola theory, i.e., the anions can indeed be assumed in the Boltzmann's balance, which has been already utilized in the Eqs. (26, 28, 34). The discrepancy between the absolute magnitudes of normalized anions density and constructed formula are ascribed to the fact that the tiny potential variation is relative and it is seated on top of the large potential barrel of electropositive halo region as illustrated in Fig. 4(a). The increment of the large potential barrel is about $15.0\,\mathrm{V}$. The function of the Boltzmann's relation is nonlinear and hence the influence of major potential cannot be deducted directly as the linear functions do.

It is known from the electromagnetics that the potential and energy are exchangeable through the electron charge. So, this tiny variation of potential in that path of Fig. 4(b) can be described by the thermodynamic unit, Kelvin. The concrete process is $\Delta = 15.514 - 15.478 = 0.036\,\mathrm{V}$ and $0.036\,\mathrm{eV} = \frac{0.036\times1.6\times10^{-19}}{1.38\times10^{-23}} \approx 417\,\mathrm{K}$. In this formula, the symbol, $\mathrm{eV}$, is called electron volt. It is an energy unit and responsible for turning the potential into the energy. The electron charge, $1\,e = 1.6\times10^{-19}\,\mathrm{C}$, and the Boltzmann constant, $K: 1.38\times10^{-23}\,\mathrm{J/K}$, are then utilized in the formula, which gives rise to the total variation of potential in the selected path of core, about $417\,\mathrm{K}$. The tiny potential barrel of electronegative core in a room temperature range is caused by the ambi-polar diffusion process of electronegative plasma. In the electropositive plasma, the ambi-polar potential is high, around ten Volts, to stop light electrons freely diffusing from the plasma and hence maintain the electrical neutrality of plasma. In the highly electronegative plasma, the ambi-polar potential is low, around several hundreds of Kelvins, since it is used to stop a mixture of small light electrons and many heavy anions escaping from the plasma. Hence, this tiny potential is defined as a weighted ambi-polar diffusion potential. Besides, regarding the room temperature property of potential that is analogous to the anions temperature, it is also called as *anion potential*, which has been revealed by other analytic theory [17]. Accordingly, the ambi-polar diffusion potential in the electropositive halo is called as *electron potential*.

### (3.1.4) Theories of double layer

### (a) Dipole model

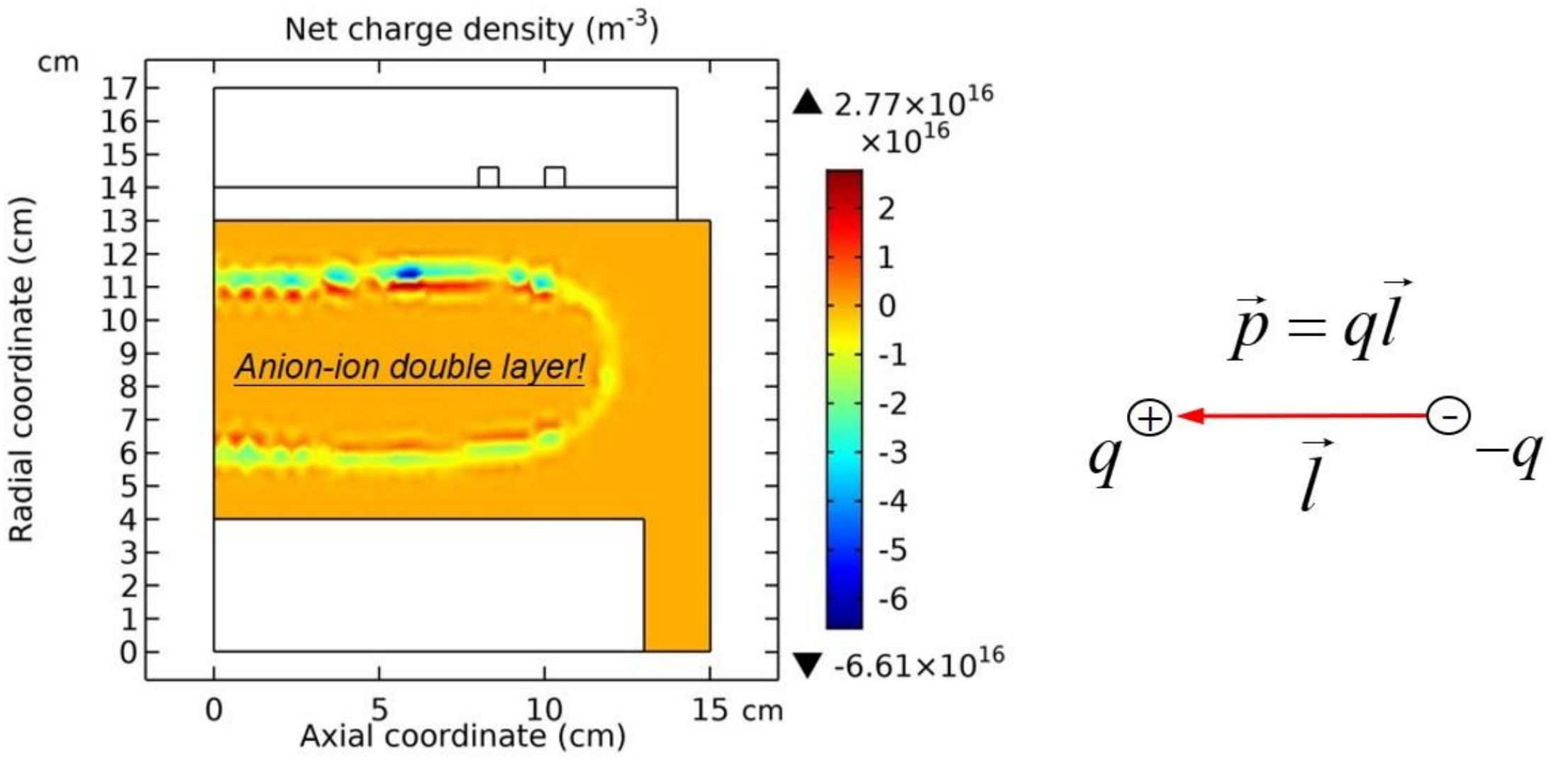

(a) (b)

Figure 5. (a) Simulated net charge density two-dimensional profile by fluid model and (b) transformed dipole model of double layer structure that is exhibited in the left panel. The discharge conditions are 300W, 10mTorr and 10% $SF_6$ content.

In Fig. 5(a), the simulated net charge density two-dimensional profile by fluid model is shown. Upon comparing Fig. 5(a) to Fig. 1, it is seen that at the interface of electronegative core plasma and electropositive halo plasma, a double layer structure, i.e., one layer of positive charge adjacent to one layer of negative charge, is appeared. In Fig. 5(b), this double layer is modelled as a dipole along the direction of plasma transport, i.e., from the core to halo. Besides, it is seen from the figure that the two layers of charge are very close to each other. So, in Fig. 6, the spatial distribution of electric field intensity of such a dipole at the limit of dipole distance tending to zero is given. It is seen that the field is strong at the dipole center but tends to zero at other points. The concrete mathematic process is presented in Eqs. (41-44). In both the Fig. 6 and the Eqs. (41-44), the point, $O$, is the dipole center and the point, $A$, is an arbitrary *far-field* point of the space. $E_O, E_A$ are the dipole fields of points, $O, A$, respectively. Herein, $l$ is the distance of dipole moment and $q$ is the dipole charge. Hence, $\vec{p} = q\vec{l}$ and it is the dipole moment. $r$ is the distance of dipole center to the point, $A$, and $\vec{r}_0$ is the corresponding unit vector. $\varepsilon_0$ is the vacuum permittivity. The field intensities at *far-field* points tend to zero because of the counteracting effect of positive and negative charges while the field intensity at the dipole center is infinitely large because of the localization effect, at the limit of $l \to 0$.

$$E_O = \frac{2q}{\pi \varepsilon_0 l^2}, (41)$$

$$\lim_{l\to 0} E_O = \lim_{l\to 0} \frac{2q}{\pi\varepsilon_0 l^2} = \infty, (42)$$

$$\vec{E}_A = \frac{1}{4\pi\varepsilon_0 r^3}\left[-\vec{p} + 3\left(\vec{r}_0 \cdot \vec{p}\right)\vec{r}_0\right], (43)$$

$$l \to 0,\ \vec{p} = q\vec{l} \to \vec{0}, \Rightarrow \vec{E}_A \sim \vec{0}. (44)$$

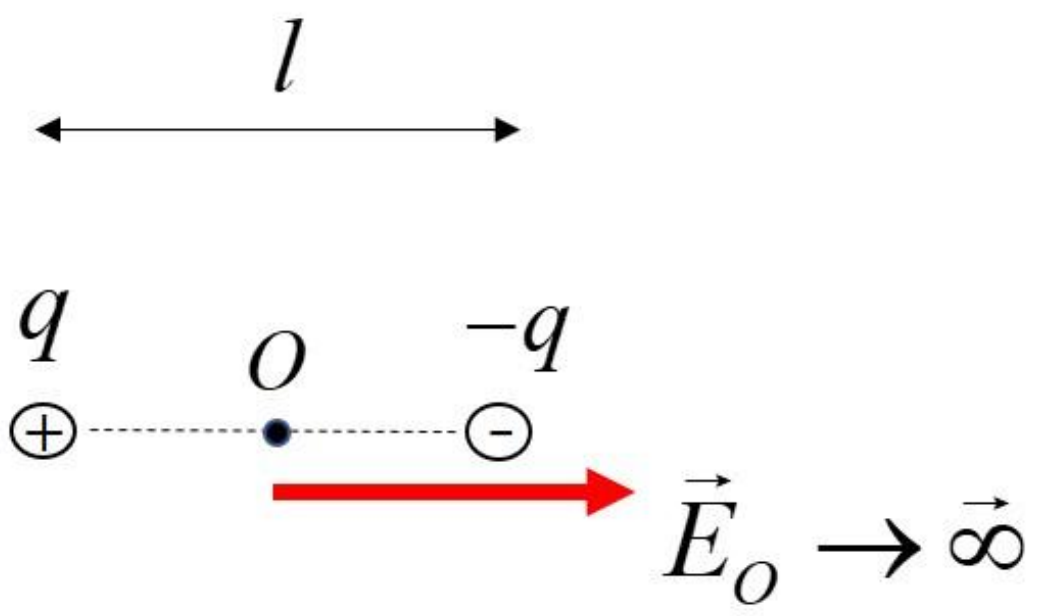

(a) Electric field intensity of dipole moment at the dipole center

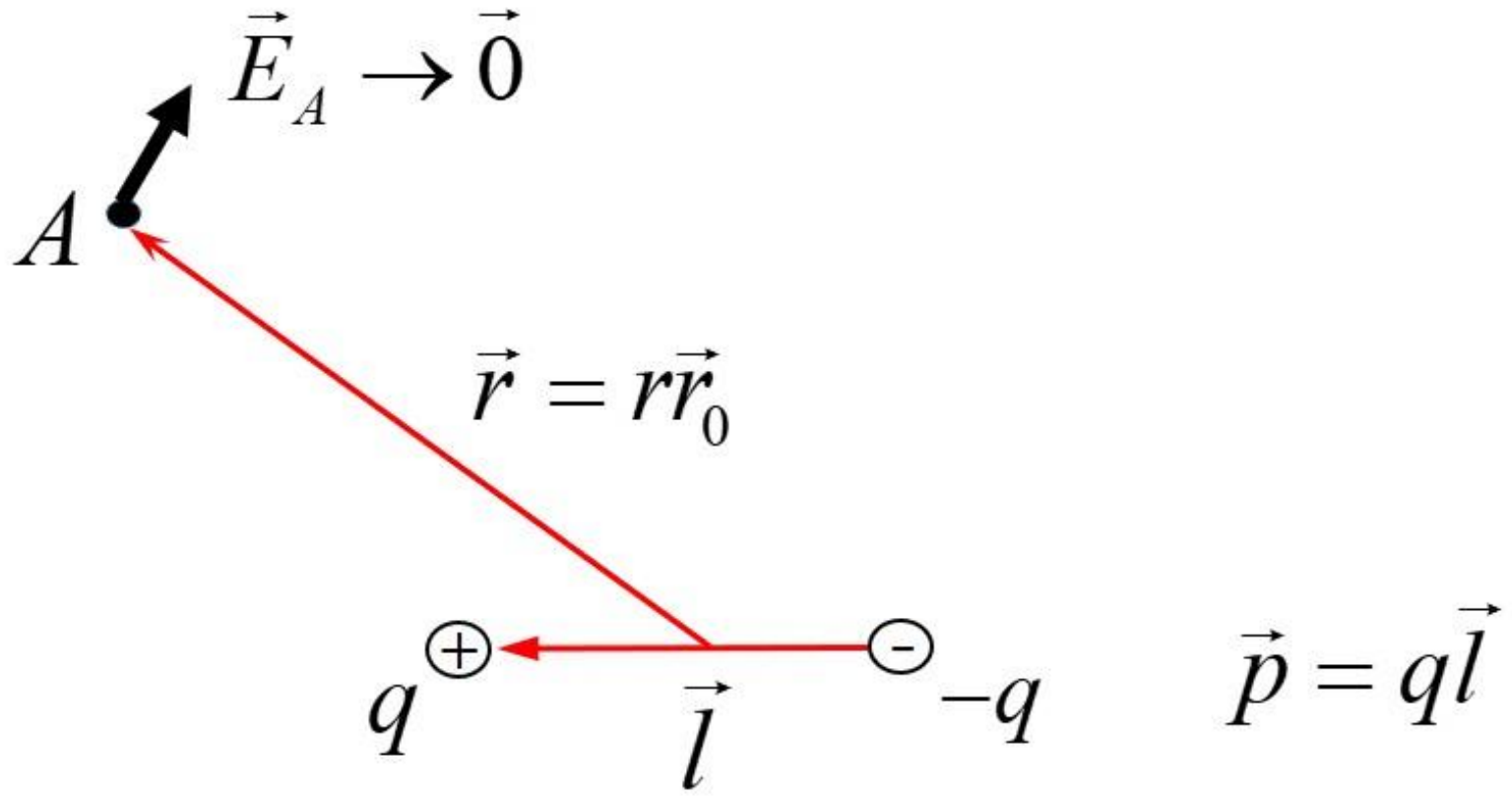

(b) Electric field intensity of dipole moment at arbitrary *far-field* point of the space

Figure 6. Spatial distribution of electric field intensity of transformed dipole model at the limit of $l \to 0$ at (a) dipole center and (b) arbitrary *far-field* point. Herein, $l$ is the distance of dipole moment.

### (b) Capacitor model

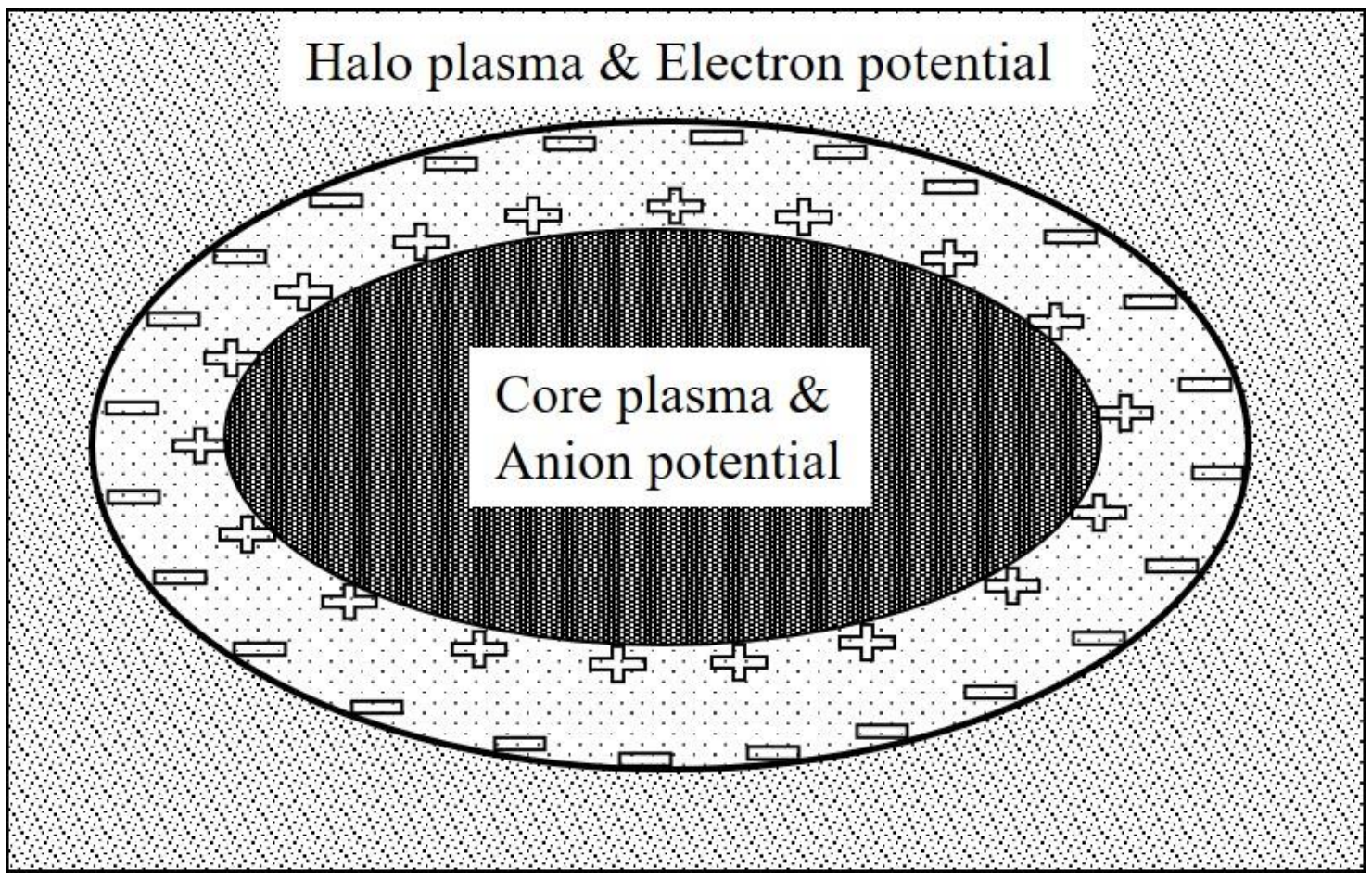

Figure 7. Transformed capacitor model of double layer

Considering the three-dimensional geometry of cylindrical chamber and the cylindric symmetry, the double layer when seen along the whole interface of electronegative core plasma and electropositive halo plasma can be transformed to a capacitor of the ellipsoid, as shown in Fig. 7. Different to the CCP source that is radio frequency vibrated, the transport of ICP source is more like a direct current (DC) plasma source in the two-dimensional radial and axial directions. The capacitor is known to be able to block the direct current, which means that the electropositive halo is separated from the electronegative core, as seen from Fig. 8 where a transformed circuit of such ICP source is given. The stratification of the $Ar/SF_6$ ICP through the capacitor model of double layer is formed by the external power source, as illustrated in Fig. 8.

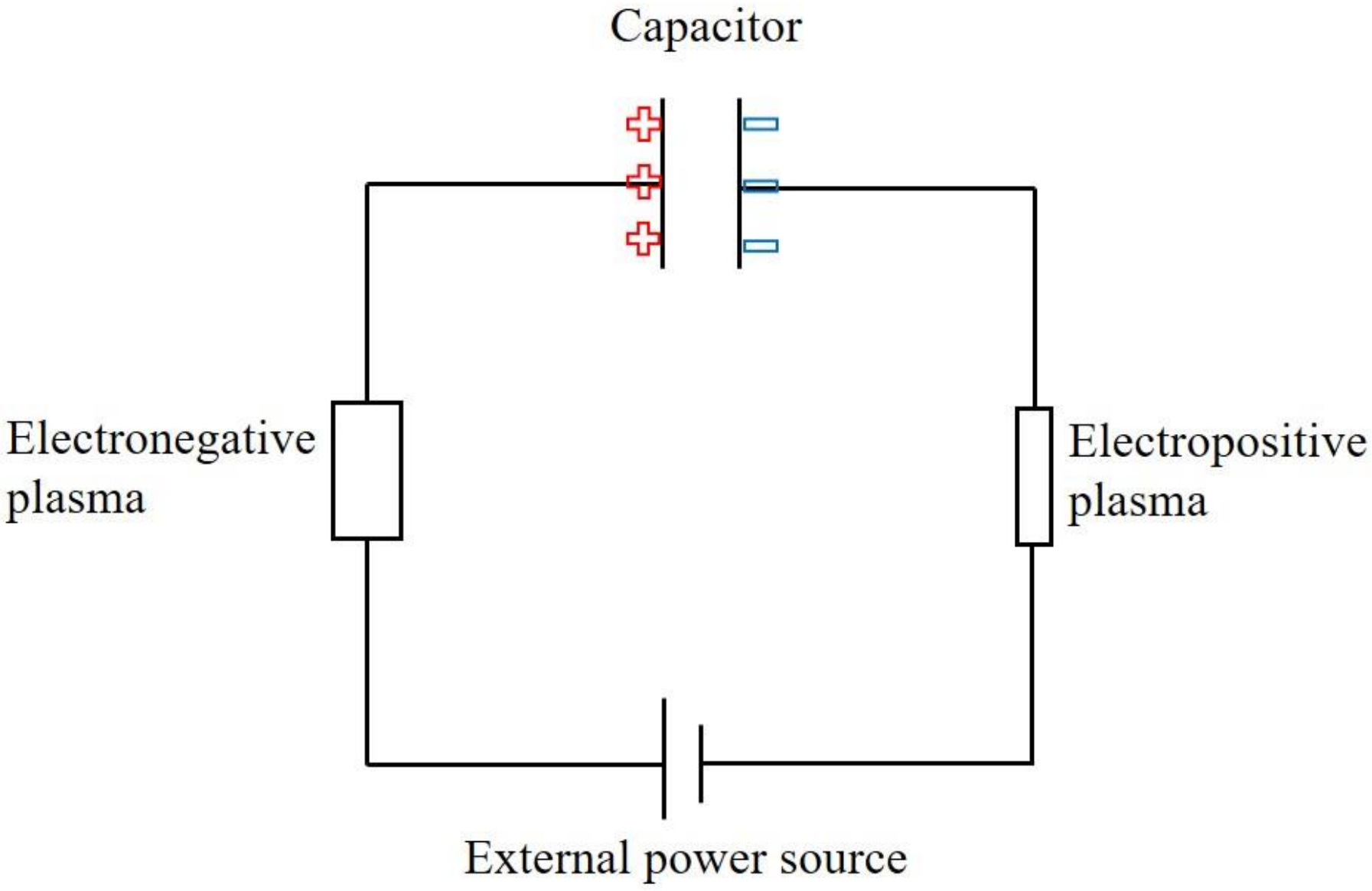

Figure 8. Transformed circuit of electronegative ICP source in the transport dominated regime.

**(c) The property of open system**

$$\alpha=\alpha_0\left(1-\frac{x^2}{l^2}\right),\ (45)$$

$$x=0,\ \alpha=\alpha_0,\Rightarrow \frac{d\alpha}{dx}\Big|_{x=0(\alpha=\alpha_0)}=\frac{-2\alpha_0 x}{l^2}\Big|_{x=0}=0,\ (46)$$

$$x=l,\ \alpha=0,\Rightarrow \frac{d\alpha}{dx}\Big|_{x=l(\alpha=0)}=\frac{-2\alpha_0 x}{l^2}\Big|_{x=l}=\frac{-2\alpha_0}{l}.\ (47)$$

In Eq. (45), the parabola profile of electronegativity that is originally listed in Eq. (40) is re-shown. In the Eqs. (46, 47), the first order derivative of electronegativity and its magnitudes at the two critical points, i.e., $x=0$ (corresponding to $\alpha=\alpha_0$) and $x=l$ (corresponding to $\alpha=0$), are shown. As seen, the axial symmetry is satisfied at the center, i.e., $\frac{d\alpha}{dx}=0$. Nevertheless, an outward flux of cation exists at the edge of electronegative core, which is expressed by the formulae of $\frac{d\alpha}{dx}=\frac{-2\alpha_0}{l}$ and $\Gamma_+=-2D_+n_{e0}\frac{d\alpha}{dx}=\frac{4D_+n_{e0}\alpha_0}{l}$ (refer to Eq. (33)). Moreover, by correlating to the Figs. 2(a), 5(a) and 6(a), it is speculated this flux is truncated by means of strong field intensity of dipole center. From the sub-Sec. (3.1.3), it is known that the inner side of double layer is the anion potential with the order of Kelvin and the outer side of it is the electron potential with the order of Volt (refer to the Fig. 7). And meanwhile, the strong field of dipole center is directed to the electropositive halo, which, together with the electron potential of halo, swiftly delivers the cations that come from the weighted ambi-polar diffusion of core to the chamber border. At the cooperative transport scheme of dipole and electron potential, the cation density is almost zero at the border of core, thus satisfying the plasma neutrality therein ($\alpha=0$). The cation fluxes are discontinuous in the region of the core and halo interface because the two adjacent types of transport, the weak one of anion potential in the core and the strong one of dipole filed acceleration, co-exist therein. This is in accord to the study of Ref. [14] that predicts two plasma fluxes at the location of double layer.

Besides for the cations, it is noted the electrons can and the anions cannot pass through the dipole since the electrons are light and thermal while the anions are heavy and cold. The electrons and cations that pass neutralize mutually and hence no net DC discharge current can pass through the double layer, which represents the role of capacitor model of it. The double layer filters the anions and the passed electrons and cations form the electropositive plasma and electron potential in halo. The anions constrained in the core are hence depleted through the recombination of self-coagulation process while the cations and electrons are depleted at the chamber surface, i.e., the standard plasma species loss mechanism of electropositive plasma source. Regarding the surface depletion of plasma species, the plasma in the transport dominated regime is an open system. Besides, the dual property of double layer, i.e., it separates the core and halo (isolation) and meanwhile it delivers the cations (connection), is called the *quantum property* of double layer. As seen, the judgement of property that it on earth separates or connect the two regions depends on the views of observer.

### (3.1.5) A summary of parabola related theories: transport dominated regime

In sum, the parabola theory, the stratification of overall discharge profile, the Boltzmann's balance of anions and anion potential, and the double layer, are belonged to the transport dominated regime since they are obtained at the low-pressure, 10 mTorr, and the transports of plasma species are fast. This is caused by the electron mobility [74], $\mu_e$, that is a function of pressure, $P$. The mobility is used when solving the electron equation of fluid model, in Eqs. (1,2). Its formula is $\mu_e P = 3.4\times 10^3$, in a unit of $\mathrm{C\cdot s^{-1} m^{-1}}$. The fast electron transport helps establish the ambi-polar diffusion potential that includes the anion and electron ones, to speed the transport of heavy ions. In the speedy transport of electronegative plasma at low pressure, the chemistry is not able to respond, as stated in Ref. [75]. That's why the recombination loss can be neglected in the continuity equation of cations in Eq. (36). This is different to the high-pressure case where the transports of plasma species are slow and the chemistry participates in forming the main discharge structure. The electronegative plasma is open system that needs the shield of chamber border in the transport dominated regime, while it is closed system that does not need the shield anymore in the self-balanced regime of high pressure; see next.

## (3.2) Transport and chemistry self-balanced regime

### (3.2.1) Ellipse theory

$$\frac{d}{dx}\left(-D_+\frac{dn_+}{dx}+n_+\mu_+E\right)=K_{iz}n_0n_e-K_{rec}n_+n_-,\ (48)$$

$$\frac{d}{dx}\left(-D_-\frac{dn_-}{dx}-n_-\mu_-E\right)=K_{att}n_0n_e-K_{rec}n_+n_-,\ (49)$$

$$n=n_{e0}\exp\left(\varphi/T_e\right),\ (50)$$

$$D_e=\mu_eT_e,\ (51)$$

$$\frac{d}{dx}\left(-D_+\frac{d}{dx}\left(n_-+n_e\right)-\mu_+\left(n_-+n_e\right)\frac{D_e}{\mu_e}\frac{1}{n_e}\frac{dn_e}{dx}\right)=K_{iz}n_0n_e-K_{rec}n_+n_-,\ (52)$$

$$\frac{d}{dx}\left(-D_+\frac{dn_-}{dx}+\mu_+n_-\frac{D_e}{\mu_e}\frac{1}{n_e}\frac{dn_e}{dx}\right)=K_{iz}n_0n_e-K_{rec}n_+n_-,\ (53)$$

$$D_-\approx D_+,\ \mu_-\approx\mu_+,\ (54)$$

$$\frac{d}{dx}\left(-2D_+\frac{dn_-}{dx}-\gamma D_+\frac{dn_e}{dx}\right)=\left(K_{iz}+K_{att}\right)n_0n_e-2K_{rec}n_+n_-,\ (55)$$

$$-D_+\frac{dn_e}{dx}-\mu_+T_e\frac{dn_e}{dx}=-D_+\frac{dn_e}{dx}-D_+\gamma\frac{dn_e}{dx}\approx-D_+\gamma\frac{dn_e}{dx}.\ (56)$$

In Eqs. (48, 49), the continuity equations of cations and anions are given. Herein, the approximated drift and diffusion fluxes are adopted for the two types of ion. The Boltzmann's balance is used for the electrons in Eq. (50) but not for the anions, which is different to the parabola theory. Using the electron's Boltzmann balance, the Einstein's formula in Eq. (51), the relation between the electric field and potential gradient, $\vec{E} = -\nabla\varphi$, and the electrical neutrality in Eq. (17), the continuity equations of cations and anions are reformed in Eqs. (52, 53), at neglecting the differences between the diffusion coefficients and mobilities of cations and anions in Eq. (54). Furthermore, the two continuity equations in Eqs. (52, 53) are added and the Eq. (55) is obtained. In this process, the approximation illustrated in Eq. (56) is used.

$$\Gamma_- = \int_0^x K_{att} n_0 n_e \cdot dx - \int_0^x K_{rec} n_+ n_- \cdot dx, \quad (57)$$

$$\Gamma_- = -D_- \frac{dn_-}{dx} - n_- \mu_- E, \quad (58)$$

$$n_- \mu_- E = -\left( \int_0^x K_{att} n_0 n_e \cdot dx - \int_0^x K_{rec} n_+ n_- \cdot dx + D_- \frac{dn_-}{dx} \right). \quad (59)$$

Next, a new procedure is used to treat the fluxes and continuity equations of cations and anions. First, by integrating the continuity equation of anions in Eq. (49), the flux of anions can be reformed in Eq. (57). In Eq. (58), the flux of anions is re-shown. Substituting the flux of Eq. (57) with Eq. (58), the Eq. (59) is obtained, in which the drift term of anions has been expressed as a combination of its source terms and the diffusion term.

$$\frac{d}{dx}\left( -D_+ \frac{dn_+}{dx} + n_+ \mu_+ E \right) = K_{iz} n_0 n_e - K_{rec} n_+ n_-, \quad (60)$$

$$n_+ \approx n_-, \; n_+ \mu_+ E \approx n_- \mu_- E, \quad (61)$$

$$D_- \frac{dn_-}{dx} \approx D_+ \frac{dn_+}{dx}, \quad (62)$$

$$\frac{d}{dx}\left[ -2D_+ \frac{dn_+}{dx} + \int_0^x K_{rec} n_+^2 \cdot dx - \int_0^x K_{att} n_0 n_e \cdot dx \right] = K_{iz} n_0 n_e - K_{rec} n_+ n_-, \quad (63)$$

$$\frac{d}{dx}\left( -2D_+ \frac{dn_+}{dx} \right) = \left( K_{iz} + K_{att} \right) n_0 n_e - 2K_{rec} n_+ n_-, \quad (64)$$

$$\frac{d}{dx}\left( -2D_+ \frac{dn_-}{dx} \right) = \left( K_{iz} + K_{att} \right) n_0 n_e - 2K_{rec} n_+ n_-. \quad (65)$$

$$D_+ \frac{dn_+}{dx} \approx D_+ \frac{dn_-}{dx}. \quad (66)$$

We then return to the continuity equation of cations in Eq. (60). At a set of approximations that are illustrated in the Eqs. (61, 62) due to the high electronegativity, the continuity equation of cations in Eq. (60) is reformed to be the Eq. (63), by substituting the term, $n_- \mu_- E$, with the Eq. (59). In Eq.

(63), the differential and integral operations are counteracted for the source terms. These terms are then shifted to the right side of equation and accordingly the Eq. (64) is formed. As seen next, for a better comparison, the continuity equation of cation in Eq. (64) is further slightly reformed in Eq. (65), at utilizing another approximation illustrated in Eq. (66).

By comparing the Eqs. (55, 65), it is seen that the only difference is the term that exists in Eq. (55), $\gamma D_+ \frac{dn_e}{dx}$. As seen future, if this term is negligible the reformed continuity equation of cations (or more precisely, the anions density) can be solved analytically and it is found that the cations/anions density is in fact an elliptic profile. So, the inequality, i.e., $\left|\frac{\gamma d^2 n_e / dx^2}{2d^2 n_+ / dx^2}\right| \ll 1$, becomes a criterion of the ellipse theory.

$$\gamma D_+ \frac{dn_e}{dx} \approx 0, n_e \sim n_{e0}, (67)$$

$$n_+ = n_{e0}(\alpha+1), n_- = n_{e0}\alpha, n_+ n_- = n_{e0}^2 \alpha(\alpha+1), (68)$$

$$-2D_+ \frac{d^2 n_+}{dx^2} = (K_{iz} + K_{att}) n_0 n_{e0} - 2K_{rec} n_+ n_-, (69)$$

$$\beta(\alpha) = -2D_+ n_{e0} \frac{d\alpha}{dx}, (70)$$

$$-\frac{1}{4D_+ n_{e0}} \frac{d}{d\alpha} \beta^2 = (K_{iz} + K_{att}) n_0 n_{e0} - 2K_{rec} n_{e0}^2 \alpha(\alpha+1), (71)$$

$$\beta(\alpha) = \beta_0 \left[(\alpha_0 - \alpha) - \frac{2K_{rec} n_{e0}}{(K_{iz} + K_{att}) n_0} \times \left(\frac{\alpha_0^3 - \alpha^3}{3} + \frac{\alpha_0^2 - \alpha^2}{2}\right)\right]^{1/2}, (72)$$

$$\beta_0 = \left[4D_+ (K_{iz} + K_{att}) n_0 n_{e0}^2\right]^{1/2}, (73)$$

$$\int_{\beta(\alpha_0)}^{\beta} d\beta^2 = \int_0^{\beta} d\beta^2 = \beta^2, \beta(\alpha_0) = 0, \frac{d\alpha}{dx}\Big|_{x=0\,(\alpha=\alpha_0)} = 0. (74)$$

Now, we deduce the ellipse theory of electronegative plasma at the above inequality. The implication of the inequality condition is illustrated in Eq. (67). Herein, the spatial variation of electron density is neglected and the electron density equals to its value at the center, $n_{e0}$, which has the same effect as the Eqs. (34, 35) of parabola theory in the Sec. (3.1.1). With the constant electron density, the cation and anion densities and their product are all expressed as a function of electronegativity $\alpha$ in Eq. (68). In Eq. (69) the reformed continuity equation of cations of Eq. (64) is re-shown and in Eq. (70) a new variable, $\beta$, that is a function of $\frac{d\alpha}{dx}$ is introduced. At the help of Eqs. (68, 70), the continuity equation of cations in Eq. (69) is further reformed to be the Eq. (71). It is seen that the newly reformed continuity equation is a first-order differential equation and its independent variable is the electronegativity $\alpha$ now, instead of $x$. By integrating this differential equation, a new expression is obtained for $\beta$ in Eq. (72), at the assistance of a parameter $\beta_0$ defined in Eq. (73). It is noticed that during this integral process, the cylindrical symmetry condition at the center, which represents the lower integral limit, is used, as shown in Eq. (74).

$$x(\alpha)=\int_{\alpha(x)}^{\alpha_0}\frac{2D_+n_{e0}}{\beta(\alpha)}d\alpha,\quad(75)$$

$$x(\alpha)=\frac{2D_+n_{e0}}{\beta_0}\int_{\alpha(x)}^{\alpha_0}\frac{d\alpha}{\left[(\alpha_0-\alpha)-\frac{2K_{rec}n_{e0}}{(K_{iz}+K_{att})n_0}\times\left(\frac{\alpha_0^3-\alpha^3}{3}+\frac{\alpha_0^2-\alpha^2}{2}\right)\right]^{1/2}},\quad(76)$$

$$x(\alpha)=\frac{2D_+n_{e0}}{\beta_0}\cdot\frac{\alpha_0}{(\eta/3)^{1/2}}\int_{\alpha(x)}^{\alpha_0}\frac{d\alpha}{(\alpha_0-\alpha)^{1/2}\left\{(-1)\times\left[\alpha^2+\left(\alpha_0+\frac{3}{2}\right)\alpha-\left(\frac{3\alpha_0^2}{\eta}-\alpha_0^2-\frac{3}{2}\alpha_0\right)\right]\right\}^{1/2}},\quad(77)$$

$$\eta=\frac{2K_{rec}n_{e0}\alpha_0^2}{(K_{iz}+K_{att})n_0},\quad(78)$$

$$\alpha=\frac{-\left(\alpha_0+\frac{3}{2}\right)\pm\sqrt{\left(\alpha_0+\frac{3}{2}\right)^2+4\left(\frac{3\alpha_0^2}{\eta}-\alpha_0^2-\frac{3}{2}\alpha_0\right)}}{2},\quad(79)$$

$$\alpha=\frac{-\left(\alpha_0+\frac{3}{2}\right)\pm\sqrt{\left(\frac{12}{\eta}-3\right)\alpha_0^2-3\alpha_0+\frac{9}{4}}}{2},\quad(80)$$

$$\begin{cases}\alpha_0+\frac{3}{2}\sim\alpha_0,\\ \eta<1,\\ \left(\frac{12}{\eta}-3\right)\alpha_0^2-3\alpha_0+\frac{9}{4}\sim\left(\frac{12}{\eta}-3\right)\alpha_0^2,\end{cases}\quad(81)$$

$$\alpha=\frac{-\alpha_0\pm\left(\frac{12}{\eta}-3\right)^{1/2}\alpha_0}{2},\quad(82)$$

$$x=\frac{2D_+n_{e0}}{\beta_0}\frac{\alpha_0}{(\eta/3)^{1/2}}\times\int_{\alpha(x)}^{\alpha_0}\frac{d\alpha}{\left[(b\alpha_0-\alpha)(\alpha_0-\alpha)(a\alpha_0+\alpha)\right]^{1/2}},\quad(83)$$

$$\begin{cases} a = \frac{1}{2} + \frac{1}{2}\left(\frac{12}{\eta} - 3\right)^{1/2}, \\ b = -\frac{1}{2} + \frac{1}{2}\left(\frac{12}{\eta} - 3\right)^{1/2}. \end{cases} \quad (84)$$

Slightly reforming the initial expression of $\beta$ in Eq. (70) and then integrating it, the Eq. (75) is obtained. At the left side of Eq. (75), the spatial coordinate has been expressed as a function of the electronegativity, i.e., $x(\alpha)$. At the right side of it, the quantity, $\beta$, in the denominator of integrated term is substituted by its new expression in Eq. (72) and then the Eq. (76) is obtained. After this substitution, it is seen that the denominator is now a third-order polynomial of electronegativity, $\alpha$. We now factorize the polynomial. At the first factorization, the Eq. (76) is reformed to be the Eq. (77) and the denominator becomes now a second-order polynomial of $\alpha$, except for the factor, $\alpha_0 - \alpha$. It is noticed that in the transformation of Eq. (76) onto Eq. (77), a new parameter, $\eta$, is introduced. It is a function of rates of recombination, ionization and attachment, as expressed in Eq. (78). As seen further, this is a very important parameter since it reveals the principle of discharge structure transition between different modes, such as the parabola, ellipse, and self-coagulation. The two roots of $\alpha$ in the second order polynomial of denominator are obtained in Eq. (79), by means of another factorization. It is then reformed in Eq. (80), for the convenience of approximation. In the Eq. (80), the term, $\alpha_0 + \frac{3}{2}$, is approximated to be $\alpha_0$, considering the high electronegativity condition. Still in the Eq. (80), under the root symbol, the term, $\left(\frac{12}{\eta} - 3\right)\alpha_0^2 - 3\alpha_0 + \frac{9}{4}$, is approximated to be $\left(\frac{12}{\eta} - 3\right)\alpha_0^2$. Namely, the first order and zero order terms of $\alpha_0$ are both ignored when comparing to the second order term, $\alpha_0^2$, again due to the high electronegativity condition. In this approximation, it is noted that the condition, $\eta < 1$, is used. This is a very important requirement since it ensures the coefficient of second order term, $\frac{12}{\eta} - 3$, is positive and the roots are both real numbers. The two approximations and the requirement, $\eta < 1$, are summarized in Eq. (81). At these approximations, the roots of $\alpha$ are reformed in Eq. (82). At the two simplified roots, the integral in Eq. (77) is reformed to be the Eq. (83) and the complete polynomial factorization of denominator is ended at introducing two new parameters, $a, b$, with their expressions illustrated in Eq. (84), respectively. In a view of mathematics, the integral of Eq.

(83), $\int_{\alpha(x)}^{\alpha_0} \frac{d\alpha}{\left[(b\alpha_0-\alpha)(\alpha_0-\alpha)(a\alpha_0+\alpha)\right]^{1/2}}$ , is an elliptical integral. This is the origin of ellipse theory since it indicates the cations and anions density profiles are elliptic at the electrical neutrality of electronegative plasma and the condition of high electronegativity. In the appendix, more details are presented for proving that the variables, $x, \alpha$, satisfy the elliptic function, $\left(\frac{x}{a'}\right)^2+\left(\frac{\alpha}{b'}\right)^2=1$ .

Herein, the parameters, $a', b'$, are supposed to be the major and minor semi- axes of an ellipse.

**(3.2.2) Closed system of ellipse**

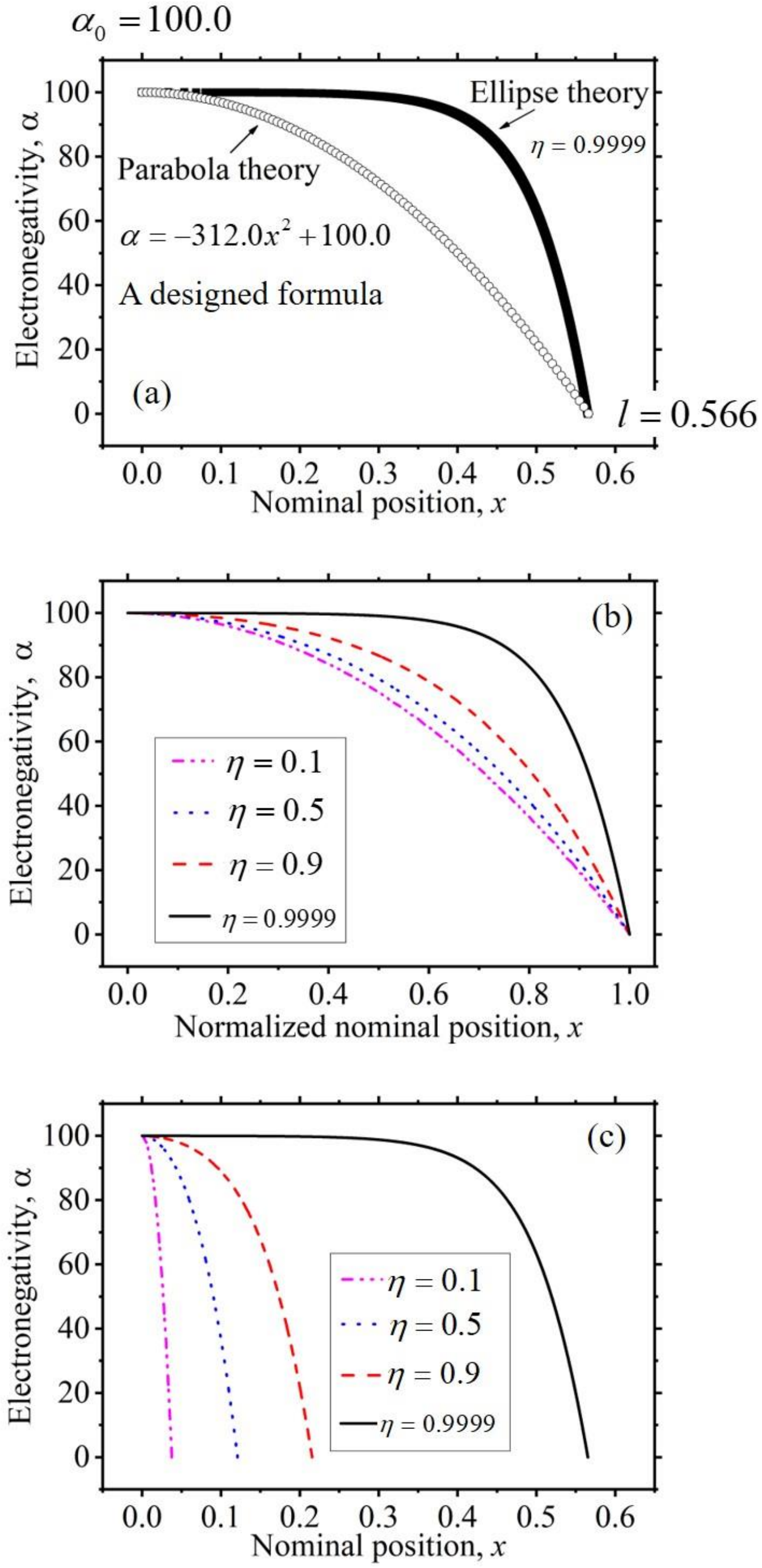

Figure 9. (a) Comparison of electronegativities between the parabola theory (refer to Eq. (40)) and ellipse theory (refer to Eq. (83)), (b) in the scope of ellipse theory, the dependence of electronegativity, $\alpha(x)$, on the parameter, $\eta$, that is illustrated in Eq. (78), at a normalized nominal position, and (c) still the variation of $\alpha$ with $\eta$, but at the realistic nominal position.

In Fig. 9(a), the electronegativities, $\alpha(x)$, of the parabola theory in Eq. (40) and of the ellipse theory in Eq. (83) are compared. It is already found that the $\alpha(x)$ profile of ellipse theory is affected by the parameter, $\eta$. In this sub-figure, the elliptic profile is given at the value of $\eta = 0.9999$. The parabolic profile, $\alpha = -312.0x^2 + 100.0$, is a designed formula that is based on the maximal nominal position of elliptic profile, 0.566, and the minimal electronegativity at that position, 0.049. Herein, the $\alpha_0$ value is set as 100.0, which is presumed to satisfy the high electronegativity requirement. It is noticed that the expression of the designed parabolic function is in accord to the Eq. (40) of parabola theory, i.e., $312.0 \approx \frac{\alpha_0}{l^2} = \frac{100.0}{0.566^2}$. Besides, it is seen that the parabolic profile corresponds to an analytic function. The elliptic profile is given by only the integral part of Eq. (83), i.e., $x = \int_{\alpha(x)}^{\alpha_0} \frac{d\alpha}{\left[(b\alpha_0 - \alpha)(\alpha_0 - \alpha)(a\alpha_0 + \alpha)\right]^{1/2}}$, and it is numerically calculated. The other terms of Eq. (83), $\frac{2D_+ n_{e0}}{\beta_0} \frac{\alpha_0}{(\eta/3)^{1/2}}$, are treated as constants and hence their influence on the profile are supposed to be ignorable. Herein, the $a, b$ parameters of the above integral are both a function of $\eta$, as illustrated in the Eq. (84).

In Fig. 9(b, c), the variations of $\alpha(x)$ against $\eta$ in the scope of ellipse theory are presented, at the normalized and realistic nominal positions, respectively. It is seen from the normalized Fig. 9(b) at the small value of $\eta = 0.1$, when the recombination is insignificant, the profile given by the ellipse theory is more like a parabola as compared to the Fig. 9(a). Upon increasing the $\eta$ value, the profile of $\alpha(x)$ becomes wide and meanwhile along the normalized nominal position, the front part of curve is fattened while the latter part of curve is steepened, which indicates that the rate of slope at the end of curve increases. It is predicted from the tendency that upon continually increasing the $\eta$ value, the slope rate will tend to be infinitely large, i.e., $\frac{d\alpha}{dx} \to \infty, \frac{dx}{d\alpha} \to 0$. As expected, these are just the characteristics that an ellipse function in Eq. (85) holds, where the major and minor semi-axes are fixed. In Eq. (86), the derivative, $\frac{d\alpha}{dx}$, is obtained from the Eq. (85). Based on the derivative, it is seen from Eq. (87) that at the point $(0, \alpha_0)$, we have $\frac{d\alpha}{dx} = 0$, which has been utilized in Eq. (74) as the symmetric condition. In Eq. (88), the other derivative, $\frac{dx}{d\alpha}$, is obtained still from the Eq. (85). Again, based on this derivative, it is seen from Eq. (89) that at the point, $(x_{\max}, 0)$, we have $\frac{dx}{d\alpha}=0$, which characterizes the curve end of high $\eta$ value in the normalized Fig. 9(b). It indicates the role of recombinations, embodied in the numerator of Eq. (78), in determining the transport of electronegative plasma and then characterizing their final density profiles. It is known that ellipse is closed curve as illustrated in Eqs. (85-89) by means of its two first order derivative values of boundaries, i.e., both of them are zeros. This is different to the truncated open curves of parabola described in Sec. (3.1.4(c)).

$$\frac{x^2}{a'^2}+\frac{\alpha^2}{b'^2}=1,\, a'=x_{\max},\, b'=\alpha_0, \quad (85)$$

$$\frac{d\alpha}{dx}=-\frac{x}{\alpha}\frac{b'^2}{a'^2}, \quad (86)$$

$$x=0,\, \alpha=\alpha_0=b', \quad \Rightarrow \quad \left.\frac{d\alpha}{dx}\right|_{x=0,\, \alpha=\alpha_0=b'}=0, \quad (87)$$

$$\frac{dx}{d\alpha}=-\frac{\alpha}{x}\frac{a'^2}{b'^2}, \quad (88)$$

$$\alpha=0,\, x=x_{\max}=a', \quad \Rightarrow \quad \left.\frac{dx}{d\alpha}\right|_{x=x_{\max}=a',\, \alpha=0}=0. \quad (89)$$

### (3.2.3) Transformed recombination flux

When $\eta=0.9999$, the ambi-polar diffusion of electronegative plasma that is caused by the positive chemical source, i.e., ionizations for cations creation, is counteracted by the recombinations. As seen from the dimension and physics analysis in the Eqs. (90-98), the recombinations term can play the role of drift term. In Eq. (90), the continuity equation of cations is reformed, by means of adding its inertial term and meanwhile transferring it from the one-dimensional form to three-dimensional form (refer to the flux expression and its divergence degree), which is designed for demonstrating the generality. In Eq. (91), a frequency of recombination is introduced and as noticed, it is a function of electronegativity. In Eq. (92), the term, $n_K_{rec}$, in the chemical sources of equation is replaced by the recombination frequency, $\nu_{rec}$. In Eq. (93) an effective drift flux, $\vec{\Gamma}_{d,eff}$, is introduced and in Eq. (94) the divergence degree of the supposed drift is presented. Herein, $\vec{E}_{eff}$ is the effective electric field and deduced from the negative chemical source, i.e., ionic recombinations. It is seen that there are two parts contained in the divergence. However, the first part, i.e., $\mu_{+}\nabla n_{+}\cdot\vec{E}_{eff}$, tends to zero in Eq. (95) since the gradient of cations density (i.e., $\nabla n_{+}$) is zero in the flat part of cations density curve (refer to Fig. 9(a)) and the effective field (i.e., $\vec{E}_{eff}$) is weakened in the dropping part of cations density curve, due to the reduced recombination effect arising from the decreasing densities. Still in Eq. (94), by utilizing the Poisson's equation, the second part of divergence, $\mu_{+}n_{+}\nabla\cdot\vec{E}_{eff}$, can be rewritten as $\mu_{+}n_{+}\frac{\rho_{eff}}{\varepsilon_0}$. Herein, $\rho_{eff}$ is the effective charge density of drifting electrical field through the recombinations. In Eq. (96) the dimension of each parameter in the above term is shown and the combination of them has the same dimension as the chemical term, $n_{+}\nu_{rec}$. So, in Eq. (97) the chemical term is intentionally shifted to the left of continuity equation and in Eq. (98) it is transformed into the divergence of effective drift flux. The ambi-polar flux is counteracted by the transformed *recombination flux*, which ultimately generates the ellipse density profile that is closed if it is drawn symmetrically.

$$\frac{\partial n_{+}}{\partial t}+\nabla\cdot\left(-D_{+}\nabla n_{+}+n_{+}\mu_{+}\vec{E}\right)=n_{e0}n_0K_{iz}-n_{+}n_K_{rec},\ (90)$$

$$\nu_{rec}=n_K_{rec},\ \nu_{rec}=\nu_{rec}\left(\alpha\right),\ (91)$$

$$\frac{\partial n_{+}}{\partial t}+\nabla\cdot\left(-D_{+}\nabla n_{+}+n_{+}\mu_{+}\vec{E}\right)=n_{e0}n_0K_{iz}-n_{+}\nu_{rec},\ (92)$$

$$\vec{\Gamma}_{d,eff}=\mu_{+}n_{+}\vec{E}_{eff},\ (93)$$

$$\nabla\cdot\vec{\Gamma}_{d,eff}=\mu_{+}\nabla n_{+}\cdot\vec{E}_{eff}+\mu_{+}n_{+}\nabla\cdot\vec{E}_{eff}=\mu_{+}\nabla n_{+}\cdot\vec{E}_{eff}+\mu_{+}n_{+}\frac{\rho_{eff}}{\varepsilon_0},\ (94)$$

$$\mu_{+}\nabla n_{+}\cdot\vec{E}_{eff}\rightarrow 0.\ (95)$$

$$[\mu_{+}]=\frac{m^2}{V\cdot s},[n_{+}]=\frac{1}{m^3},[\rho_{eff}]=\frac{C}{m^3},[\varepsilon_0]=\frac{C}{m\cdot V},$$
$$[\mu_{+}n_{+}\frac{\rho_{eff}}{\varepsilon_0}]=\frac{1}{m^3\cdot s}=[n_{+}\nu_{rec}].\quad (96)$$

$$\frac{\partial n_{+}}{\partial t}+\nabla\cdot\left(-D_{+}\nabla n_{+}+n_{+}\mu_{+}\vec{E}\right)+n_{+}\nu_{rec}=n_{e0}n_0K_{iz},\ (97)$$

$$\frac{\partial n_{+}}{\partial t}+\nabla\cdot\left(-D_{+}\nabla n_{+}+n_{+}\mu_{+}E\right)+\nabla\cdot\vec{\Gamma}_{d,eff}=n_{e0}n_0K_{iz},\ (98)$$

### (3.2.4) A loose self-balance

Besides, in the realistic Fig. 9(c), it is still seen that upon increasing the $\eta$ value, the profile width is strikingly expanded, which agrees quantitatively with the fluid simulation in Fig. 10, at increasing the pressure. It is seen in Fig. 10 that the fluid simulation presents the density profile in complete discharge region, i.e., including the electronegative core, electropositive halo, and the tiny sheath at the border of ICP chamber. In addition, the simulation exhibits the density profile between the two borders of chamber while the analytic theory exhibits the density profile from the symmetric center to the border of electronegative core. Both the parabola and ellipse theories define that the electronegativity at the core profile border is zero, which is known to be as well the border between the electronegative core and electropositive halo. In Fig. 10, upon increasing the pressure, the simulated axial profiles of cations and anions densities change from a parabola to an ellipse, in agreement with the prediction of analytic theory in Fig. 9(b), and meanwhile the width of electronegative core region that is highly electronegative and electrically neutral expands. Accordingly, the electropositive halo behind the double layer that is also neutral is shrunk. As mentioned in Sec. (3.1.4), the electronegative plasma in the parabola regime is an open system because the dipole model of double layer and the electropositive halo cooperatively communicate the core plasma and chamber border. In the ellipse regime, the persistently shrinking double layer and halo indicate their insignificant roles in sustaining the plasma. In a word, it is now more like a self-balanced closed system. As seen further, this is a type of loose balance between the transport (limited by the ambi-polar diffusion) and chemistry that generate the definite distribution of density profile (i.e., ellipse) since the positive chemical source still dominates the discharge, i.e. $\eta < 1$. In the next section, it will be shown that when $\eta > 1$ and the recombination loss dominates in certain species, e.g., anions, there will be a tight self-balance between the transport (limited then by the free diffusion) and chemistry; see the Sec. (3.3.5). It is called self-coagulation behavior and can generate a mass-point shape of density distribution (i.e., coagulated).

As compared to Fig. 9(b, c), it is seen in Fig. 10(d) the idealized ellipse profile in the middle part of curve is not flat, but slightly increases as directing to the RF field heating region of ICP source, i.e., under the dielectric window and coil of rf current. As seen next, this is due to the self-coagulation behavior that leads to the radial bump of ions as well in Fig. 2(b).

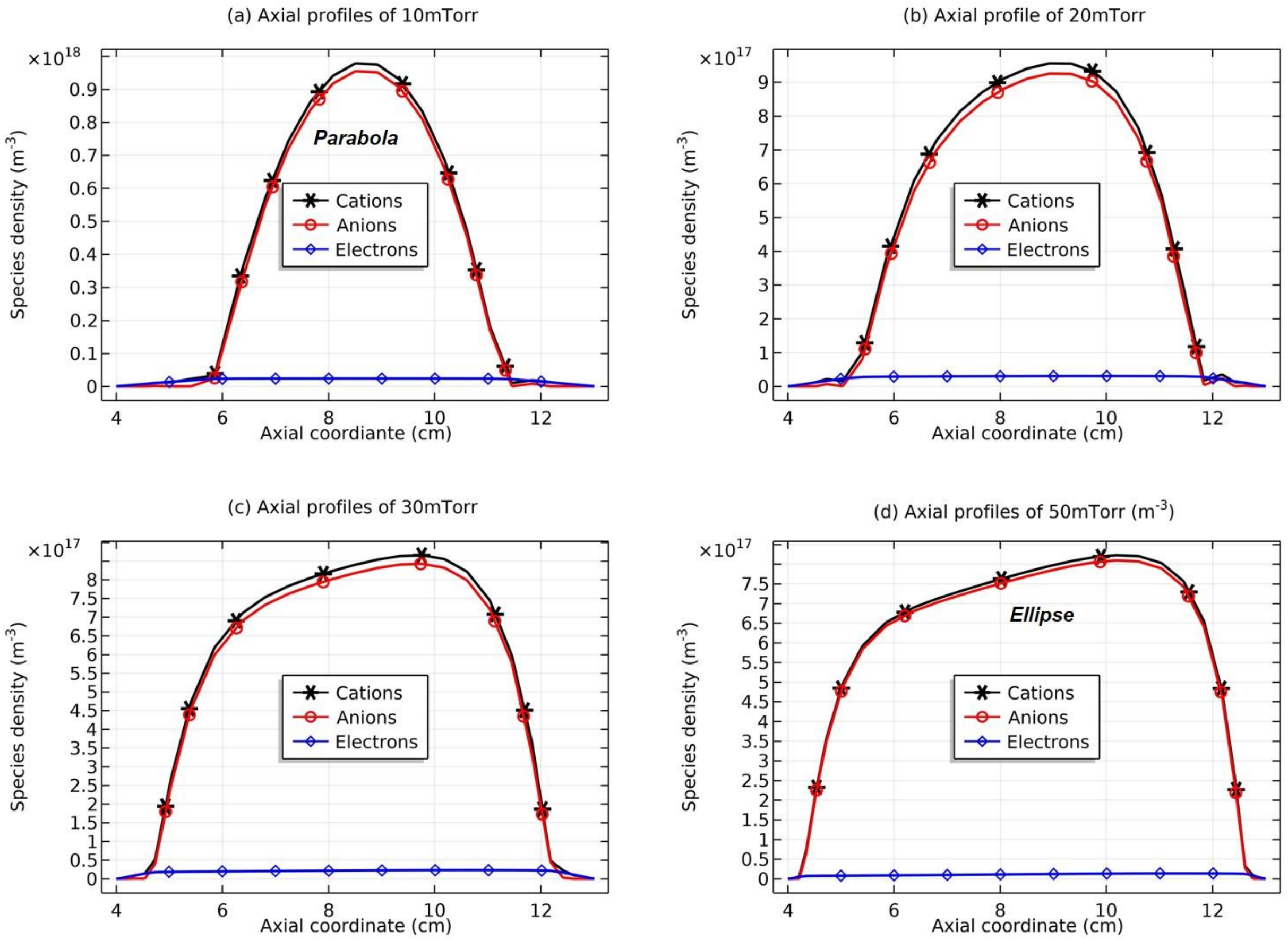

Figure 10. Simulated axial profiles of species densities in Ar/$SF_6$ inductively coupled plasma, at (a) 10mTorr, (b) 20mTorr), (c) 30mTorr, and (d) 50mTorr, by fluid model. The discharge power is 300W and the reactive $SF_6$ concentration is 10%.

### (3.3) Chemistry dominated regime
### (3.3.1) Net negative chemical source

In Fig. 11, the simulated total anions density and their net chemical source two-dimensional profiles by fluid model are shown, at the pressures of 10mTorr, 50mTorr and 90mTorr, respectively. It is seen that for the three considered pressures, at the locations of anions density peaks the chemical sources are negative, i.e., the recombinations dominate over the attachments ($\eta > 1$). As seen further, this is one of the important requirements for the self-coagulation of anions. Besides, it is still seen that at low pressure, 10mTorr, the spatial stratification is obvious and the self-coagulation is happened at the edge of electronegative core. At increasing the pressure to 50mTorr and 90mTorr, the stratification is almost disappeared and the expanding core profile turns into ellipse from parabola. Meanwhile, the location of self-coagulation event is shifted, toward the coil, under which the RF-field heating is strong. In sum, the formed profiles are superpositions of basic environment, parabola or ellipse, and refined mass-point type of astro-structure. These superpositions of two levels explain the bump that is seated over the radial parabola in Fig. 2(b) and the slight ascending tendency of ellipse in Fig. 10(d).

At the high pressures, the transports of all plasma species, cations, anions, and electrons, are slow and meanwhile the high concentration of background gas molecules, such as $SF_6$, leads to the strong electrons depletion via attachments. It is expected from the simulation that the attachments effect is so strong that the electrons created via ionizations of Ar and $SF_6$ cannot be dispersed throughout the chamber, but are localized under the coil. It is known that the thresholds of ionizations are high, around 10eV generally, and at the high-pressure discharges with small diffusion coefficients the ionizations are concentrated in the strong field region, which can then balance the strong depletion of electrons. At the far-field regions, i.e., wide elliptic profile except for the self-coagulated astro-structure, this balance cannot be completed. So, since the electrons have to be localized, so do the cations and anions. This partially explains the shift of self-coagulation location to the coil against the pressure.

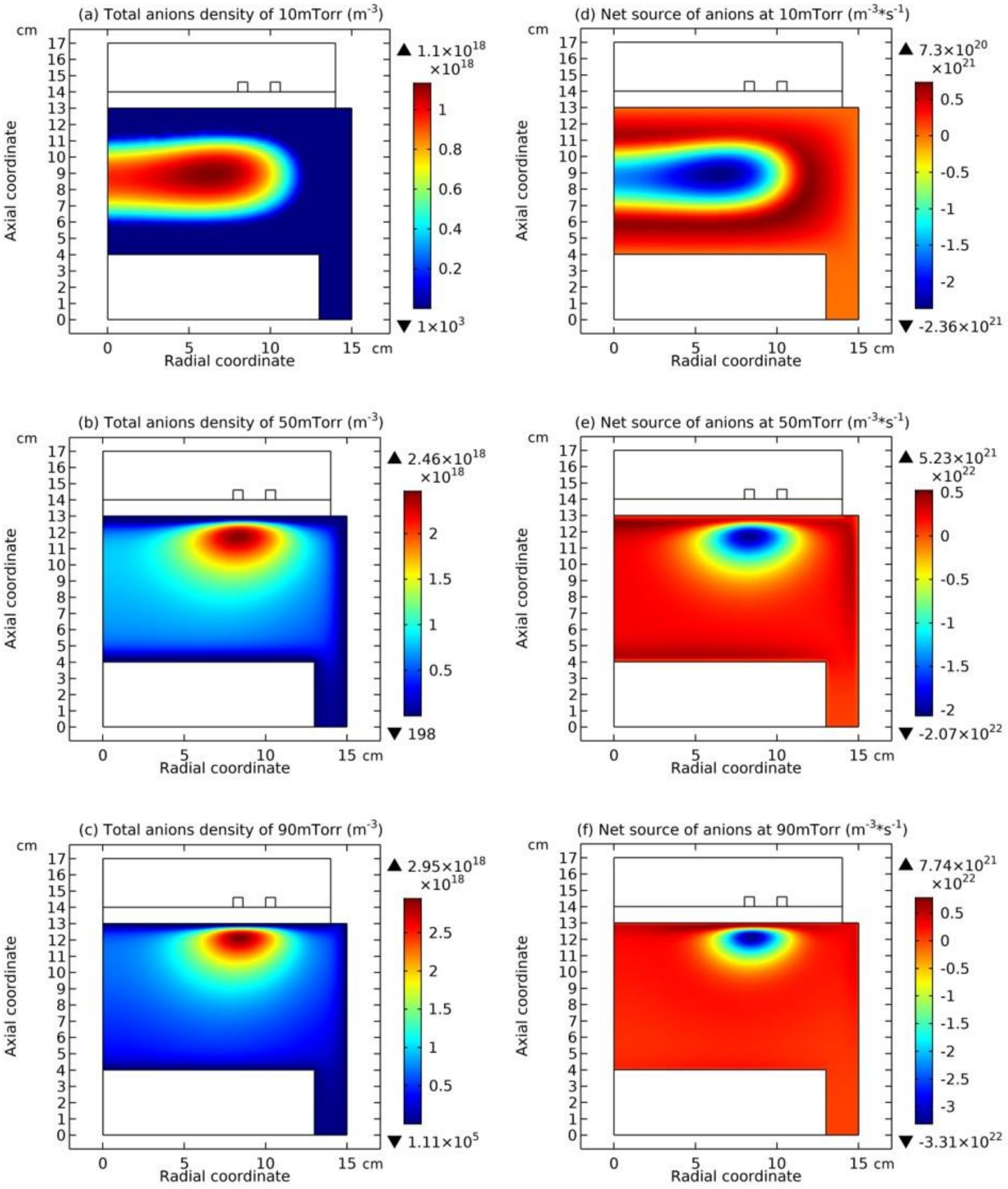

Figure 11. Simulated total anions density profiles of (a) 10mTorr, (b) 50mTorr, and (c) 90mTorr, respectively, and corresponding net source of anions of (d) 10mTorr, (e) 50mTorr, and (f) 90mTorr, respectively, by fluid model. The discharge power is 300W and the ratio of reactive $SF_6$ in the gas mixture is 10%.

### (3.3.2) Free diffusion transport component

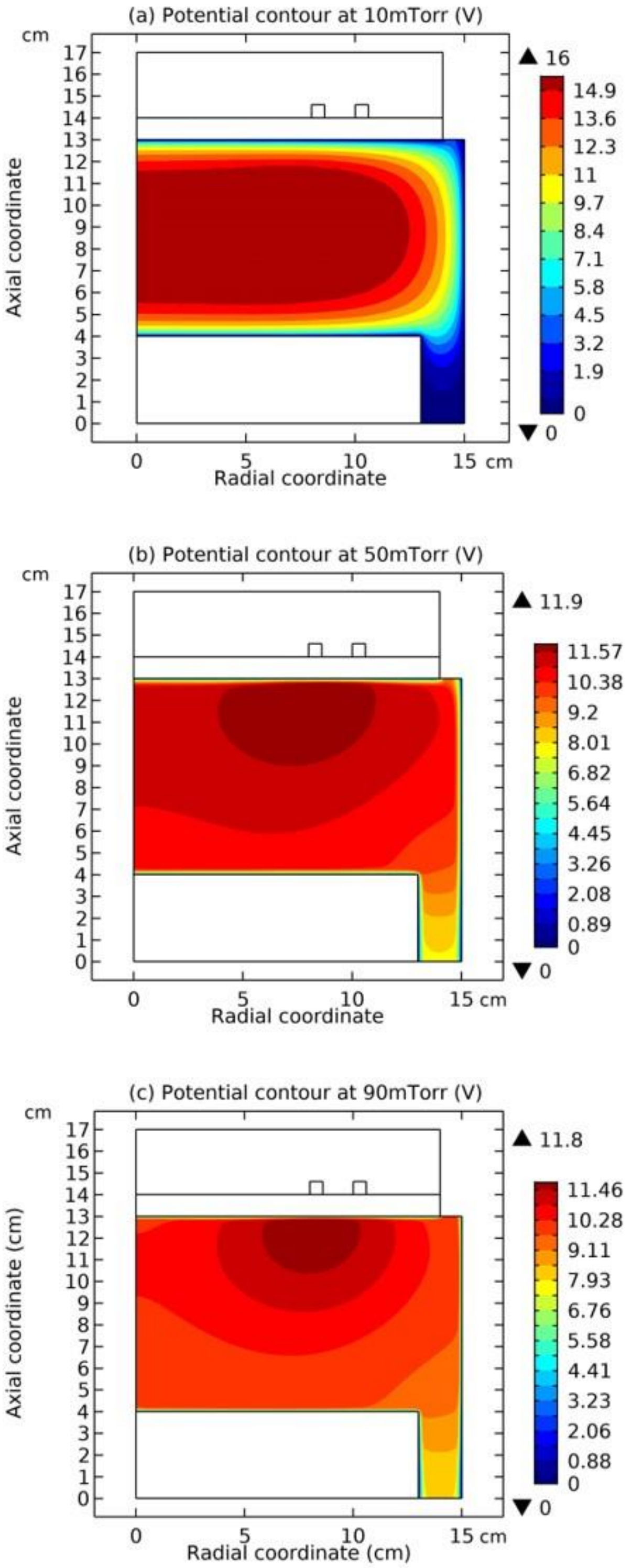

Figure 12. Simulated plasma potential contours of (a) 10mTorr, (b) 50mTorr, and (c) 90mTorr, respectively, by fluid model. The discharge power is 300W and the ratio of reactive $SF_6$ in the gas mixture is 10%.

In Fig. 12, the simulated plasma potential contours by fluid model are presented, at the pressures of 10mTorr, 50mTorr and 90mTorr, respectively. At the low pressure, 10mTorr, the discharge is in the parabola regime and the potential in the core is anions potential, as analyzed in Sec. (3.1.3). In Fig. 12(a), the potential is plotted in the order of Volt (refer to the figure legend). So, the potential is seen unchangeable in the core. In Fig. 12(b,c), the potentials in the self-coagulation locations are also unchangeable. Therefore, it is concluded that besides for the negative chemical source shown in Fig. 11, the free diffusion transport component is still needed for the self-coagulation occurrence since the astro-structures are all laid on the tops of electrons potentials. Herein, the influence of anions potential gradient at 10mTorr in Fig. 4(b) is neglected since it is in the room temperature range and the weak ambi-polar diffusion coefficient is only two times of free diffusion, as illustrated in the Eq. (33).

The self-coagulation of anions at 10mTorr in Fig. 11(a) is happened at the core edge (not the center), since herein the electrons potential barrel pushing effect in Fig. 12(a) to the anions is the strongest, which is then a geometric effect. The self-coagulation and double layer structure are co-existed since the self-coagulation can absorb many anions that are pushed inward by the electrons potential barrel and then it is easy for the plasma self-consistently forming the negative and positive charge layers, through the attracting effect between the minor and polarly opposite charges left. The simulations show that the macroscopic parabola and ellipse discharge structures bases are always accompanied by the astro-structures. And when the plasma is not successful in forming the astro-structure, its electrons potential barrel is collapsed and the anions potential barrel then controls the discharge, in which the plasma density becomes eight orders smaller.

### (3.3.3) Self-coagulation dynamics of Anions

The steady state continuity equation of anions that consists of free diffusion flux and negative source term (represented by the recombinations) is expressed in Eq. (99).

$$-D_-\nabla^2 n_- = -n_- n_+ k_{rec} = -n_- \nu_{rec}. \quad (99)$$

Slightly reforming the Eq. (99) and meanwhile introducing a parameter, $k_-$, a quasi-Helmholtz equation is constructed in Eq. (100).

$$\nabla^2 n_- - n_- \frac{\nu_{rec.}}{D_-} = \nabla^2 n_- - n_- k_-^2 = 0,$$
$$\nabla^2 n_- - n_- k_-^2 = 0. \quad (100)$$

Herein, $k_-^2 = \frac{\nu_{rec.}}{D_-}$. For simplicity, in Eq. (101), all the subscripts of quantities are omitted.

$$\nabla^2 n - nk^2 = 0. \quad (101)$$

In Eq. (102), this quasi- Helmholtz equation is reformed by the method of separation of variables in the cylindrical coordinate system, at the assumption of azimuthal symmetry.

$$\begin{aligned}
&\frac{1}{\rho}\frac{\partial}{\partial\rho}\left(\rho\frac{\partial n}{\partial\rho}\right)+\frac{\partial^2 n}{\partial z^2}-k^2 n=0,\\
&n(\rho,z)=R(\rho)Z(z),\\
&Z''+v^2 Z=0,\\
&\frac{d^2R}{d\rho^2}+\frac{1}{\rho}\frac{dR}{d\rho}-\left(k^2+v^2\right)R=0.
\end{aligned}\tag{102}$$

Herein, $v^2$ represents the eigenvalues.

By means of utilizing the homogeneous boundary conditions of the axial and ordinary differential equation, $Z''+v^2Z=0$, the eigenvalues of $v^2$ and the eigen functions, $Z_m(z)$, are acquired in Eq. (103).

$$\begin{aligned}
&{v_m}^2=m^2\pi^2/l^2,\\
&Z_m=\sin(m\pi z/l),\\
&\mathrm{Z}=\sum_{m=0}^{\infty}c_m Z_m=\sum_{m=0}^{\infty}c_m\sin(m\pi z/l).
\end{aligned}\tag{103}$$

As noticed, the radial and ordinary differential equation, $\frac{d^2R}{d\rho^2}+\frac{1}{\rho}\frac{dR}{d\rho}-\left(k^2+{v_m}^2\right)R=0,$ is one zero-order *imaginary* Bessel equation, because of the property of *negative* source. Considering that the density value is limit at the axial center, the imaginary Bessel function, not the Hankel function, is adopted. Then, we get the expression of $R(r)$ as follow in Eq. (104).

$$R=d_m I_0(\sqrt{k^2+{v_m}^2}\rho)=d_m I_0(\sqrt{k^2+m^2\pi^2/l^2}\rho).\tag{104}$$

Finally, we obtain the expression of $n(\rho,z)$ that is a product of $R(\rho)$ and $Z_m(z)$ in Eq. (105).

$$\begin{aligned}
n(\rho,z)&=R(\rho)Z(z)=\sum_{m=0}^{\infty}c_m\sin(m\pi z/l)\cdot d_m I_0(\sqrt{k^2+{v_m}^2}\rho)\\
&=\sum_{m=0}^{\infty}a_m\sin(m\pi z/l)\cdot I_0(\sqrt{k^2+m^2\pi^2/l^2}\rho).
\end{aligned}\tag{105}$$

Next, some special mathematic skills, i.e., the limit idea, are used in Eq. (106) and a delta distribution that is independent on the spatial coordinates is obtained, which represents the self-coagulation idea.

$$
\begin{aligned}
n(\rho,z) &= R(\rho)Z(z) = \sum_{m=0}^{\infty} a_m \sin(m\pi z/l)\cdot I_0(\sqrt{k^2+m^2\pi^2/l^2}\rho) \\
&= \lim_{m\to\infty}\left[a_m \sin(m\pi z/l)\cdot\infty\right] = \lim_{m\to\infty}\left[a_m \sin(m\pi z/l)\cdot \lim_{z\to 0}\frac{1}{z}\right] \\
&= \lim_{z\to 0}\left[\lim_{m\to\infty} a_m \sin(m\pi z/l)\cdot\frac{1}{z}\right] = \lim_{z\to 0}\left[\lim_{m\to\infty} a_m \cdot \frac{\sin(m\pi z/l)}{z\pi/l}\cdot\frac{\pi}{l}\right] \\
&= \lim_{\zeta\to 0}\left[\lim_{m\to\infty} a_m' \cdot \frac{1}{\pi}\frac{\sin(m\zeta)}{\zeta}\right] = \lim_{\zeta\to 0}\left[a_\infty' \lim_{m\to\infty}\frac{1}{\pi}\frac{\sin(m\zeta)}{\zeta}\right] \\
&= a_\infty' \lim_{\zeta\to 0}\delta(\zeta).
\end{aligned} \qquad (106)
$$

In the above deduction, the property of imaginary Bessel function, $\lim_{x\to\infty} I_0(x)\to\infty$, is utilized, which represents the condition of $\lim_{m\to\infty} I_0(\sqrt{k^2+m^2\pi^2/l^2}\rho)\to\infty$. The condition when $m\to\infty$ is chosen as the final solution. Other $m$ values give rise to either vibrating solutions ($m>1$), which are not suitable to describe the density positive property, $n>0$, or solution with the profile that is relatively smooth, i.e., $m=1$, not in accord with the simulation result, localized profile. The limit, $\lim_{z\to 0}$, is moved in front of another limit, $\lim_{m\to\infty}$, because at $m\to\infty$, the value of $z$ in the expression of $\sin(m\pi z/l)$ has no meaning, except at the point, $z=0$. In addition, only the limit result, i.e., the infinite $\infty$ given by $\lim_{m\to\infty} I_0(\sqrt{k^2+m^2\pi^2/l^2}\rho)$, is important, not the process, since when $m\to\infty$, $\lim_{m\to\infty}\sin(m\pi z/l)$ is not existed. So, the invented limit $\lim_{z\to 0}\frac{1}{z}$, although holding different evolving mathematic behavior with the imaginary Bessel, is used to replace the infinite given by the imaginary Bessel limit.

## (3.3.4) Correlations to the parabola and ellipse theories

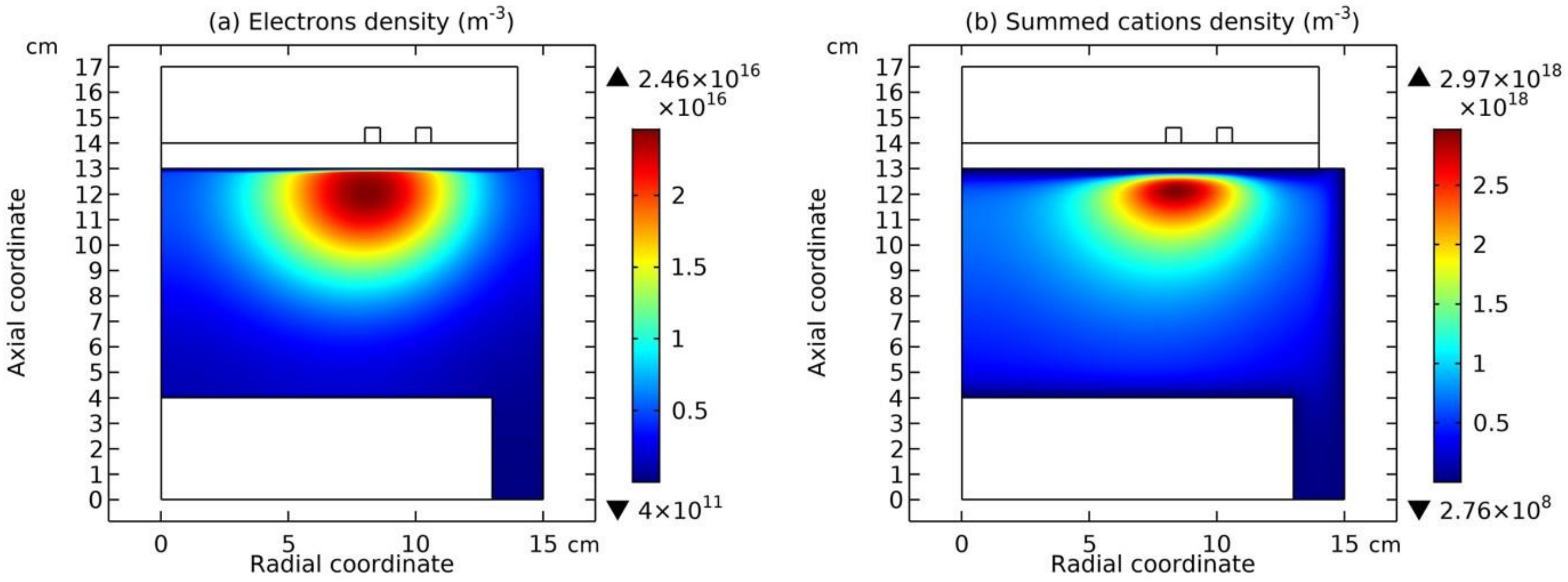

Figure 13. Simulated profiles of (a) electrons density and (b) summed cations density at 90mTorr, by fluid model. The discharge power is 300W and the ratio of reactive $SF_6$ in the gas mixture is 10%.

Three correlations need to be clarified. First, the ellipse is not flat in Fig. 10(d) and the parabola is not parabolic in the radial direction in Fig. 2(b), because of the influence of self-coagulations of ions. Next, in the simulation it is noticed that the self-coagulation cannot happen alone, but is parasitic on either the parabola or ellipse regime, since the two regimes pave the ways for the occurrence of self-coagulation, i.e., providing free diffusion and the high concentrations of cations and anions.

Still, the electrons are light and the ambi-polar potential thereby halts their free diffusion. The cations are heavy, but they carry positive charge, which is easy for them to travel along potential gradient to the chamber border. The anions are heavy and meanwhile negatively charged, and hence it is not easy for them to be delivered to the chamber surface. So, only the anions, neither of the electrons and cations, have the strong self-coagulation behavior in the localized region (Note: the peripheral electrons at 90mTorr have the weak self-coagulation, which leads to the new transport of electrons in Sec. (3.4)). At the high electronegativity, the cations distribution is convergent to the anions to keep the neutrality. The electrons distribution is less influenced and hence smoother than the ions. This fact has been validated by the fluid simulation in Fig. 13, where the density profiles of electrons and cations at 90mTorr is plotted. So, the inequality of ellipse theory stated in Sec. (3.2.1) is reasonable at the considered discharge conditions. Regarding this truth, the flat-top theory [7] that was constructed when the inequality is not satisfied, is omitted herein.

### (3.3.5) A tight self-balance with respect to the inertia

$$\frac{\partial n_+}{\partial t} + \nabla \cdot \vec{\Gamma}_+ = n_{e0} n_0 K_{iz} - n_+ n_- K_{rec}, (107)$$

$$n_{e0} n_0 K_{iz} - n_+ n_- K_{rec} \sim 0^+, (108)$$

$$\vec{\Gamma}_+ = -D_+ \nabla n_+ + \mu_+ n_+ \vec{E} \sim \mu_+ n_+ \vec{E}, (109)$$

$$\nabla \cdot \vec{\Gamma}_+ = \nabla \cdot \left( \mu_+ n_+ \vec{E} \right) = \mu_+ n_+ \nabla \cdot \vec{E} + \mu_+ \vec{E} \cdot \nabla n_+ \sim \mu_+ \vec{E} \cdot \nabla n_+ . (110)$$

In Eq. (107), the unsteady-state continuity equation of cations with the drift-diffusion flux, $\vec{\Gamma}_+$, is presented. On one hand, as mentioned in the ellipse theory of Sec. (3.2), the recombination rate of chemical source is almost equal to the ionization rate illustrated by the value of $\eta = 0.9999$. So, the summed source terms of cations tend to a value of positive zero, $0^+$, in Eq. (108). On the other hand, it is seen from Figs. 9 and 10 that when the ionization is almost counteracted by the recombination and the self-coagulation influence is excluded as imagined the density profile of cations is almost flattened, except for the fractional decrease in the thin region of curve edge. It means that the free diffusion proportional to the density gradient is negligible. So, the component of flux given by the free diffusion is omitted in Eq. (109). The divergence of drift flux is then calculated in Eq. (110). As seen further, the term related to the divergence of electric field is omitted since the uniformly distributed cations are treated as continuum media and the non-source field is assumed here. This is different with the Eq. (94) of Sec. (3.2.3), where the term of electric field divergence is specifically retained. It is because the physical essence of the chemical sink of recombination to cations that causes the internal transport is analyzed there. The left term of flux divergence in Eq. (110) is shown to be the product of three physical quantities, i.e., the mobility, the non-source electric field and the cation density gradient. It is noted that the inclusion of density gradient here in the flux divergence is not contradicted to the assumption of uniform cation media since it, together with the inertia, represents the wave. Concretely, when a disturbance is imposed onto the continuum media and wave is transmitted in it. Neither the simulation nor the theory of present article is focused on the wave dynamics and the continuum assessment of cations that can transmit the wave just depicts the potential property of plasma in the scope of ellipse theory (see next).

$$\frac{\partial n_+}{\partial t} + \mu_+ \vec{E} \cdot \nabla n_+ = 0, (111)$$

$$\mu_+ \vec{E} = \frac{u_+}{E} \cdot E \vec{e}_0 = u_+ \vec{e}_0 = \vec{u}_+, (112)$$

$$\frac{\partial n_+}{\partial t} + \vec{u}_+ \cdot \nabla n_+ = 0, (113)$$

$$\vec{u}_+ \sim u_+ \vec{e}_x, \nabla n_+ \sim \frac{dn_+}{dx} \vec{e}_x, (114)$$

$$\frac{\partial n_+}{\partial t} + u_+ \frac{\partial n_+}{\partial x} = 0, (115)$$

$$\frac{\partial n}{\partial t}+u\frac{\partial n}{\partial x}=0, \text{(116)}$$

$$n\sim \exp\left[i\left(\omega t-kx\right)\right], u=\frac{\omega}{k}, \text{(117)}$$

$$\frac{\partial^2 n}{\partial t^2}=u^2\frac{\partial^2 n}{\partial x^2}. \text{(118)}$$

Substituting the flux divergence already operated into the Eq. (107), a new expression is obtained for the continuity equation in Eq. (111). In Eq. (112), the product of mobility and electric field vector is treated as a velocity vector, by considering the meaning of mobility, i.e., the velocity magnitude obtained by the acceleration of unit electric field intensity. The new continuity equation of Eq. (111) is then further reformed in Eq. (113). In Eq. (114), the velocity vector is confined in one-dimensional space, e.g., along the $\vec{e}_x$ direction and the density gradient is assumed to be along this direction as well. Accordingly, the dot product of Eq. (113) turns to be $u_+\frac{\partial n_+}{\partial x}$ shown in Eq. (115). In Eq. (116) the subscripts of two variables are omitted and in Eq. (117) the wave format of density variable is assumed and meanwhile the velocity is expressed as $u\sim\frac{\omega}{k}$, which represents the wave velocity, actually. It is easy to validate that the wave format of density variable satisfies both the Eqs. (116) and (118), i.e., the continuity and wave equations, respectively.

$$\frac{\partial n_-}{\partial t}+\nabla\cdot\vec{\Gamma}_{-,f}=-n_-n_+K_{rec}, \text{(119)}$$

$$\frac{\partial n_-}{\partial t}=-\nabla\cdot\vec{\Gamma}_{-,f}-n_-n_+K_{rec}=D_-\nabla^2 n_- - kn_-, \text{(120)}$$

$$D_-\nabla^2 n_- - kn_- =0, \text{(121)}$$

$$\frac{\partial n_-}{\partial t}=0. \text{(122)}$$

In Eq. (119), the unsteady-state continuity equation of anions with the free diffusion flux, $\vec{\Gamma}_{-,f}$, and pure negative chemical source that describe the two necessary conditions for the self-coagulation is presented. In Eq. (120), the anion continuity equation is reformed and the quasi- Helmholtz equation components are collected. Letting the right side of reformed continuity equation equal to zero the quasi- Helmholtz equation is built, in Eq. (121). It describes the self-coagulation theory that has been illustrated in Sec. (3.3.3). Correspondingly, the left side of reformed continuity equation in Eq. (120), i.e., the inertial term, should be zero as well shown in Eq. (122). It implies when the self-coagulation is occurred the inertia of anions in this process disappears. This is different with the picture of ellipse theory that is illustrated in Eqs. (107-118), i.e., the inertia and advection are self-balanced. At holding the inertia, when a disturbance is imposed to the plasms of ellipse regime the plasma forms either the Langmuir (caused by electrons) or acoustic (caused by ions) oscillations [54]. These oscillations represent the collective interaction of plasma and at certain conditions the electron and ion *waves* are formed. Nevertheless, in the self-coagulation regime, the inertia is lost and the plasma does not hold the collective interaction of plasma anymore. In a word, the plasma of self-coagulation regime behaves somehow like *particle*. Considering the significant difference with respect to the inertia, the sustaining mechanism of plasma in the self-coagulation regime is defined as a tight self-balance.

**(3.3.6) Interdisciplinary meanings of discharge structure hierarchy**
**(a) Astro-structure evolved from the coagulated body in the parabola**

The charge density profile of transport dominated regime is given in Fig. 5(a) and upon comparing this figure to the Fig. 11(a) it is found that there is no sheath around the coagulated structure in the parabola profile of electronegative core. The coagulated and dispersed plasmas are connected in the parabola profile and the coagulated body consumes the ions of surrounding plasma through the recombination. As mentioned in sub-Sec. (3.3.3), the coagulated body of parabola is given by the self-balance between the free diffusion and recombination loss, which is analogous to the formations of astro-structures that are also given by the self-balances. As far as we know, the white dwarf star, the neutron star and the fixed star are formed by the balances of the universal gravitation with the electron degeneracy pressure, the neutron degeneracy pressure, and the radiation pressure [45-47], respectively. The free diffusion of self-coagulation represents the thermal pressure and it plays the role of the pressures of electron degeneracy, neutron degeneracy and radiation of astronomy. Besides, the recombination loss represents the universal gravitation since they both have the nonlinear quadratic and dissipative terms, i.e., $-n_+ n_- k_{rec}$ and $-\frac{Gm_1 m_2}{r^2}$. Herein, $n_+$ and $n_-$ are the densities of cations and anions, and $k_{rec}$ is the recombination rate coefficient of them. $G$ is the universal gravitation constant, $m_1$ and $m_2$ are the masses of two objects in the field of universal gravitation, and $r$ is the distance between them. The meaning of dissipation is that the nonlinear mechanisms block the occurrence of certain events, like the gaseous discharge that generates species and the inflations of stars that deliver the mass outward. It is believed that once the nonlinear dissipative mechanisms dominate the continuums that are investigated tend to form the singularities in the otherwise dispersed density profiles, like the coagulated bump and the celestial bodies respectively found in the laboratory and universe. In the inertial Ar ICP, the nonlinear and dissipative term of energy is represented by the inelastic collision energy loss of electron-impact, which is expressed as $-\varepsilon_{th} n_e k_i(T_e) n_0 \sim -\varepsilon_1 \varepsilon_2$ at the relations of $\varepsilon_{th} n_e \sim \varepsilon_1, k_i(T_e) \sim \varepsilon_2$. This may explain the electron temperature bumps in the profiles given by the fluid simulations in Ref. [4]. The dissipative meaning of the collisional energy loss term is that it blocks the energy deposition of Ohm heating process illustrated in Sec. (2). Herein, $\varepsilon_{th}$ is the threshold of inelastic collision, $n_e$ is the electron density, $k_i(T_e)$ is the ionization rate coefficient that is a function of electron temperature $T_e$, and $n_0$ is the density of target neutral atoms of collisions. The coagulated structures given by the nonlinear and dissipative mechanism analyzed here is similar to the soliton model given by the Korteweg de Vries equation (abbreviated as KdV equation) that carries the nonlinear advective term that is quadratic [76].

**(b) The earth structure model from the coagulation and stratification**

It is seen that the coagulated body, the electronegative core, the electropositive halo and the sheath of chamber border are distributed from the center to the border of a stratified profile at low pressure. As analyzed, the order of parabola and stratification dominated discharge structure corresponds well to the earth structure that consists of the inner core with the solid metal, the outer core with the liquid metal, the mantle with the metasilicate, and the crust with the rock [49]. The *solid* inner core of earth with the constituents of oxidized iron and nickel corresponds to the *coagulated* ion-pair plasma of electronegative core. The *liquid* outer core of earth with the constituents of oxidized iron and nickel as well corresponds to the *dispersed* ion-pair plasma of electronegative core. The mantle with the constituent of *chemically unreactive* metasilicate corresponds to the electropositive halo that is essentially *inertial* plasma as well. The crust of earth with the *rock* constituent corresponds to the sheath at the chamber border that surrounds the stratified and coagulated structures and holds the *robust* potential. The ingredient, texture and chemistry of the discharge structure of low pressure and the earth structure of geophysics are similar, which reveals that precursor of our earth is probably the electronegative and laboratory plasma that is stratified and meanwhile coagulated.

**(c) For the nuclear structure evolved from the coagulation and ellipse**

The coagulated and dense body in the ellipse is treated as the neutron [50] that holds most of the nuclear mass. As analyzed in Ref. [26], the coagulated body inside the ellipse is given by the ambi-polar self-coagulation of ions, i.e., the anions are self-coagulated and the cations are absorbed by the anions that coagulate to satisfy the electrical neutrality (embodying the plasma collective interaction), and the ambi-polar self-coagulation potential is formed since the masses of cations and anions are the similar. So, the coagulated body is concisely neutral, which is in accord to the property of neutron. The elliptic and sparse background plasma is treated as the proton [51] that is electropositive. As mentioned in the next Sec. (3.4), the quasi-chemical potential is constructed to sustain the plasma in the self-balanced regime of high pressure and the chemical potential has the similar structure as the ambi-polar diffusion potential before it collapses. The potential is a barrel and so the electric field given by the potential gradient is directed to the chamber border. In the two-dimensional cylindrical coordinates, the radial and axial electric field components are both positive and the charge density of the potential barrel is positive as well, which is predicted by the Poisson's equation as stated in Eqs. (123-125). It is noted that the potential of either the ambi-polar diffusion potential or the quasi-chemical potential is linear mechanism that is used to keep the plasma's neutrality. This is different to the nonlinear mechanism of blue sheath that is described below.

$$\nabla \cdot \vec{E} = \frac{\rho}{\varepsilon_0}, \quad (123)$$

$$\frac{\partial E_z}{\partial z} + \frac{1}{r}\frac{\partial}{\partial r}\left(rE_r\right) = \frac{\rho}{\varepsilon_0}, \quad (124)$$

$$E_z > 0, E_r > 0, \rightarrow \rho > 0. \quad (125)$$

In our work of Ref. [26] related to the coagulation, it is found that the blue sheath surrounds the coagulated body in the ellipse, which then represents the mesotrons [52, 53] that confines both the neutron and proton in the nucleus. The Debye's potential [54] of blue sheath has the same expression as the Yukawa's potential [55, 56], $\psi(r)=\frac{\alpha}{r}e^{-\frac{r}{R}}$, that describes the mesotron and represents the short-range nuclear force. Herein, $\alpha$ and $R$ are constants. The Debye's shield potential formulae that represents the blue sheath, i.e., the nonlinear mechanism and the charge separation region in the plasma, is given in the Appendix B. As mentioned in Ref. [26], the blue sheath is formed by the anions when their directional velocity exceeds the Bohm's velocity threshold, at which the plasma neutrality is broken. This is different to the ambi-polar diffusion process which is aimed at sustaining the plasma's neutrality at establishing the weak potential barrel. Besides, since the electropositive halo is suppressed, the residual electrons of this region are absorbed by the electropositive chemical potential, which represents the extranuclear electrons of atom model [57].

**(d) For the wave-particle duality**

The coagulated body in the ellipse represents the particle model and the ellipse represents the wave model, as mentioned in the sub-Sec. (3.3.5). The new model for the wave-particle duality can explain the diffraction of electrons of Ref. [58]. When the slit is small enough, the intermolecular force of slit material can be treated as the strong electromagnetic interactions that are imposed onto the tiny coagulated body, i.e., particle, which then destroys the self-coagulation of plasma that is based on the free diffusion. Hence, the wave property of *electrons* is exhibited, which have been modeled as the mixtures of coagulated body and dispersed ellipse reported here.

### (3.4) New transports of electrons at high pressure

### (3.4.1) Collapse of electrons potential

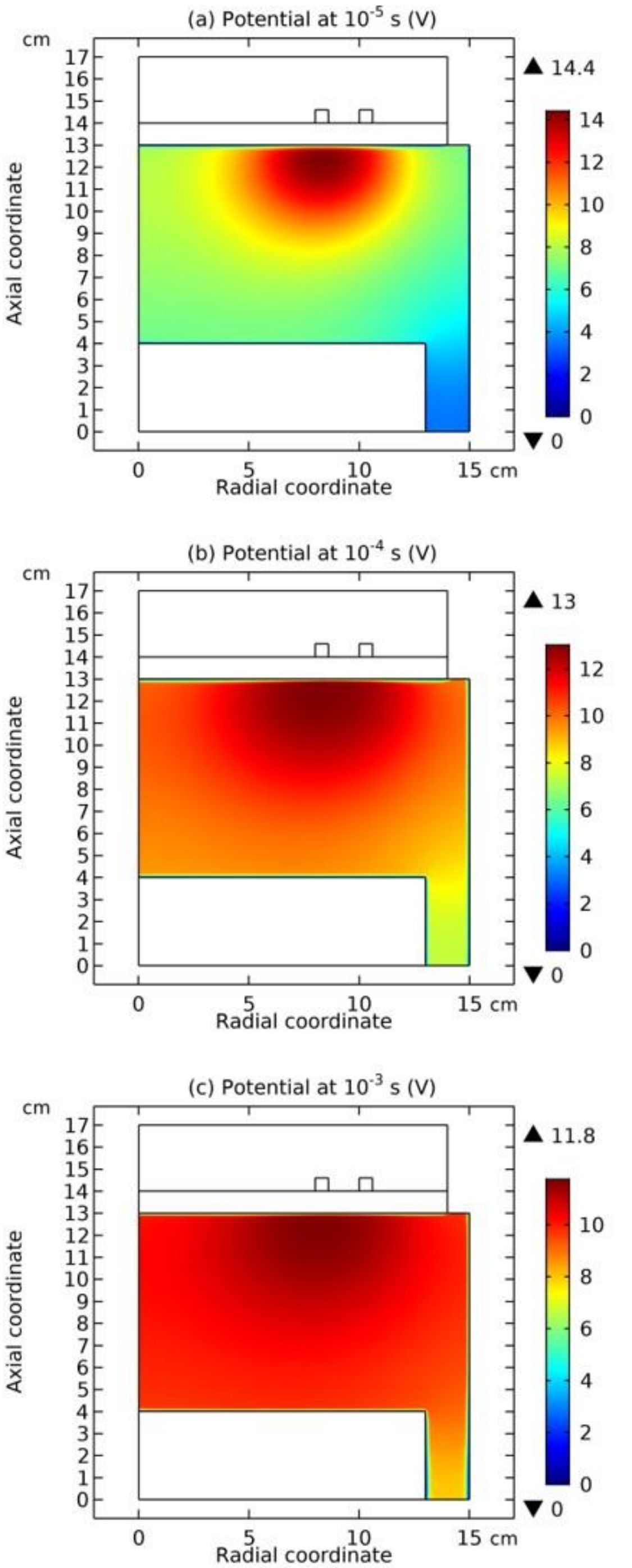

Figure 14. Plasma potential two-dimensional profiles at different simulated times, (a) $10^{-5}$ s, (b) $10^{-4}$ s and (c) $10^{-3}$ s, respectively, given by fluid model at 90mTorr. The discharge power is 300W and the ratio of reactive $SF_6$ in the gas mixture is 10%.

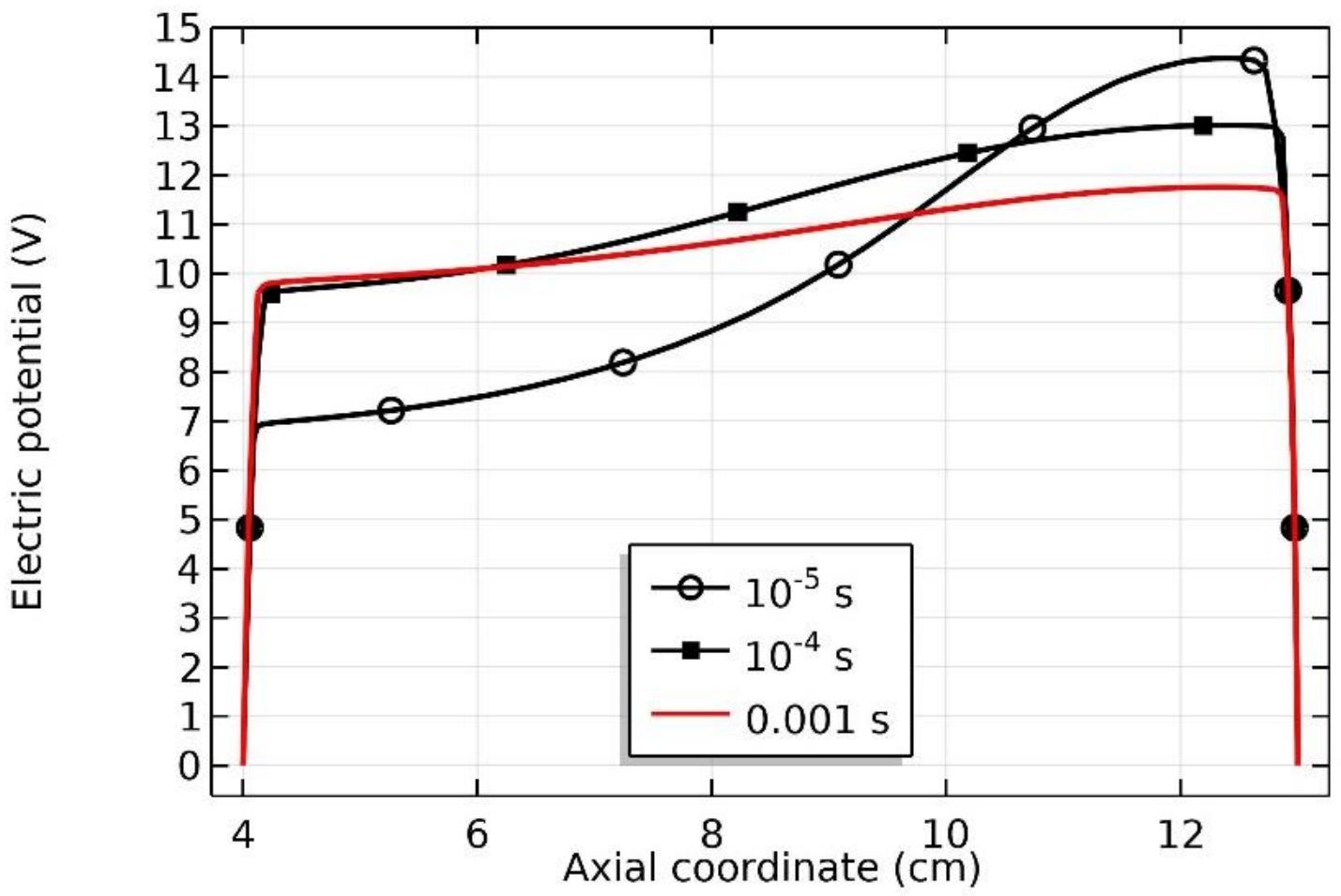

Figure 15. Plasma potential axial profiles at different simulated times, (a) $10^{-5}$ s, (b) $10^{-4}$ s and (c) $10^{-3}$ s, respectively, given by fluid model at 90mTorr. The discharge power is 300W and the ratio of reactive $SF_6$ in the gas mixture is 10%.

In Fig. 14, the simulated plasma potential two-dimensional profiles of 90mTorr by fluid model are shown at different simulated times, $10^{-5}$ s, $10^{-4}$ s and $10^{-3}$ s, respectively. Accordingly, the axial profiles of it at the three same times are shown in Fig. 15. The radial position, 8.5 cm, is selected so that these axial profiles cut through the anions self-coagulation area under the coil (refer to Figs. 14 and 11(c,f)). As mentioned in Sec. (3.3.1), at the simulation initial, such as $10^{-5}$ s in Fig. 14(a), the electron density is localized under the coil due to the strong attachments depletion. So, the electrons potential is also localized therein. It is noticed that at the initial time, the potential and electrons density still satisfy the Boltzmann's balance. Upon increasing the simulated time, as seen from Figs. 14(b,c), the predominant electrons potential gradient is swiftly weakened. Specifically, in Fig. 15, except the steep border sheath potential drops at the curves ends, the bulk potential drop is seen to descend from ~8eV at $10^{-5}$ s to ~2eV at $10^{-3}$ s. This collapsed potential in Fig. 14(c) is neither the anions potential (in a room temperature range) nor the electrons potential (~10eV, refer to the potential gradient of Fig. 12(a) in the electropositive halo). The anions and electrons potentials are caused by ambi-polar diffusions, at the high electronegativity and zero electronegativity (i.e., electropositive plasma), respectively. In a word, they are transport dominated potentials. The reasons for forming the collapse of plasma potential herein are similar to the ellipse and self-coagulation theories, i.e., the chemistry of electrons start to play roles by means of either forming a transformed *attachment* flux or self-coagulation. As seen in the Fig. 16(c) of next section, the electrons chemistry of 90mTorr at the steady state is a mixture of positive source in the localization center and negative source in the periphery of localization area. As analyzed, on one hand, the strong positive source existed in the rf- field area is used to sustain the overall complex discharge, and so the collapsed potential cannot tend to the room temperature magnitude. On the other hand, the one order smaller but wide spread negative source in the periphery of rf- field area (see Fig. 16(c)) can chemically constrict the diffusion of electrons that are generated from the positive source through depletion, and so the potential has not to be that high anymore. This self-balance process is more or less chemically controlled and the collapsed potential formed is thereby a quasi- *chemical* potential.

**(3.4.2) Weak self-coagulation of peripheral electrons**

In Fig. 16, the simulated two-dimensional profiles of net chemical sources of electrons by fluid model are shown at different pressures, 10mTorr, 50mTorr and 90mTorr, respectively. In the picture of Fig. 16(a) the rainbow legend is normally exhibited, while in the pictures of Fig. 16(b,c) the legends are truncated for clearly demonstrating the negative sources of electrons in the peripheries of localization areas. More details about the truncations can be found in the figure caption. It is seen that at low pressure, 10mTorr, the net source of electrons is predominantly positive, with its negative source magnitude four orders lower than the positive source. However, at the high pressures, 50mTorr and 90mTorr, obvious negative sources are existed in the periphery, which is only one order lower than the positive source. As mentioned in Sec. (3.4.1), the negative source of electrons leads to the collapse of potential and hence forms the type of *chemical* potential. Since the potential is not ambi-polar anymore, it is then belonged to the quasi- free diffusion. Referring to the self-coagulation theory illustrated in Sec. (3.3.3), weak self-coagulation behavior is then caused by the free diffusion and peripheric negative source of electrons discovered herein. As seen next, this chemical behavior will change the dynamics and lead to a new statistic balance of electrons, non- Boltzmann type.

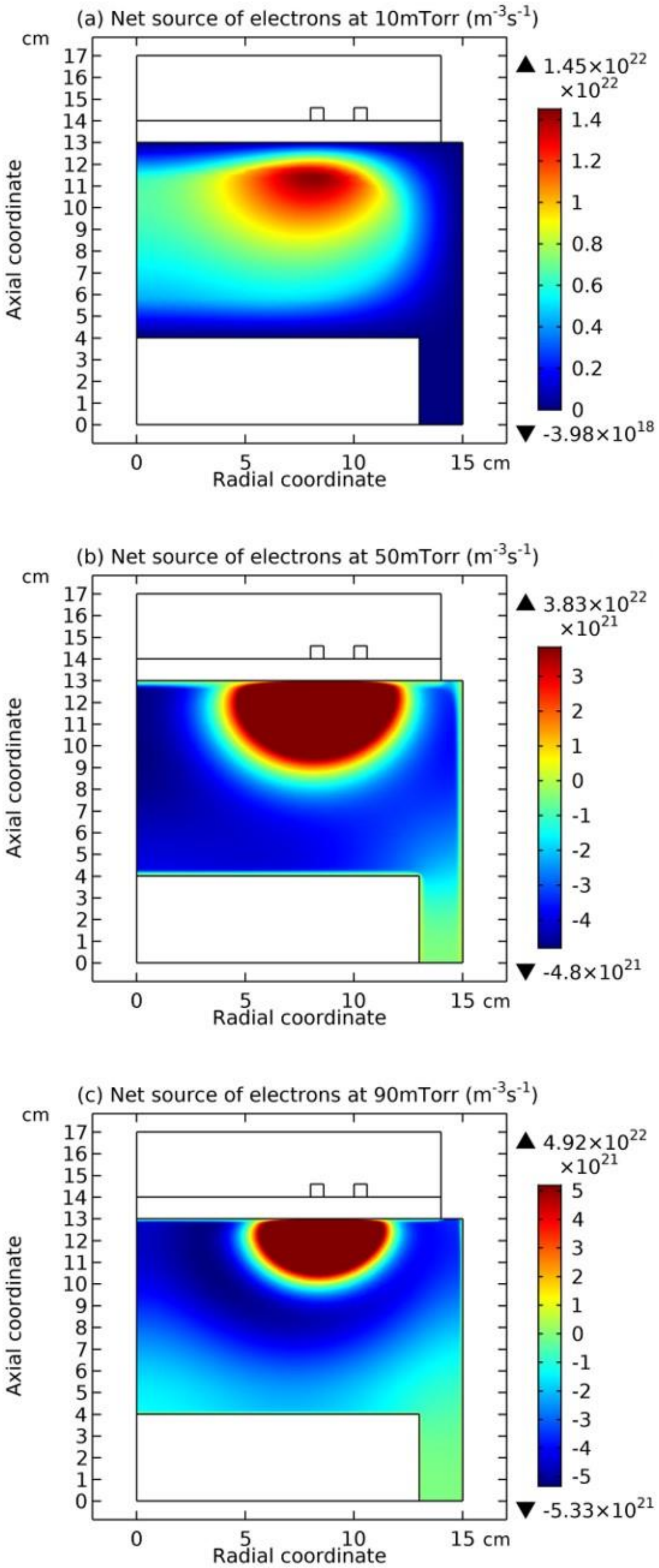

Figure 16. Simulated two-dimensional profiles of net chemical sources of electrons at (a) 10mTorr, (b) 50mTorr, and (c) 90mTorr, respectively, by fluid model. In panel (a), the rainbow legend of 10mTorr source is normally exhibited. Nevertheless, in panels (b) and (c), the highest levels of rainbow legends of 50mTorr and 90mTorr sources are truncated, at $3.8\times10^{21}$ $m^{-3}s^{-1}$ and $5.2\times10^{21}$ $m^{-3}s^{-1}$, respectively, for clearly demonstrating the negative parts of sources. It is noticed

that their realistic maxima are $3.83\times10^{22}$ m$^{-3}$s$^{-1}$ and $4.92\times10^{22}$ m$^{-3}$s$^{-1}$, respectively. The legend untruncated figures of 50mTorr and 90mTorr net electrons sources are shown in the supplementary material. The discharge power is 300W and the ratio of reactive $SF_6$ in the gas mixture is 10%.

### (3.4.3) Its non- Boltzmann balance

Along with the potential collapse, the electrons density profile smooths up in Fig. 17. This is because the ambi-polar electrons potential is strong and constricts the electrons by physical factor at $10^{-5}$ s. Nevertheless, after the weak self-coagulation of peripherical electrons is occurred, the electrons are mainly constricted by chemical factor, at $10^{-4}$ s and $10^{-3}$ s. The chemical interaction is gentle and so the electrons density profile is less localized. In Fig. 18, the axial profiles of net negative electrons sources out of the coil current rf- heating region are displayed, thus validating the chemical scheme. It is still seen from Fig. 18 that even though the peripherical net source is negative, e.g., at $10^{-5}$ s, the self-coagulation will not happen when the ambi-polar potential is still dominated (i.e., before collapse), again indicating the importance of free diffusion in the chemical behavior. Moreover, in Fig. 19(a,b) the electrons deviate from the Boltzmann's balance both in the periphery and under the coil when the self-coagulation is happened. Herein, the potential exponential function is constructed as in the Fig. 4(d), but at a tunable electrons temperature in a unit of electron volt. The axial electrons density profile is obtained by dividing its right-terminal density value, which is then not normalized but has the value of one at the curve right end. The process is the electrons temperature is consistently tuned until the two maxima of exponential function and axial electrons density profile converge. The final electron temperatures that are tuned until satisfying the above requirement are found to be, 3.56 eV, 3.73 eV, and 3.66 eV, respectively, as illustrated in the figure. Different to Fig. 4(c,d) where only the relative changes can be analyzed, the absolute changes can be referred to herein.

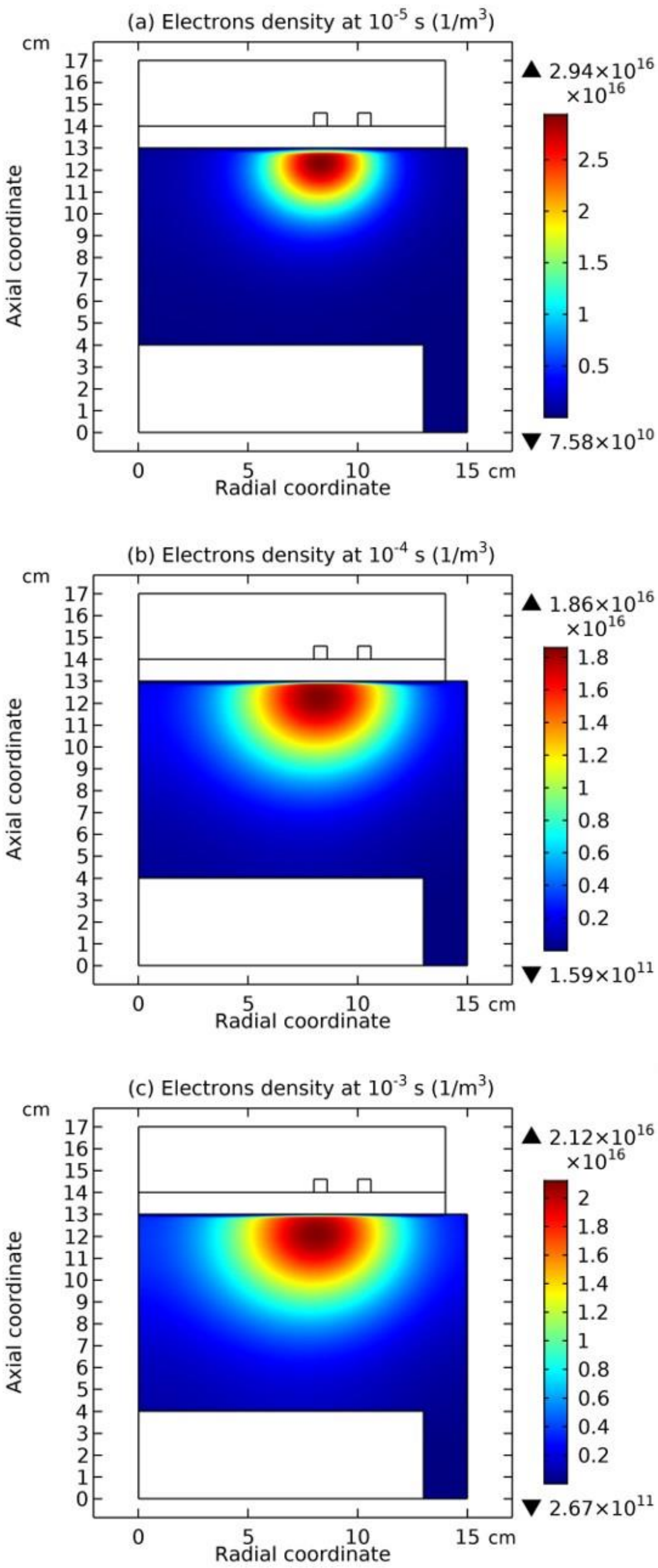

Figure 17. Simulated electrons density two-dimensional profiles of 90mTorr by fluid model, at (a) $10^{-5}$ s, (b) $10^{-4}$ s, and (c) $10^{-3}$ s, respectively. The discharge power is 300W and the ratio of reactive $SF_6$ in the gas mixture is 10%.

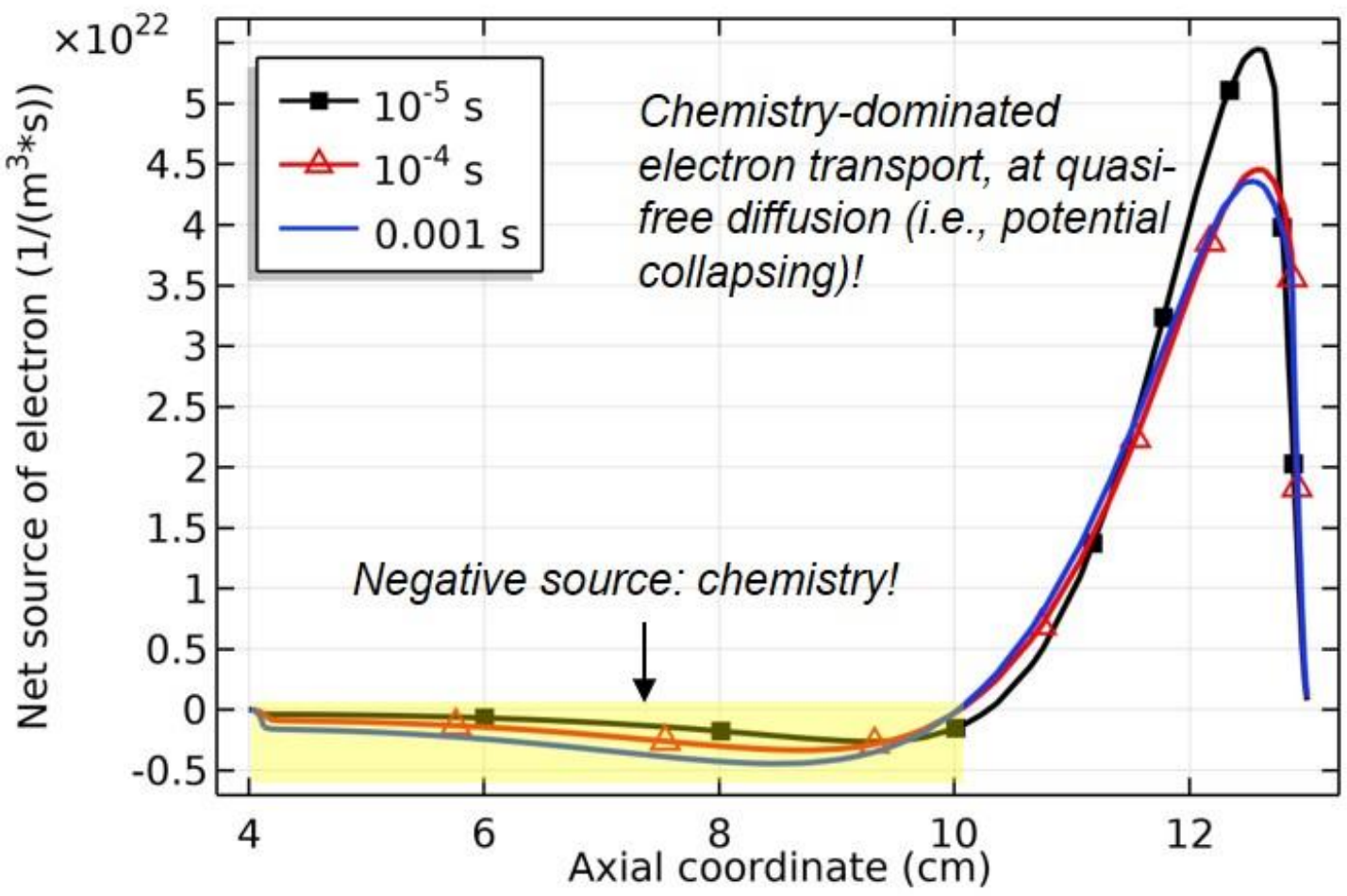

Figure 18. Simulated axial profiles of net sources of electrons at different times, by fluid model at 90mTorr. The discharge power is 300W and the ratio of reactive $SF_6$ in the gas mixture is 10%.

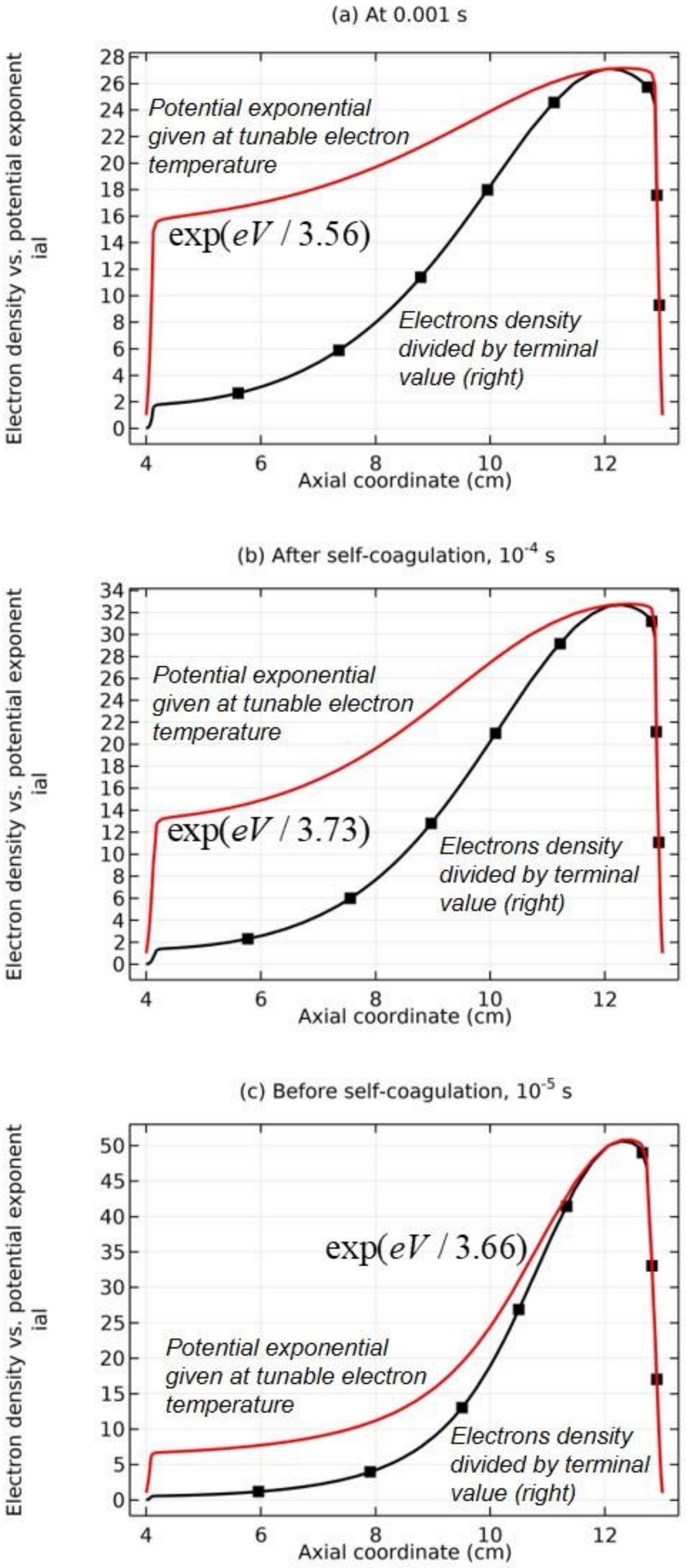

Figure 19. Simulated electrons density and the Boltzmann's balance of 90mTorr at different times, (a) $10^{-3}$ s, (b) $10^{-4}$ s, and (c) $10^{-5}$ s, respectively, by fluid model and along the axial direction. Herein, the self-coagulation is meant to the electrons. The electrons density is divided by its right terminal value and the potential exponential function constructed at tunable electron temperature represents the Boltzmann's balance. The electron temperatures that are tuned until the two maxima of electrons density and exponential function converge are listed in the figure. The discharge power is 300W and the ratio of reactive $SF_6$ in the gas mixture is 10%.

## IV. Conclusions and further remarks

The discharge structure of highly electronegative and radio frequency inductively coupled Ar/$SF_6$ plasma is built, by means of fluid model simulation. At low pressure, the chemistry significance is covered by the ambi-polar diffusion of plasma species, which is caused by the light electrons and accelerates the transport of heavy plasma species. The plasma of transport dominated regime is stratified into the electronegative core and electropositive halo, by means of double layer. The profiles of ions in the core region are parabolic and the anions therein satisfy the Boltzmann's balance. Together with the electron Boltzmann's balance, the plasma potential gradient of core is zero in an order of electron Volt, but is significant in an order of room temperature. The potential in this room temperature range is thereby defined as anion potential. The double layer can be modelled as capacitor macroscopically and as dipole microscopically. Macroscopically, the capacitor model of double layer blocks the mass exchange of electronegative core plasma and electropositive plasma of halo. Microscopically, the dipole model of double layer instead delivers the cations from the core to halo. The plasma of transport regime is an open system since the double layer and electropositive halo lead to the surface loss of plasma species.

At high pressures, the chemical effect becomes important since the transport of plasma species are slow. As analyzed, the recombination reaction is transformed into drift flux, hence counteracting the ambi-polar diffusion flux. The self-balance existing in the high pressure and electronegative plasma leads to the ellipse profiles of ions. Regarding the close feature of ellipse, the plasma of chemistry and transport balanced regime is thought to be closed system that does not need the shield of chamber border anymore. This idea is supported by the simulation that gives the shrunk double layer and halo, and the persistently expansive cores.

The simulation shows that the coagulated and astro- structures are inlayed in both the parabolic and elliptic ion profiles. By analyzing the simulation dynamics, the free diffusion and negative chemical source of anions are selected to construct the quasi-Helmholtz equation. The self-coagulation theory is built since the delta-type solution is obtained at solving this equation, analytically. Since the free diffusion is not specifically belonged to the plasma transport (which is ambi-polar type, usually), the self-coagulation discharge structure is thereby defined as chemistry dominated regime. This self-balance between the free diffusion and negative chemical source is a tight type due to the loss of inertia.

At the high pressure (90mTorr), the mixture of elliptic profile and coagulated anion inlayed therein implies the wave and particle duality of quantum mechanics, since the loose self-balance of ellipse holds the collective interaction of plasma but the tight self-balance of self-coagulation losses it. The former can be treated as continuum media where *wave* can be generated and transported in case of a disturbance while the latter is only treated as *particle* due to the loss of inertia. Besides, the interdisciplinary meanings of discharge structure hierarchy of electronegative plasma with the astrophysics, geophysics, and nuclear and atomic physics are described.

At high pressures, the electrons have their own self-coagulation in the periphery of coagulated body and the Boltzmann's balance of electrons is deviated. In this process, the novel chemical potential is created.

## Appendix A: Theory of the first kind of ellipse integral

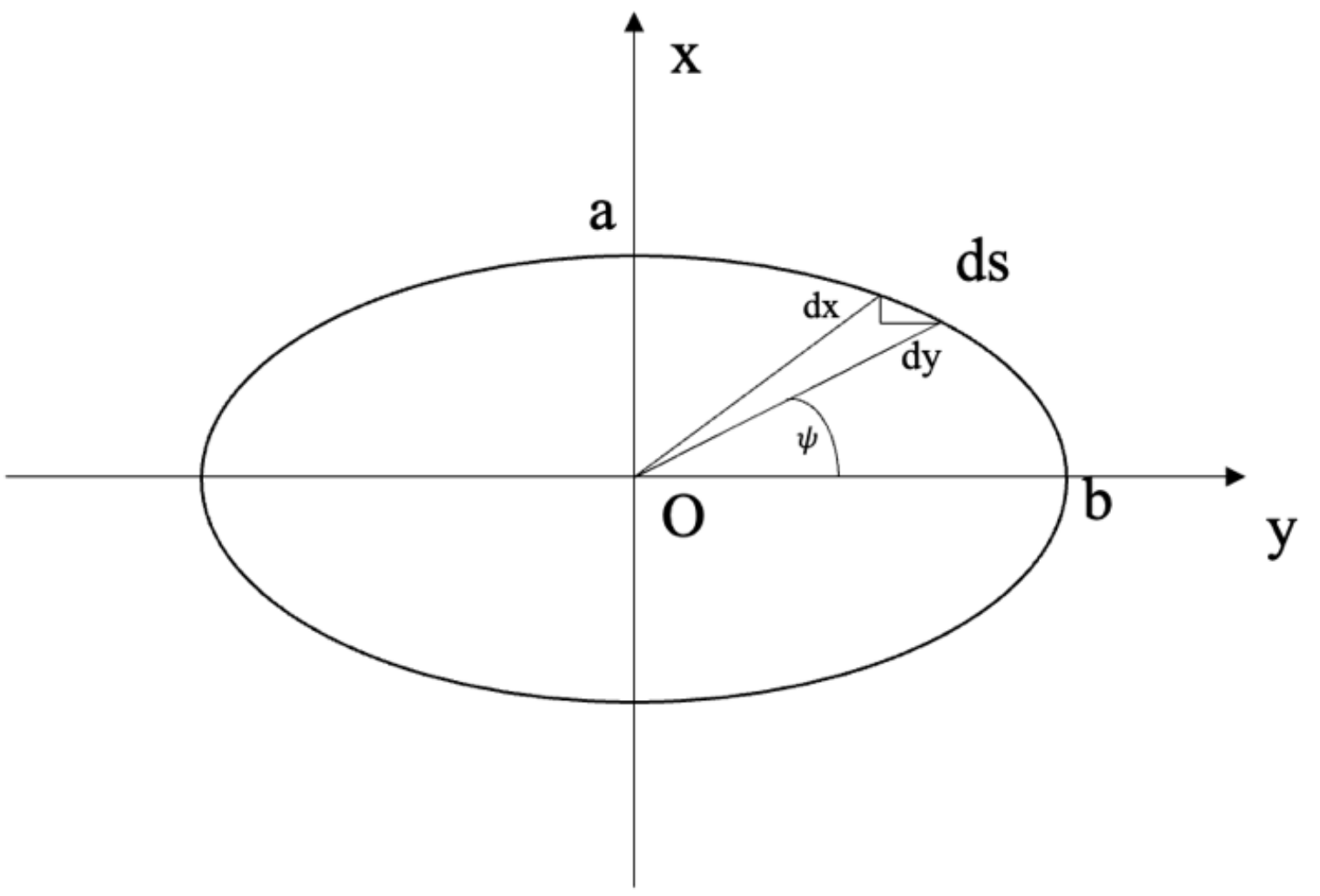

Figure A1. Schematic of ellipse and its coordinates, parameters and the infinitesimal arc length.

Consider an ellipse with its coordinates shown in Fig. A1. The parametric equations are given by

$$\begin{cases} x = a\sin\psi \\ y = b\cos\psi \end{cases}.\text{(A1)}$$

Then, the trajectory equation of this ellipse can be written as

$$\left(\frac{x}{a}\right)^2 + \left(\frac{y}{b}\right)^2 = 1.\text{(A2)}$$

From the ellipse, we have the infinitesimal arc length written as

$$(ds)^2 = (dx)^2 + (dy)^2.\text{(A3)}$$

It is stressed that we keep deducing the formulae in the first quadrant of the Cartesian coordinate. So, we have the following relations, $dx < 0, dy > 0$ and $x > 0, y > 0$.

Substituting Eq. (A1) into Eq. (A3) and taking the square root, we get

$$ds = \sqrt{a^2\cos^2\psi + b^2\sin^2\psi}\,d\psi.\text{(A4)}$$

The following trigonometric identity is used.

$$\sin^2\psi + \cos^2\psi = 1.\text{(A5)}$$

And the ellipse eccentricity is set as

$$k^2 = 1 - \left(\frac{b}{a}\right)^2.\text{(A6)}$$

Then substituting Eq. (A6) into Eq. (A4), we get

$$ds = a\sqrt{1 - k^2\sin^2\psi}\,d\psi.\text{(A7)}$$

Based on Eq. (A7), we can construct a new formula in Eq. (A8) and the right side of it represents the first kind of elliptic integral[ref].

$$\frac{1}{a}\frac{ds}{1-k^2\sin^2\psi}=\frac{d\psi}{\sqrt{1-k^2\sin^2\psi}}.\text{(A8)}$$

Substituting the Eqs. (A1-A3) into the left side of Eq. (A8), we get

$$\frac{a\sqrt{d^2x+d^2y}}{a^2\left(\frac{y}{b}\right)^2+b^2\left(\frac{x}{a}\right)^2}=\frac{d\psi}{\sqrt{1-k^2\sin^2\psi}}.\text{(A9)}$$

At this point, differentiating Eq. (A2) gives

$$dy=-\frac{b^2}{a^2}\frac{x}{\sqrt{b^2-\frac{b^2}{a^2}x^2}}dx.\text{(A10)}$$

Herein, it is stressed again that $dx<0$. Substituting the Eq. (A10) into the left-side numerator of Eq. (A9) and meanwhile eliminating the variable $y$ of the left-side denominator with Eq. (A2), we get

$$\frac{a\sqrt{d^2x+\frac{b^4}{a^4}\frac{x^2}{b^2-\frac{b^2}{a^2}x^2}d^2x}}{a^2\left[1-\left(\frac{x}{a}\right)^2\right]+b^2\left(\frac{x}{a}\right)^2}=\frac{d\psi}{\sqrt{1-k^2\sin^2\psi}}.\text{(A11)}$$

Substituting Eq. (A6) into the left side of Eq. (A11) and meanwhile simplifying its complicated expression, we get

$$a\frac{-dx\sqrt{1+\frac{b^2}{a^2}x^2\frac{1}{a^2-x^2}}}{a^2-x^2+\frac{b^2}{a^2}x^2}=a\frac{-dx\sqrt{\frac{a^2-k^2x^2}{a^2-x^2}}}{a^2-k^2x^2}=a\frac{-dx}{\sqrt{\left(a^2-x^2\right)\left(a^2-k^2x^2\right)}}.\text{(A12)}$$

Then, the Eq. (A11) is rewritten in Eq. (A13).

$$\frac{-adx}{\sqrt{\left(a^2-x^2\right)\left(a^2-k^2x^2\right)}}=\frac{d\psi}{\sqrt{1-k^2\sin^2\psi}}.\text{(A13)}$$

Herein, the left side of Eq. (A13) represents another expression of the first kind of elliptic integral[ref]. From Eq. (A2), the relationship between the variables, $x$ and $y$, can be obtained, and the left side of Eq. (A13) can be reformed in total with the variable, $y$. Then, we get

$$\frac{ady}{\sqrt{\left(b^2-k^2y^2\right)\left(b^2-y^2\right)}}=\frac{d\psi}{\sqrt{1-k^2\sin^2\psi}}.\text{(A14)}$$

By further utilizing the key term in Eq. (A8), i.e., $\frac{1}{a}\frac{ds}{1-k^2\sin^2\psi}$, and meanwhile correlating the Eqs. (A1, A8, A14), we get

$$\frac{ds}{a^2-k^2x^2}=\frac{dy}{\sqrt{\left(b^2-k^2y^2\right)\left(b^2-y^2\right)}}.\text{(A15)}$$

Next, the Eq. (A15) is integrated and Eq. (A16) is obtained. The left-side integral of Eq. (A16) is a curvilinear integral and hence unrelated to the variable, $x$, except for the integrating upper limit. It is integrated from an arbitrary point of the ellipse (when the arc length is zero) until the long half axis in the first quadrant, which finally gives rise to an integrated arc length, $s(x)$. The right-side integral is a one-dimensional integral of the variable, $y$. It is integrated from an arbitrary $y(x)$ (which is also a function of $x$) to the long half axis, $b$.

$$\frac{\int_0^{s(x)}ds'}{a^2-k^2x^2}=\int_{y(x)}^{b}\frac{dy'}{\sqrt{\left(b^2-k^2y'^2\right)\left(b^2-y'^2\right)}}.\text{(A16)}$$

The left side of Eq. (A16) can be written into a new function, $f(x)$, in Eq. (A17).

$$f(x)=\frac{\int_0^{s(x)}ds'}{a^2-k^2x^2}.\text{(A17)}$$

Now, the Eq. (A16) is correlated to the original integral listed in Eq. (A18) that needs to be authenticated to be an ellipse.

$$x=\frac{2D_+n_{e0}}{\beta_0}\frac{\alpha_0}{(\eta/3)^{1/2}}\times\int_{\alpha(x)}^{\alpha_0}\frac{d\alpha}{\left[\left(b\alpha_0-\alpha\right)\left(\alpha_0-\alpha\right)\left(a\alpha_0+\alpha\right)\right]^{1/2}}.\text{(A18)}$$

Herein, $\beta_0=\left[4D_+\left(K_{iz}+K_{att}\right)n_0n_{e0}^2\right]^{1/2}$, $K_{iz}, K_{att}$, $D_+$ and $\eta$ are all a function of $x$. The denominator under the square root on the right side of Eq. (A18) is a cubic power of α, but this does not affect the fact that this integral represents the first kind of elliptic integral, because the elliptic integral has a general form in Eq. (A19).

$\int F(x,y)dx$, where $y=\sqrt{Ax^4+4Bx^3+6Cx^2+4Dx+E}$, and $A=0$.(A19)

$$\frac{x\beta_0(x)\left[\frac{\eta(x)}{3}\right]^{\frac{1}{2}}}{2D_+(x)n_{e0}\alpha_0} = \int_{\alpha(x)}^{\alpha_0} \frac{d\alpha}{\left[(b\alpha_0-\alpha)(\alpha_0-\alpha)(a\alpha_0+\alpha)\right]^{1/2}}. \text{(A20)}$$

The Eq. (A18) is reformed in Eq. (A20). Upon comparing the Eq. (A20) to the Eq. (A16), it is seen that $\alpha(x)$ can indeed be represented by means of $y(x)$, which is just an ellipse function, as illustrated in Eq. (A2).

**Appendix B: The formulae of Debye's shield potential [77] in electronegative plasma.**

It is well known that the potential of a point charge, $q$, in the vacuum environment is the Coulomb's potential, as illustrated in Eqs. (B1, B2). Herein, the potential zero point is selected at the position that is infinitely far away from the point charge, which can be taken as the boundary conditions for solving this equation.

$$\nabla^2\phi(r) = 0, \text{(B1)}$$

$$\phi(r) = \frac{q}{4\pi\varepsilon_0 r}. \text{(B2)}$$

When in the electronegative plasma, besides for the point charge, the probable charge density of plasma needs to be considered. So, the above Laplace equation is transformed into the Poisson's equation, as illustrated in Eq. (B3), where $n_e(r)$, $n_-(r)$, and $n_+(r)$ represent the electron density, anion density and cation density, respectively.

$$\nabla^2\phi(r) = -\frac{\rho}{\varepsilon_0} = \frac{e}{\varepsilon_0}\left[n_e(r)+n_-(r)-n_+(r)\right]. \text{(B3)}$$

Herein, the electron, anion and cation densities are assumed to satisfy the Boltzmann's balance, as illustrated in Eqs. (B4-B6), where $n_e^{(0)}$, $n_-^{(0)}$, and $n_-^{(0)}$ represent the electron density, anion density, and cation density at the position of $\phi = 0$, i.e., infinitely far field. $T_e$ is the electron temperature and $T_i$ represents both the cation and anion temperatures at the assumption of thermal balance of ions. $K_B$ Is the Boltzmann constant.

$$n_e\left(r\right)=n_e^{(0)}\exp\left(\frac{e\phi}{K_BT_e}\right), \text{(B4)}$$

$$n_-\left(r\right)=n_-^{(0)}\exp\left(\frac{e\phi}{K_BT_i}\right), \text{(B5)}$$

$$n_+\left(r\right)=n_+^{(0)}\exp\left(-\frac{e\phi}{K_BT_i}\right). \text{(B6)}$$

It is shown that the self-coagulation of electrons happens far away from the coil and hence it does not influence the assumption herein near the coil. Both the cation and anion are in the regime of self-balance and so their Boltzmann balance assumptions are reasonable. In addition, the blue sheath that represent the Debye's potential is occurred in the event of self-coagulation and in this process only the free diffusion transport component is considered, which verifies the sparse or the idealized gas condition, $e\phi \ll K_BT_e, K_BT_i$. At the sparse condition, the exponential functions of the above three Boltzmann balances are expanded to be the Taylor's series and only the zero and one-order terms are kept, and then we obtain the approximated expressions for the electron, the anion and the cation densities, as illustrated in Eqs. (B7-B9)

$$n_e\left(r\right)=n_e^{(0)}\left(1+\frac{e\phi}{K_BT_e}\right), \text{(B7)}$$

$$n_-\left(r\right)=n_-^{(0)}\left(1+\frac{e\phi}{K_BT_i}\right), \text{(B8)}$$

$$n_+\left(r\right)=n_+^{(0)}\left(1-\frac{e\phi}{K_BT_i}\right). \text{(B9)}$$

Substituting the above three approximated densities into the Poisson's equation of Eq. (B3) and meanwhile considering the electrical neutrality, $n_+^{(0)}=n_e^{(0)}+n_-^{(0)}$ we obtain,

$$\frac{1}{r^2}\frac{d}{dr}\left(r^2\frac{d\phi}{dr}\right)=\frac{\phi}{\lambda_D^2}, \text{(B10)}$$

$$\lambda_D=\left(\frac{\varepsilon_0K_BT_eT_i^2}{n_e^{(0)}e^2T_i^2+n_-^{(0)}e^2T_eT_i+n_+^{(0)}e^2T_eT_i}\right)^{1/2}. \text{(B11)}$$

Herein, the introduced constant, $\lambda_D$, in Eq. (B11) is called the Debye's shield length. The general solution of Eq. (B10) is

$$\phi(r) = \frac{A}{r}\exp\left(-\frac{r}{\lambda_D}\right) + \frac{B}{r}\exp\left(\frac{r}{\lambda_D}\right). \text{(B12)}$$

Herein, the coefficients, $A$ and $B$, are determined by the boundary conditions, i.e., $\phi = 0$ at $r \to \infty$ and $\phi = \frac{q}{4\pi\varepsilon_0 r}$ at $r \to 0$. Then, the final definite solution of the Poisson's equation is given as the Debye's potential in Eq. (B13).

$$\phi(r) = \frac{q}{4\pi\varepsilon_0 r}\exp\left(-\frac{r}{\lambda_D}\right). \text{(B13)}$$

When comparing the Eqs. (B2, B13), it is seen that the Debye's potential decays faster than the Coulomb, due to the shield of plasma to the point charge. Besides, $\lambda_D$ represents the length of charge separation in the plasma, i.e., the appearance of blue sheath and its width.

## Acknowledgement

The authors are thankful to Profs. Lichtenberg and Lieberman since they have established the parabola and ellipse theories in the 90s years of last century. Our self-consistent fluid simulation of $Ar/SF_6$ ICP source predicted well the parabolic and elliptic profiles of ions that are given by the theoretic analysis. Moreover, the simulation generated the coagulated pictures of ions and thereby led to the self-coagulation theory. The systematic and theoretic framework of parabola, ellipse and self-coagulation is correlated by three conditions of the parameter defined in Eq. (78), i.e., $\eta \ll 1$, $\eta \to 1$, and $\eta \gg 1$.